# NONUNIFORM STATES IN FERROELECTRICS AND ANTIFERROELECTRICS. STATES WITH NEGATIVE INTERPHASE ENERGY.

V. M. Ishchuk, Science & Technology Center "Reaktivelektron" of the National Acad. of Sci. of Ukraine, Donetsk, Ukraine

V. L. Sobolev, Department of Physics, South Dakota School of Mines & Technology, Rapid City, SD 57701 USA

**Abstract**

Results of detailed investigations of stability of phases in substances with the small difference in the energies of the ferroelectric and the antiferroelectric types of dipole ordering are presented. It is shown that interaction of locally separated domains of the ferroelectric and antiferroelectric phases stabilizes a nonuniform state of the samples. The analysis of physical phenomena caused by the presence of domains of coexisting ferroelectric and antiferroelectric phases in different dipole ordered substances is presented. The consideration of peculiarities of behavior of systems with nonuniform state of coexisting phases under the action of changing external thermodynamic parameters (temperature, pressure, electric field, and chemical composition) is given.

# INTRODUCTION

Condensed media physics often deals with nonuniform substances that contain the domains of two coexisting phases, for instance, the substances characterized by the first order phase transitions (PT). Inside the hysteresis region between the lability boundaries of each of the phases (in the diagram of phase states), which participate in the PT, initiation of the stable state out of the metastable one occurs by nucleation, and within a certain interval of thermodynamic parameters the domains of the two phases coexist. Nowadays the presence of the interphase boundary (the boundary between the domains of coexisting phases) is considered to increase the free energy of the discussed system. In other words, the energy of the interphase boundary is positive (though this fact has not been proved so far).

However, a situation when the said rule seems to be violated is known at present. This example concerns type II superconductors placed in magnetic field. Within a certain field interval (between the first and the second critical fields) the normal (resistive) and superconducting phases co-exist [1, 2]. The nonuniform state of type II superconductors is caused by negative surface energy of the boundary between the two phases. However, such a phenomenon is observed only in magnetic field and is a consequence of the energy redistribution between the sample and the field: the energy of the nonuniform system decreases, whereas the density of the magnetic field energy rises over the whole space. The resulting energy of the system "superconductor – magnetic field" increases, and the superconductor's nonuniform state is maintained due to the energy of the source of electric current, which creates magnetic field. The evidence of such rise of total energy is the fact that the superconductor regains its uniform state when the magnetic field source is switched off (i.e. in the absence of the field).

A natural question arises during consideration of the above example: whether there exist the substances able to create the field, which will lead to the effects observed for type II superconductors (i.e. the nonuniform state)? This question is to be answered as follows. Such substances do exist – these are magnetic and ferroelectric (FE) materials at the temperatures lower than the Curie point, and their state is advantageous from the viewpoint of energy. Spontaneous appearance of the order parameter leads to advent of a field conjugate to the order parameter (scattering field). Here it should be noted that such fields do not penetrate ferroelectrics: they are shielded by the charges leaking on the surface, however, for our consideration this is not essential. Among the mentioned substances, there exist such compounds for which the two-phase state of domains of coexisting phases may exist under certain conditions (certain values of thermodynamic parameters, i.e. temperature, field intensity, pressure etc.). These are the substances, which undergo ferromagnetic-antiferromagnetic or ferroelectric-antiferroelectric PT of the first order. Therefore, it seems very interesting to consider the influence of the field generated by the order parameter of the ferromagnetic or ferroelectric phase on the general state of the substance. For definiteness, let us dwell on a physical system characterized by the PT between the ferroelectric and antiferroelectric (AFE) phases in dipole ordered state.

Ii is shown in Chapter 1 that the stable (from the viewpoint of energy) non-uniform quasi-two-phase state may be realized in a substance under two conditions. First, if the difference between the free energies of the AFE phase and the FE phase is small within a wide interval of changes of external parameters. Second, if the interphase interaction of the domains of the phases is taken into account. Uniform state of any phase (FE or AFE) will be metastable at these values of thermodynamic parameters. The consideration of the problem is made on the base of phenomenological Landau theory of phase transitions. The relations between phenomenological parameters, which determine the stability the two-phase state are established.

The second chapter is devoted to the substances in which wide hysteretic regions and the discussed nonuniform states manifest themselves. The factors leading to the phase diagrams, which provide the possibility of realization of the stable two-phase state are found in the frames of the Landau theory of phase transitions. The discussion of the phase diagrams of these substances subjected to the action of factors of different physical nature is presented.

The third chapter contains experimental results supporting theoretical conclusions presented in the first and second chapters. The diagrams of the phase states for the $PbZr_{1-y}Ti_yO_3$ (PZT) based solid solutions are presented. The extended hysteresis regions of the FE-AFE phase transformations are present in these diagrams. The "composition-temperature" (Y-*T*), "pressure-temperature" (*P-T*), and "temperature-electric field" (*T-E*) phase diagrams for the above-mentioned solid solutions are obtained.

Experimental results obtained by the means of transmission electron microscopy that allow identifying the stability of the nonuniform two-phase state are also presented.

The fourth chapter covers the discussion of the AFE → FE transitions caused by DC field. It is shown that the non-uniform state of the coexisting FE and AFE phase domains is realized in the bulk of the substance. Its geometrical properties are analogous to those of the intermediate state in antiferromagnets or superconductors. However, the nature of such a state is quite different: it is the interaction of the domains of the coexisting phases.

The chapter five contains consideration of the processes, which occur near the interphase boundaries. Local mechanical stresses arising in the said region are shown to lead to local decomposition of the solid solution and to the processes of long-duration relaxation. The experimental results obtained on different substances for which the difference between the free energies of FE and AFE states is relatively small are presented.

The chapter six contains results of consideration of the paraelectric (PE) phase transitions near the triple FE-AFE-PE point. The model of these transitions taking into account the process of local decomposition of the solid solution near the FE-AFE boundary is discussed. The presented experimental results point to the presence of the two-phase FE+AFE states not only at low temperatures, but also at temperatures essentially higher than the Curie point.

In the seventh chapter, the properties of the substances that have an insignificant difference between the free energies of the FE and AFE states in the presence of the in AC field are considered. It is shown that at low and medium frequencies their dielectric properties are defined by the dynamics of the interphase boundary. The equation describing the dynamics of the interphase boundary is obtained and solved. It is demonstrated that the relaxation dynamics of domain boundaries is responsible for dielectric constant dispersion. The Vogel-Fülcher relation describing the behavior of dielectric constant at the phase transition into PE state near the triple FE-AFE-PE point is obtained.

The eighth chapter contains analysis of variety of experimental results associated with the so-called relaxor state of ferroelectrics. These results are analyzed based on the point of view that takes into account the non-uniform state of coexisting phases developed in this review. Special attention is paid to the analysis of the phase diagrams for the mentioned class of materials obtained at extreme values of external thermodynamic parameters such as pressure and electric field.

In the ninth chapter the discussion of properties of solid solutions that have one component being ferroelectric and another one being antiferroelectric is presented. These substances are often referred to as dipole glasses. As a rule, analyzing behavior of such substances one does not take into account the following factors:

– stability of the two-phase state due to negative energies of the interphase boundaries,

– local decomposition of the solid solutions near the said boundaries.

Experimental phase diagrams of such materials are studied in Chapter 9. It is shown that not all the substances which are referred to as "glasses" belong to them in fact.

The presentation of material of this review is given in such a way that in all cases theoretical models are immediately compared with available experimental data.

## 1. THEORY OF NONUNIFORM STATE

The thermodynamic potential of the infinite two-phase system may be presented as:

$$\varphi = \xi_1 \varphi_1 + \xi_2 \varphi_2 + W_{int}, \tag{1.1}$$

where $\varphi_1$ and $\varphi_2$ are the thermodynamic potentials of the phases; $\xi_1$ and $\xi_2$ are the volume shares occupied by each of the phases; and $W_{int}$ describes the interaction between the domains of these phases.

The energy of the interaction between phases can be expressed in the following form [3]:

$$W_{int} = (W_{1,2} + W_{2,1}), \tag{1.2}$$

$$W_{1,2} = \xi_2 \sum_{i,j} [E_{\eta_{1,i}}(x_2, y_2, z_2) \eta_{2,j}(x_2, y_2, z_2)]; \tag{1.2a}$$

$$W_{2,1} = \xi_1 \sum_{i,j} [E_{\eta_{2,j}}(x_1, y_1, z_1) \eta_{1,i}(x_1, y_1, z_1)]. \tag{1.2b}$$

In these formulas: $\eta_{1,i}$ and $\eta_{2,j}$ are the order parameters of the first and second phases, respectively (*i* and *j* are the numbers of the order parameters of these phases); $E_{\eta_{1,i}}(x_2, y_2, z_2)$ are the fields, induced in the domains of the second phase by the nonzero values of the order parameters of the first phase, $E_{\eta_{2,j}}(x_1, y_1, z_1)$ are the fields in the domains of the first phase, initiated by the nonzero values of the order parameters of the second phase. Based on the physical nature of the order parameters (or the coexisting phases), it is evident that not all the products entering in the expressions (1.2a) and (1.2b) are nonzero. This is a consequence of the symmetry of the order parameters and fields, which influence the order parameters. The only terms that are nonzero in the expressions (1.2a) and (1.2b) are those that transformed according to the representations containing the identity one.

The dependencies of the fields $E_{\eta_{1,i}}(x_2, y_2, z_2)$ and $E_{\eta_{2,i}}(x_1, y_1, z_1)$ on the order parameters, which generate them, have a complex form. To determine these dependencies one must take into consideration the physical nature of the order parameters, the particular shape of the domains, as well as the spatial distribution of the order parameters over the domain volume. This problem has to be solved by means of a self-consistent procedure, taking into account the fact that the fields change the spatial dependence of the order parameters, which, in their turn, define the fields.

Such a problem does not seem to be solved exactly at present and, therefore, for finding the solution various approximations are to be used. We shall restrict ourselves by the expansion into a series of the powers of $\eta_{\alpha,i}$:

$$E_{\eta_{1,i}}(x_2, y_2, z_2) = \xi_1 C_{1,i}(x_2, y_2, z_2)\eta_{1,i} + \cdots \tag{1.3a}$$

$$E_{\eta_{2,j}}(x_1, y_1, z_1) = \xi_2 D_{2,j}(x_1, y_1, z_1)\eta_{2,j} + \cdots \tag{1.3b}$$

Note that the considered fields are rather long-range and smoothly vary in space.

In the case when external fields are present, they may be taken into account in (1.2) in the usual way.

Consequently, the study of the behavior of the system of interacting coexisting phase domains is reduced to the investigation of the thermodynamic potential (1.1) with $W_{int}$ in the form (1.2), in view of the expressions (1.3) under the condition $\sum_\alpha \xi_\alpha = 1$ $(\alpha = 1,2)$. Our next step is the conventional procedure of minimization for the nonequilibrium thermodynamic potential

$$\varphi_\lambda = \varphi - \lambda\left(\sum_\alpha \xi_\alpha - 1\right), \tag{1.4}$$

where $\lambda$ is the indeterminate Lagrange multiplier. This procedure leads to the system of equations for the determination of the equilibrium values of the order parameters:

$$\xi_\alpha(\partial\varphi_\alpha/\partial\eta_{\alpha,i} + E_{\eta_{\alpha',i}}) = 0, \quad (\xi_\alpha \neq 0), \tag{1.5}$$

$$\varphi_\alpha + \eta_{\alpha,i}E_{\eta_{\alpha',i}} = \lambda = const, \quad \alpha,\alpha' = 1,2., \tag{1.6}$$

As seen from Eq. (1.6), the condition for the coexistence of the thermodynamically equilibrium structure of domains of coexisting phases is the equality of their thermodynamic potentials, with the account of the fields (1.3), but not the equality of the "bare" thermodynamic potential.

As one can see from (1.6) there is a peculiarity of the considered multiphase structure connected with the spatial dependence of the coefficients in (1.3). This peculiarity is observed inside the region separating the domains of the coexisting phases, which is wider than the usual domain boundary. The fields $E_{\eta_{\alpha,i}}$ are spatially dependent, and their intensity decreases while moving inside the domains of the other phase. Therefore, in these boundary regions the thermodynamic potentials differ from those, which characterize the inner regions of the domains. This fact testifies that the phase state inside the region separating the domains will not be similar to that in the inner regions of the adjacent domains. Physical considerations allow us to conclude that this state is transient between the states of the adjacent domains.

The simplest density of the nonequilibrium thermodynamic potential for particular example of the ferroelectric (FE) and the antiferroelectric (AFE) phases has the form [4, 5, 6]:

$$\varphi = \frac{\alpha_1}{2}P^2 + \frac{\alpha_2}{4}P^4 + \frac{\beta_1}{2}\eta^2 + \frac{\beta_2}{4}\eta^4 + \cdots + \frac{A}{2}P^2\eta^2. \qquad (1.7)$$

Here $\alpha_2$, $\beta_2$ and $A$ are positive values ($A > \sqrt{\alpha_2 \beta_2}$), $\alpha_1$ and $\beta_1$ may change their sign at the temperatures $T_{c,f}$ and $T_{c,af}$, respectively:

$$\alpha_1 = \alpha_0(T - T_{c,f}); \qquad \beta_1 = \beta_0(T - T_{c,af}). \qquad (1.7a)$$

While considering the thermodynamic potential in the form (1.7), it is to be taken into account that AFE PT is a structural transition into the state with the order parameter $\eta$, and the latter interacts with the polarization $P$, which is the order parameter of FE state [7]. If $T_{c,f} > T_{c,af}$, the expression (1.7) describes PT between paraelectric (PE) and FE states at varying the temperature, if $T_{c,f} < T_{c,af}$, (1.7) describes PE-AFE transition.

For definiteness let us consider the case when the FE phase (phase 1) is more stable in energy than the AFE one ($T_{c,f} > T_{c,af}$) and investigate the peculiarities of the behavior of the system.

To write down the thermodynamic potential for each of the phases let us refer to Fig.1.1 in which the schematic equipotential lines of the nonequilibrium thermodynamic potential (1.7) on the $(P - \eta)$ are presented. This schematic map of equipotential lines corresponds to the temperature interval within which there exist local minima, corresponding to possible low-temperatures states of the system. The minimum corresponding to the FE phase has the coordinates ($P_{1,0}$, 0), where $P_{1,0}^2 = -(\alpha_1/\alpha_2)$. The equilibrium value of the energy of this state is $\varphi_{1,0} = -(\alpha_1^2/(4\alpha_2))$. The minimum corresponding to the AFE state has the coordinates $(0, \eta_{2,0})$, $\eta_{2,0}^2 = -(\beta_1/\beta_2)$. The equilibrium value of the energy of this state is $\varphi_{2,0} = -(\beta_1^2/(4\beta_2))$.

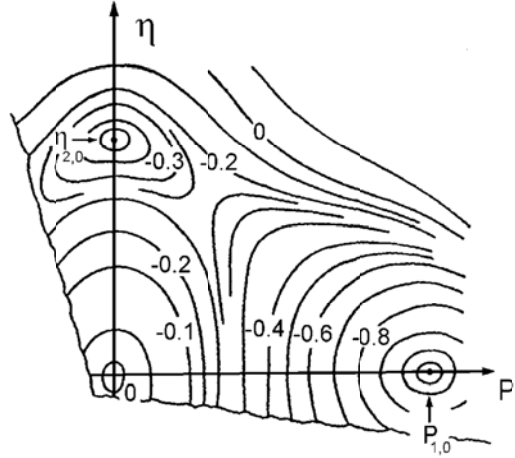

Fig.1.1. Schematic map of equipotential lines for the thermodynamic potential (1.7) for the temperatures $T < T_{c,af} < T_{c,f}$

Now the nonequilibrium potentials for each of the phases may be written (up to the quadratic terms) as:

$$\varphi_1 = \varphi_{1,0} + U_1(P_1 - P_{1,0})^2 + V_1\eta_1^2; \tag{1.8a}$$

$$\varphi_2 = \varphi_{2,0} + U_2(\eta_2 - \eta_{2,0})^2 + V_2 P_2^2. \tag{1.8b}$$

The coefficients $U_i$ and $V_i$ may be expressed in terms of the coefficients of the expansion (1.7). The most important for our consideration is the fact that they are positive.

In accordance with the relations (1.2) and (1.3), taking into account the restrictions imposed by the order parameters symmetry (see the remark accompanying the expressions (1.2)), the interphase interaction energy has the following form:

$$W_{int} = (\xi_2 P_2) E_{P_1} + (\xi_1 \eta_1) E_{\eta_2} = (\xi_2 P_2)(\xi_1 C_1 P_{1,0}) + (\xi_1 \eta_1)(\xi_2 D_2 \eta_{2,0}). \tag{1.9}$$

Now, according to (1.1), (1.8) and (1.9), the density of the nonequilibrium thermodynamic potential of the two-phase system is to be written as:

$$\varphi = \xi_1\left(\varphi_{1,0} + U_1(P_1 - P_{1,0})^2 + V_1\eta_1^2\right) + \xi_2\left(\varphi_{2,0} + U_2(\eta_2 - \eta_{2,0})^2 + V_2 P_2^2\right)$$

$$+ (\xi_2 P_2)(\xi_1 C_1 P_{1,0}) + (\xi_1 \eta_1)(\xi_2 D_2 \eta_{2,0}), \qquad \xi_1 + \xi_2 = 1. \tag{1.10}$$

The use the conventional procedure of minimization with respect to $P_1$, $\eta_1$, $P_2$, and $\eta_2$ allows us to obtain the equilibrium values of these parameters:

$$P_1 = P_{1,0}, \quad \eta_1 = -\frac{1}{2}V_1\xi_2 D_2\eta_{2,0}; \quad \eta_2 = \eta_{2,0}, \quad P_2 = -\frac{1}{2}V_2\xi_1 C_1 P_{1,0}. \tag{1.11}$$

The substitution of the expressions (1.11) into (1.10) gives the density of equilibrium energy for the considered two-phase system:

$$\tilde{\varphi} = \xi_1\left(\phi_{1,0} - \frac{\xi_2^2}{4V_1}D_2^2\eta_{2,0}^2\right) + \xi_2\left(\phi_{2,0} - \frac{\xi_1^2}{4V_2}C_1^2 P_{1,0}^2\right). \tag{1.12}$$

The comparison of expressions (1.12) and (1.8) shows, that the phase interaction in the form (1.9) lowers the energy of the system. As a result, this two-phase state may acquire a larger energy stability than a single phase state, if the difference of $(\tilde{\varphi} - \varphi_{1,0})$ is negative. The latter condition is fulfilled in the vicinity of the FE-AFE stability line. In this case $\varphi_{1,0} \cong \varphi_{2,0}$, $\xi_1 = \xi_2 = 1/2$, and the absolute energy advantage due to the formation of the nonuniform two-phase state will be the following:

$$\Delta W = -\frac{1}{32}\left(\frac{D_2^2}{V_1}\eta_{2,0}^2 + \frac{C_1^2}{V_2}P_{1,0}^2\right). \tag{1.13}$$

As seen from (1.3), the fields $E_1$ and $E_2$ and, consequently, the coefficients $C_1$ and $D_2$, are the functions of the coordinates. They decrease as the distance from the interphase boundary increases. Therefore, the long-range interaction fields favour the stabilization of nonuniform state to the greatest degree. The stability of the nonuniform state is also raised by low values of the parameters $V_1$, $V_2$, $U_1$ and $U_2$.

In the considered case of the coexistence of the FE and AFE phases, the long-range fields have electric (dipole) and elastic nature. Since in each real crystal there exist free charges, crystal lattice defects (e.g. dislocations), the radius of action of these fields is finite, and corresponds to the distance equal to several tens of lattice parameters. Therefore, the energy advantage of the two-phase state (13) is ensured by the regions separating the domains of the phases in the sample's volume.

For a case more general than (1.9), the expression for the interphase interaction energy has the form:

$$W_{int} = (\xi_2 P_2)(\xi_1 C_1 P_1) + (\xi_1 \eta_1)(\xi_2 D_2 \eta_2). \tag{1.14}$$

In this case [3]:

$$\tilde{\varphi} = \xi_1 \left[ \varphi_{1,0} - \frac{\xi_2^2 D_2^2 \eta_{2,0}^2}{4V_1} - \frac{(\xi_1 \xi_2)^2 C_1^4 P_{1,0}^2}{16 \cdot U_1 V_2^2} \right] + \xi_2 \left[ \varphi_{2,0} - \frac{\xi_1^2 C_1^2 P_{1,0}^2}{4V_2} - \frac{(\xi_1 \xi_2)^2 D_2^4 \eta_{2,0}^2}{16 \cdot U_2 V_1^2} \right]. \tag{1.15}$$

$$\Delta W = -\frac{1}{32}\left( \frac{D_2^2}{V_1} \eta_{2,0}^2 + \frac{C_1^2}{V_2} P_{1,0}^2 \right) - \frac{1}{512}\left( \frac{C_1^4}{U_1 V_2^2} P_{1,0}^2 + \frac{D_2^4}{U_2 V_1^2} \eta_{2,0}^2 \right). \tag{1.16}$$

The comparison of the latter representation with (1.12) shows that all the conclusions, which may be made on the base of the expression for the interaction energy in the form (1.9), are also entirely valid for the case when the interaction energy has the form (1.14).

We have considered the problem of stability for the nonuniform state of the coexisting FE and AFE phases. The negative value of $\Delta W$ in (1.13) shows in fact that the considered interphase boundary possesses a negative surface energy. Such negative energy will lead to division of the volume of the sample into unlimited number of very small domains (thus ensuring the largest area of the interphase boundaries and, consequently, the largest energy advantage).

In this chapter, we concentrated our attention on the co-existence and the interaction of phases described by different order parameters. The presence of different phases and their interaction have been experimentally confirmed by means of X-ray diffraction and transmission electron microscopy (TEM) methods. However, he should emphasize that the notions of phase and phase state are defined and introduced for rather large finite volume of substance (theoretically, for the volume which tends to infinity). This means that division of a finite-volume sample into the domains of coexisting phases cannot continue indefinitely and this process will end at a certain stage. The latter is defined by the least size of the phase domains at which the phases still exist. At further decrease of the domain size, the phases disappear and one cannot speak of the existence of the phases, of interphase interaction and the energy of interphase boundaries.

In the present chapter, we made a simplified consideration of the problem. In particular, the effects connected with the conditions of continuity for elastic medium at the interphase boundary have not been taken into account. The interphase domain wall (IDW) separates the domains with FE and AFE states, for which elementary cells have different size. In connection with the above-said two, distinctive features should be emphasized. First, it has been shown in [8] that at conjugation of crystal planes of phases with close crystal structures (this situation corresponds to the case of FE and AFE phase coexistence considered here) that have different interplanar distances, the intervals between possible dislocations have to be of order of several tens of *nm*. Second, the studies of the coexisting domains of the FE and AFE phases by TEM showed that linear sizes of the said domains are of order of several tens of nanometers [9, 10], and there are no dislocations in interphase domain boundaries.

This signifies that the crossing of the IDW (from one phase to the other) happens by means of continuous conjugation of the atom planes (free of breaks and dislocations). Such a coherent IDW structure leads to an increase of the elastic energy. This increase is the more essential the larger is the difference in the configuration volumes of the FE and AFE phases.

Such effect brings in the positive contribution into the surface energy density for the boundaries separating ordinary domains in ferroelectrics [11, 12]. This elastic energy makes weaker the condition of the nonuniform state existence. The series of solid solutions with compositions $Pb_{0.90}(Li_{1/2}La_{1/2})_{0.10}(Zr_{1-y}Ti_y)O_3$ was studied as an example [13]. It has been shown that changing the difference in interplanar distances for the FE and AFE phases (by varying the solid solution composition – *y*) one can pass from the mechanism, which provides elastic strictional blocking during the creation of domains of metastable phase, to the mechanism, which provides inhomogeneous state of domains of the coexisting FE and AFE phases.

Higher order terms of (1.3) may lead to some other physical effects competing with (1.9) and (1.13), in particular, to the effect of strictional blocking. The latter prevents the appearance of domains of a new dipole ordered phase in the volume of the preceding phase. These higher-order terms can also contribute modification of the condition of existence of the nonuniform state.

Thus, the analysis of the behavior of the system described by the thermodynamic potential (1.1) with the interphase interaction (1.2) shows that the stable state of the system may be inhomogeneous (see Eq. (1.6) and the commentary to it). For the substances characterized by ferroelectric-antiferroelectric PT and described by the potential (1.7) it is shown that the said inhomogeneous state may be considered as the state of coexisting domains of the FE and the AFE phases with negative interphase boundary energy.

The possibility of existence of the inhomogeneous states with negative interphase boundary energy leads to two significant conclusions. First, this is the possibility of a simple, elegant, and self-consistent physical explanation of different aspects of known phenomena, which have not been adequately treated so far. Second, it may lead to the prediction (and subsequent experimental verification) of earlier unknown phenomena in different physical processes.

## 2. EFFECTS CAUSED BY ANHARMONICITY OF THE ELASTIC ENERGY

As one can see from (1.13) and (1.16), the smaller are values of the expansion coefficients $U_1$, $U_2$, $V_1$ and $V_2$, the higher is the contribution of the energy advantage due to the formation of the inhomogeneous two-phase state into the free energy of the system under consideration. Small values of the coefficients $U_1$, $U_2$, $V_1$ and $V_2$ correspond to gently sloped potential surfaces for each of the ordered low-symmetry states of the system. Gentle slopes in potential energy lead to the phase transitions (the first order ones) with the extended hysteresis regions separating FE and AFE states. In what follows, we will consider physical situations and concrete materials with such hysteresis regions. We will also analyze the manifestation of inhomogeneous states of coexisting domains of the FE and AFE phases in the experiment.

The phase diagrams and phase transitions for the system with the FE-AFE phase transition are described by the model of two interacting order parameters (for FE and AFE states) on base of the nonequilibrium thermodynamic potential of the following form [4,6]:

$$\Phi = \Phi_P + \Phi_\eta + \Phi_{P\eta} + \Phi_u + \Phi_{Pu} + \Phi_{\eta u};  \qquad (2.1)$$

$$\Phi_P = \frac{\alpha_1}{2}P^2 + \frac{\alpha_2}{4}P^4, \qquad \Phi_\eta = \frac{\beta_1}{2}\eta^2 + \frac{\beta_2}{4}\eta^4, \qquad \Phi_{P\eta} = \frac{A}{2}P^2\eta^2.$$

The last two terms have the form:

$$\Phi_{Pu} = Qu_{ll}P^2, \qquad \Phi_{\eta u} = \lambda u_{ll}\eta^2.$$

The elastic energy is usually considered in the harmonic approximation:

$$\Phi_u = \frac{K}{2}u_{ll}^2 + P_{ext}u_{ll.}$$

Here the following notations are used: $P_{ext}$ is the external hydrostatic pressure, $Q$ and $\lambda$ are the striction coefficients, $u_{ll}$ is the spur of deformation tensor.

Then the conventional procedure consists in the minimization of (2.1) with respect to $u_{ll}$:

$$Ku_{ll} = \sigma; \qquad \sigma = -(QP^2 + \lambda\eta^2 + P_{ext}) \qquad (2.2)$$

and the thermodynamic potential assumes the following form:

$$\Phi = \frac{\alpha'_1}{2}P^2 + \frac{\alpha'_2}{4}P^4 + \frac{\beta'_1}{2}\eta^2 + \frac{\beta'_2}{4}\eta^4 + \cdots + \frac{A'}{2}P^2\eta^2. \qquad (2.3)$$

The phase diagrams for the potential (3) drawn in the $\alpha'_1$, $\beta'_1$-plane and in the "pressure-temperature" plane are presented in Fig.2.1a. These diagrams are in a good agreement with the experimental ones obtained for $PbZrO_3$ [14] and PZT with small content of Ti [15]. The peculiarity of both the calculated and the experimental diagrams is the presence of linear lability boundaries for all the phases.

In [16] the temperature dependences for the probability of the occurrence of the Mossbauer effect in $PbZrO_3$ and $PbTiO_3$ are investigated. Based on these dependences the authors estimated the anharmonicity degree for the elastic potentials, which describe the oscillations of the ions occupying the B-sites of the perovskite crystal lattice in these compounds. It turned out that in $PbZrO_3$ the elastic potential is practically harmonic whereas in $PbTiO_3$ the deviation from harmonicity is large. The considered phenomenon is the so-called "low temperature" anharmonicity caused by the discrepancy between the size of the ions and the volume they occupy in the crystal lattice.

Similar discrepancy between ion size and the volume, which it occupies in the crystal lattice, is proper to other substances with the perovskite structure when ions of different radii are placed in equivalent lattice sites. Parameters of the crystal lattice in the said substances are determined by the "driving" the lattice towards average radius of ions. However, ions with smaller radius have "freedom" among its crystal surrounding. Naturally, in this case low-temperature anharmonicity should be taken into account.

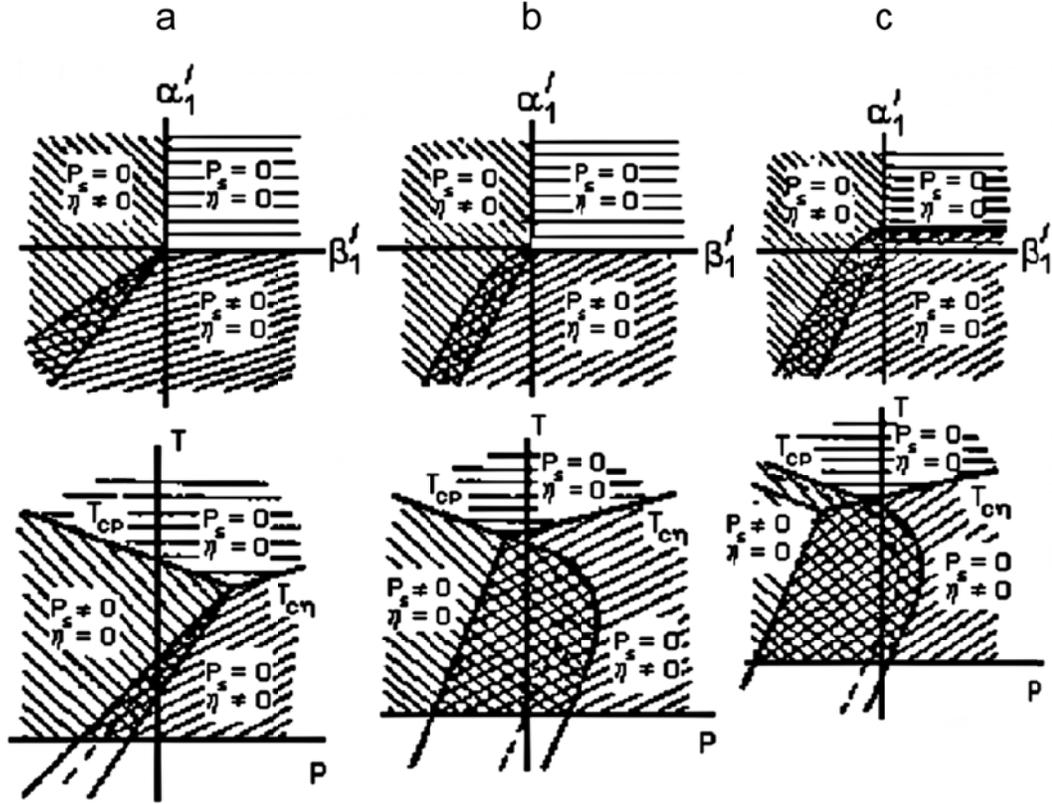

Fig.2.1. Phase diagrams described by the potential (2.3) at different values of expansion coefficients: a- $A' > 2\sqrt{\alpha_2'\beta_2'}$; $A' < 0$; b - $A' > 2\sqrt{\alpha_2'\beta_2'}$; $\alpha_2' > 0$; $\alpha_3' \neq 0$; c - $A' > 2\sqrt{|\alpha_2'|\beta_2'}$; $\alpha_2' < 0$; $\alpha_3' \neq 0$. The diagrams in the expansion coefficients' coordinates of the potential (2.3) are in the upper row. The lower row contains "pressure-temperature" diagrams.

Therefore, the terms of higher orders in deformation are to be taken into account in the expansion (2.1) for investigation of PT and phase diagrams:

$$\Phi_u = \frac{K}{2}u_{ll}^2 + \frac{G}{3}u_{ll}^3 + \frac{D}{4}u_{ll}^4 + P_{ext}u_{ll}. \tag{2.4}$$

Then the equilibrium deformation value is defined by the expression:

$$Ku_{ll} + Gu_{ll}^2 + Du_{ll}^3 = \sigma; \qquad \sigma = (QP^2 + \lambda\eta^2 + P_{ext}), \tag{2.5}$$

or, in the approximation quadratic in $\sigma$

$$u_{ll} = s_1\sigma + s_2\sigma^2 \tag{2.6}$$

($s_1$ and $s_2$ are the linear and nonlinear elastic compliance, respectively).

The substitution of the expression (2.6) into (2.1) leads to the appearance of the terms characterized by higher orders in $P$ and $\eta$ as compared with the ones of the initial expression due to the dependence of σ on $P^2$ and $\eta^2$. Thus, taking into account the elastic energy anharmonicity results in the complication of the expression for the thermodynamic potential, and this must lead to a more sophisticated form of the phase diagrams. For example, Fig.2.1b and 2.1c present the form of the phase diagrams for the substances which undergo FE-AFE phase transition when only the term of the sixth order in polarization, i.e. $\alpha'_3 P^6$, is taken into account.

Now discuss the effects which may take place in the case when the degree of anharmonicity increases (in this case the character of the phase diagrams in the $\alpha'_1$, $\beta'$-plane transforms from the type Fig.2.1a into the types 2.1b or 2.1c). This situation is shown in Fig.2.2. Note that all our considerations will use PZT-based solid solutions as example, since it is only for the latter that the sufficient amount of necessary experimental data may be found. However, these considerations remain valid for any oxide system with the perovskite structures. Moreover, we suggest that the effects under discussion may take place in the other types of crystal structures.

The upper row of Fig.2.2 presents the qualitative evolution of the phase diagrams $P_{ext}$-$T$, which may take place in the case when the degree of the elastic energy anharmonicity increases. In PZT-based solid solutions the said value rises with the growth of the content of Ti which substitutes Zr. Therefore the transition from the diagram 2.2a to the diagram 2.2c is realized in the PZT as the AFE state region in the "Ti-composition − temperature" diagram widens towards higher contents of Ti.

The ion substitutions for both A- and B-sites of the perovskite crystal lattice are equivalent to the action of some "effective" hydrostatic pressure [17]. Thus, one can build the "composition-temperature" diagrams, which correspond to the $P_{ext}$-T diagrams shown in the upper of Fig.2.2. These diagrams (2.2d-2.2f) are presented in the middle row of Fig.2.2. Here we have taken into account the fact that the increase of the content of Zr in PZT leads to the growth of the "effective" pressure, whereas the increase of the content of Ti lowers it. The temperature – electric field (T-E) phase diagrams corresponding to the pressure – temperature ($P_{ext}$-$T$) diagrams of the upper row are shown in the lowest row of Fig.2.2.

The phase diagrams shown in the first column of Fig.2.2 (a, d, g) practically coincide with the known phase diagrams of PZ and PZT with a small content of Ti (when the degree of anharmonicity is small).

The phase diagrams which may be obtained using the thermodynamic potential (2.1) for the case of a weak elastic energy anharmonicity are given in the second column of Fig.2.2. The difference of these diagrams from those shown in the first column consists only in a slight disturbance of the phase boundary linearity. PT in the substances where such diagrams are realized will slightly differ from those of the substances characterized by the diagrams from the first column.

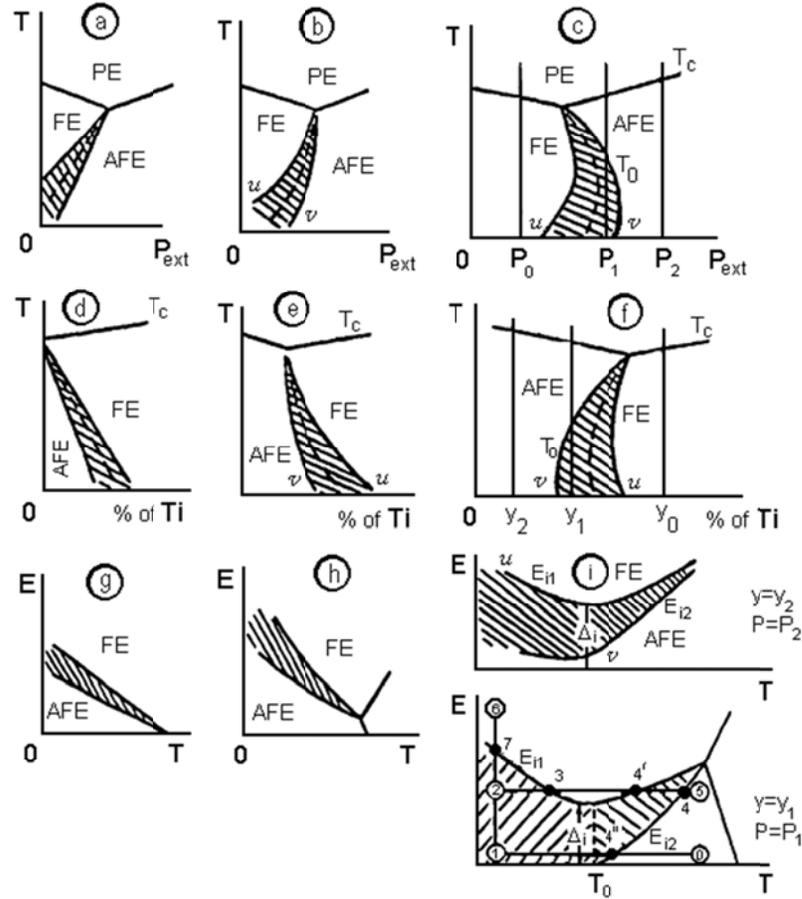

Fig.2.2. Calculated phase diagrams for substances undergoing FE-AFE transformations for different degree of anharmonicity [18].

The phase diagrams shown in the third column are the most interesting, since in this case the variation of the diagrams assumes a qualitative character. Let us first consider the $P_{ext}$ - $T$ diagram in Fig.2.2c. The most essential (and distinctive) feature of this diagram is the changing sign of the slope of the lability boundaries for the FE and AFE states, when moving along these boundaries. It is just such a character of the phase boundaries that leads to entirely new peculiarities of FE-AFE phase transformations. The first manifestations of the said peculiarities in experiments have given rise to both surprise and incomprehension. The point is in the following. The diagram in the Fig.2.2c actually represents three different phase diagrams that may belong to three different solid solutions same series of PZT-based compounds having different contents of Ti. The points $P_0$, $P_1$, and $P_2$ are essentially the origins of coordinates for the three solid solutions with the Ti-contents equal to $y_0$, $y_1$, and $y_2$, respectively, shown in Fig.2.2f. In the solid solution with the composition $y_0$, the only FE-PE transition always takes place as the temperature changes, if no external pressure being applied. In the solid solution with the composition $y_2$, the same conditions induce AFE-PE transition. For the solid solution with the composition $y_1$, the PT behavior is quite different. This solid solution is characterized by the trajectory $P_{ext} = P_1 = const$ in Fig.2.2c. This line passes through the hysteresis region as temperature changes; therefore the succession of PT that takes place during the change of the temperature depends on the sample's prehistory. If no electric field is applied then AFE $\rightarrow$ PE phase transition arises at the point $T_C$. The lability boundary for the FE phase (described by the curve $T_0(P_{ext})$), does not manifest itself. However, when this phase is induced by applying electric field below the line $T_0(P_{ext})$, it does not disappear after switching of the field. As the sample's temperature rises the following PT succession is observed: FE $\rightarrow$ AFE $\rightarrow$ PE.

For the solid solutions described by the $P_{ext}$ - $T$ diagrams of the type shown in Fig.2.2c, the "composition-temperature" diagrams (Fig.2.2f) also differ qualitatively from those presented in Fig.2.2d. First of all, in the diagram Fig.2.2f the slope of the FE states' region boundary is opposite in sign as compared with the one we see in Fig.2.2d. This is connected with the change of the slope of the FE states' boundary in the $P_{ext}$ - $T$ diagram shown in Fig.2.2c. The second distinctive feature is the appearance of a region of induced states marked by hatching in Fig. 2.2f. For the solid solutions corresponding to this region, both the phase states and the succession of PT depend on the samples' prehistory. The succession of PT at the point $T_C$ prior to the action of the electric field is the following: AFE $\leftrightarrow$ PE. After the application of the field with an intensity exceeding some threshold critical value at $T < T_0$, the FE $\rightarrow$ AFE $\rightarrow$ PE succession of PT is observed.

Now consider the peculiarities of the $T$ - $E$ diagrams. These diagrams are presented in the lowest row of Fig. 2.2. Fig. 2.2i shows the $T$ - $E$ diagrams for the solid solutions which are characterized by the $P$ - $T$ diagrams of 2.2c type. To understand the behavior of these $T$ - $E$ diagrams one has to return to the $P_{ext}$ - $T$ diagrams of 2.2c type. Consider, for example, the solid solution with the composition $y_1$ which is characterized by the line $P_{ext} = P_1$ in the $P_{ext}$ - $T$ diagram. The application of electric field raises the stability of the FE phase with respect to the AFE one, the lability boundaries for the FE and AFE states (curves *v* and *u*, respectively) are being shifted to the right (see Fig.2.2c). At the temperatures $T < T_0$ the phase transition into FE state (i.e. the electric field induces the FE) will take place at the field intensity $E_{i1}(T)$ which corresponds to the intersection of the phase boundary *u* shifted by the electric field with the straight line $P_{ext} = P_1$ (see Fig. 2.2c). The boundary *u* in the $P_{ext}$ -$T$ diagram is substantially nonlinear, therefore the phase boundary $E_{i1}(T)$ in the *T-E* diagram 2.2i is also nonlinear. The form of this boundary is similar to that of the boundary *u* in the $P_{ext}$ - $T$ diagram 2.2c. Moreover, the presence of the minimum field $\Delta_i$ is a consequence of the fact that the slope of the boundary *u* changes its sign on the $P_{ext}$ - $T$ diagram. Therefore, the slope of the dependence $E_{i1}(T)$ with respect to the temperature axis changes as well. As the field's intensity lowers, the phase boundaries *u* and *v* are shifted to the left (Fig.2.2c); the transition from FE to AFE state will take place in the field with the intensity $E_{i2}(T)$, when the phase boundary *v* intersects the straight line $P_{ext} = P_1$. It is obvious that the reversible transition may occur only at $T > T_0$. Such change in the stability of the AFE and FE states in the $P_{ext}$ - $T$ diagram manifests itself by the presence of the regions of both the irreversible and reversible FE phase induction in the $T$ - $E$ diagram 2.2i. The former exists at $T < T_0$ and is preserved after switching off the field; the latter arises at $T > T_0$ and appears when the field is switched off. As seen from Fig.6i, there is the so-called "field gap" $\Delta_i$ in the T-E diagram 2.2i. For $E < \Delta_i$ the phase transition into FE states never takes place at any temperatures.

If the FE phase is induced either in the presence of hydrostatic pressure or in the solid solution with the compositions varying from $y_1$ to $y_2$, then the curves $E_{i1}(T)$ and $E_{i2}(T)$ in the $T$ - $E$ diagram are shifted in the way shown in the upper part of Fig. 2.2i. As it follows from this diagram, there exist such pressure or Ti concentration that starting from them the FE-AFE transitions in electric field will be only reversible.

Let us now discuss the peculiarities of phase transitions, which are directly defined by the form of the $T$ - $E$ diagrams of 2.2i type. If the samples are cooled along the thermodynamic path 0-1-2 they are in the AFE state at the point 2. If now the samples are heated along the path 2-5 one shall observe AFE → FE and FE → AFE phase transitions at the points 3 and 4, respectively. The reverse movement along the path 5-2 will lead to only the AFE → FE PT taking place at the point $4'$. No phase transitions occur at the points 4 and 3. On cooling along the thermodynamic path 0-1-2-6 the AFE-FE PT takes place at the point 7; the reverse movement along the same path results in the appearance of the FE-AFE PT at the point $4''$.

The above analysis makes obvious the nature of the distinction between "zero field cooling" and "field cooling" conditions. It consists in the fact that a substance falls within the hysteresis region in the phase diagram. Being in the said hysteresis region determines the dependence of the phase transitions sequence upon the specimens prehistory.

To conclude this section, let us discuss the influence of anharmonicity on phase states based on the X - $T$ diagram, where X is the content of substituting ion, which gives rise to the appearance of anharmonicity. As mentioned above, the substituting ions must be the ones that substitute the A-site of the perovskite crystal lattice and has a smaller ionic radius (or $A'A''$-complex with a mean ionic radius smaller than that of the A-ion). For PZT solid solutions, for example, such substitution means the increase of "effective" hydrostatic pressure with the growth of X [3, 29], and the displacement of the boundary separating the regions of FE and AFE states in the "Ti-content-temperature" phase diagram towards the larger content of Ti. At the same time, the form of the said diagram changes from type Fig. 2.2d to that of Fig 2.2f with the increase of X. Let us choose a particular Ti content (e.g. $y_1$) such that at X = 0 the corresponding PZT solid solution is located far from this boundary (between regions of FE and AFE states) in the FE region of the "Ti-content-temperature" diagram. Then let us denote the considered solid solutions by the symbol X/1-$y_1$/$y_1$ (1-$y_1$/$y_1$ is the symbol of ordinary PZT). Since at X = 0 the solid solution is located in FE region on phase diagram and it undergoes FE-PE phase transition under change of the temperature. This PT corresponds to the point 1 in the diagram in Fig. 2.3. With the growth of X the "Ti content-temperature" diagram changes from the type Fig. 2.2d to the type 6e, and FE-AFE boundary is displaced to the right. However, the solid solution X/1-$y_1$/$y_1$ is still within the FE region and undergoes FE-PE transition, which corresponds to the point 2 in the diagram in Fig. 2.3. The diagram 6e transforms into the diagram 6f at further increase of X. Now (and it is very significant) the FE-AFE boundary approaches to the solution X/1-$y_1$/$y_1$ and the latter is in hysteresis region of the "Ti content-temperature" diagram. If the sample has not been previously put in an electric field, it is in the AFE state at low temperatures, and at changing the temperature it undergoes AFE-PE phase transition (point 3 in Fig. 2.3). After the action of electric field $E > E_{i1}$ at $T < T_0$ FE state is induced in the sample's volume. Under subsequent heating (at $E = 0$) the FE → AFE phase transition will take place at temperature $T_0$ (the point $3'$ in Fig. 2.3) and then AFE → PE transition (the point 3). At further increase of X the FE-AFE boundary in "Ti content-temperature" diagram becomes essentially shifted to the right of the point $y_1$. Now the solid solution X/1-$y_1$/$y_1$ is located within the AFE region of the diagram, and therefore it undergoes the AFE-PE phase transition only (the point 4 in Fig. 2.3).

The same change of state takes place at the substitution of the ions in the B-site by the complex ($B'B''$) with a mean ionic radius exceeding that of the B-ion (of course, the radii of $B'$ and $B''$ must be different).

The authors hope that the readers have already understood that in the case of A-ion the diagram shown in Fig. 2.3 is characteristic of e.g. X/65/35 PZT in which La or Sr are substituted for Pb. In the case of B-ion the diagram in Fig. 2.3 is characteristic of lead titanium in which titanium is substituted by the complex ($Mg_{1/3}Nb_{2/3}$).

All these types of phase diagrams discussed above well be considered for particular substances in the experimental part where the experimental phase diagrams and their evolution at subsequent substitution of the main ions will be analysed.

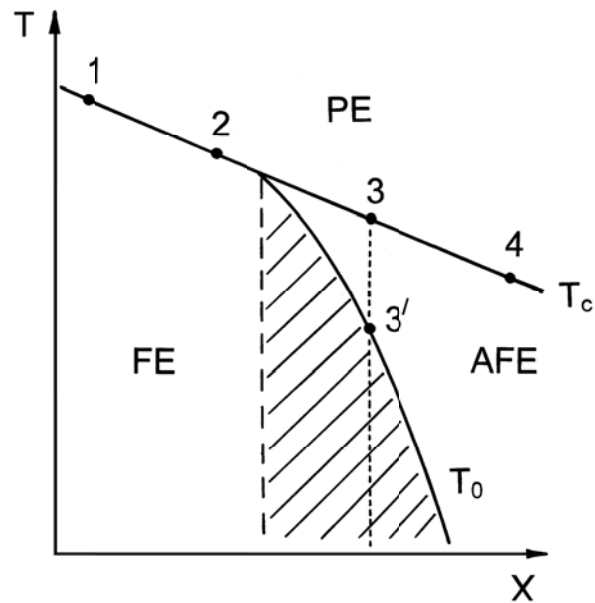

Fig. 2.3. Schematic phase diagram for X/1-$y_1$/$y_1$ series. The dashed region corresponds to hysteresis region in Fig. 2.2.f (or 2.2.c).

## 3. EXPERIMENTAL RESULTS ON PHASE DIAGRAMS FOR SUBSTANCES WITH NONUNIFORM STATES

### 3.1. Peculiarities of Y-*T*, *T-E* and *P-T* phase diagrams

It was mentioned in Ch.2 (more detailed consideration will be given in ch.4; see also [6]) that the substitution of ions with smaller ionic radii for lead in PZT increases the energy stability of the AFE state in relative to the FE state. As a consequence the boundary separating the regions of the FE and AFE ordering in the phase diagram "Ti-content - temperature" (Y - *T* diagram) is shifted towards the higher Ti concentrations. The role of anharmonic terms in the crystalline potential grows at the same time. Therefore, the Y – *T* phase diagram must change from the type presented in Fig.2.2a to that in Fig.2.2c. The Y – *T* diagrams for PZT-based solid solutions obtained in the process of successive step-by-step substitution of $(Li_{1/2}La_{1/2})$, Sr, and La for Pb are shown in the (Fig. 3.1a). The compositions of these solid solutions can be presented by the following formulas $Pb_{1-x}(La_{1/2}Li_{1/2})_x(Zr_{1-y}Ti_y)O_3$ (PLLZT), $(Pb_{1-x}Sr_x)(Zr_{1-y}Ti_y)O_3$ (PSZT), and $(Pb_{1-3x/2}La_x)(Zr_{1-y}Ti_y)O_3$ (PLZT), respectively, As one can see the character of changes in the PZT phase diagram corresponds to that predicted in Ch.2.

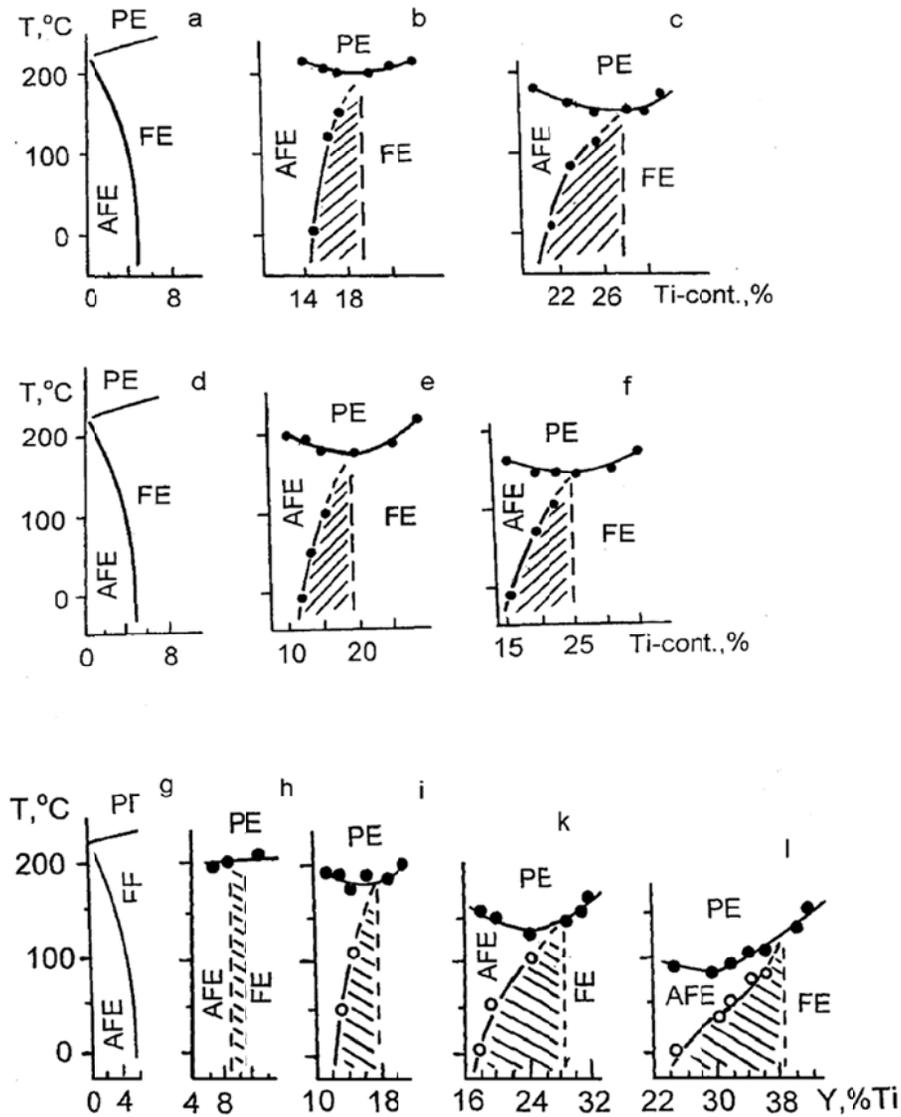

Fig 3.1a. "Composition-temperature" phase diagrams for PZT based solid solutions (X/100-Y/Y) with substitutions of Sr, ($Li_{1/2}La_{1/2}$), and La for lead. All diagrams have been obtained after induction of the FE phase by electric field [6, 18]. Sr (X),%: a - 0; b - 10; c - 20. ($La_{1/2}Li_{1/2}$) (X),%: d - 0; e - 10; f - 15. La (X),%: g - 0; h - 2; i - 4; k - 6; l - 8.

The X – *T* diagrams for PSZT, PLLZT, and PLZT solid solutions that are presented in Fig 3.1b also correspond to the model diagrams in Fig.2.2.

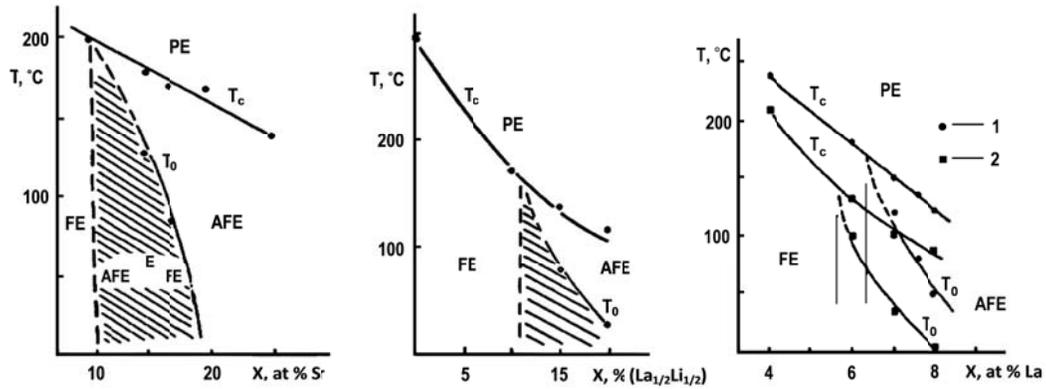

Fig.3.1b. "Composition-temperature" X/80/20 phase diagrams for PZT based solid solutions [18] PSZT, PLLZT, and PLZT (for two Ti-contents, 1 - X/75/25, 2 - X/65/35)

The Y – T phase diagrams contain are broad intermediate regions (marked by dashes in fig. 3.1a) between the regions of FE and AFE states. The same intermediate regions are also present in the X - T diagrams (Fig.3.1b). According to X-ray analysis data [13] and TEM investigations [9, 10], these regions are characterized by coexisting domains of the FE and AFE phases. The phase state of the samples with compositions located within these regions depends on the previous history of samples. Electric field with an intensity exceeding the critical value $E_{i1}$ induces macroscopic FE state. This sate is stable under heating up to the temperature $T_0(y)$. The dependence $E_{i1}(T)$, i.e. the phase $T$ - $E$ diagram, for the 10/85/15 PLLZT and 4/87/13, 4/86/14, 6/82/18, 6/80/20 and 6/65/36 PLZT solid solution is shown in Fig.3.2 (here and further, to denote the composition of the solid solution obtained from PZT X/1-Y/Y where the first number X corresponds to the percentage of the element substituting lead, whereas the second (1-Y) and the third Y numbers denote content of zirconium and titanium, respectively). As seen from this diagram, there exists a minimal critical field of FE phase induction $E_{i,min} \cong 13\ kV/cm$. In weaker fields FE phase cannot be induced, and the macroscopic state of the samples remains unchanged (AFE) at any temperature changes.

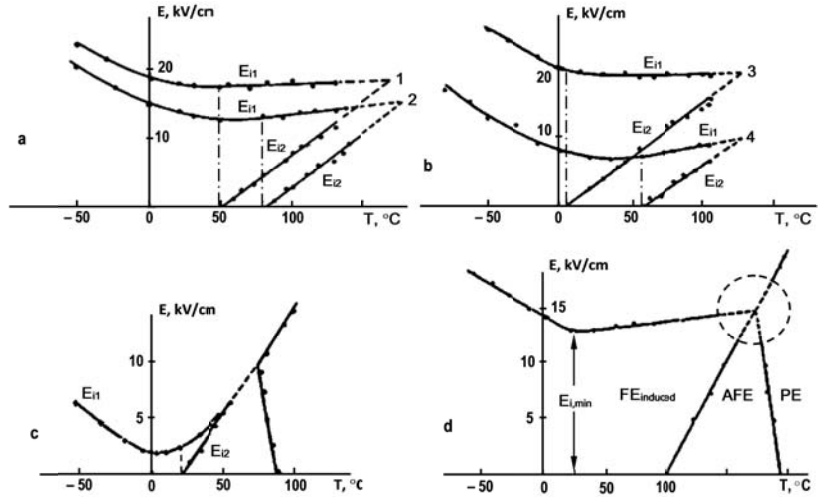

Fig.3.2. T-E phase diagrams for PLZT solid solutions with 4% (a), 6% (b), and 9% (c) content of La and for PLLZT 10/85/15 solid solution (d) [6, 18, 19, 22]. Content of Zr/Ti: 1 - 87/13; 2 - 86/14; 3 - 82/18; 4 - 80/20; 5 - 65/35.

On increasing the content of titanium in 10/100-Y/Y PLLZT solid solutions $E_{i,min}$ reduces. The equality $E_{imin} = 0$ corresponds to the transition to the region of spontaneous FE states in the $Y$-$T$ diagram of PLLZT. Based on the $E$-$T$ diagrams of the type shown in Fig.2.2 it will be easy to understand the difference in the behavior of the considered solid solutions for ZFC and FC regimes of investigation. In the first case the solid solutions undergo the PE-AFE phase transition when temperature changes from the values ($T > T_C$) to lower ones. In the second case the changes of the solid solution properties follow succession of transitions: PE-AFE-FE at cooling, provided that the electric field intensity satisfies the condition $E > E_{i,min}$. It is quite natural that the investigation results for these two cases will be absolutely different. Such a problem was discussed in detail in Ch.2. Experimental Y-$T$ and $T$-$E$ diagrams of PLLZT, PLZT and PSZT solid solutions [6, 13, 19] correspond to the phase diagrams predicted by the model of FE-AFE phase transformations (Fig.2.2).

As demonstrated in Ch.2 (see also [20]), the form of the Y-$T$ phase diagrams typical for PLZT with the La-content x ≥ 4 and PLLZT with x = 10%, and 15% (Fig.3.1) results from that of specific $P$-$T$ diagrams presented in Fig.2.2c. Experimental $P$-$T$ diagrams for PLZT solid solutions of the 6/100-Y/Y series shown in Fig.3.3 were obtained for the first time in [21]. It is easy to see that their character completely corresponds to the one predicted by the model considered in Ch.2. Moreover, it should be noted that the experimental $P$-$T$ diagrams for lead zirconate with the region of FE states and of PZT (0% content of lanthanum) known from the literature correspond to the calculated *P-T* diagrams presented in Fig.2.2a.

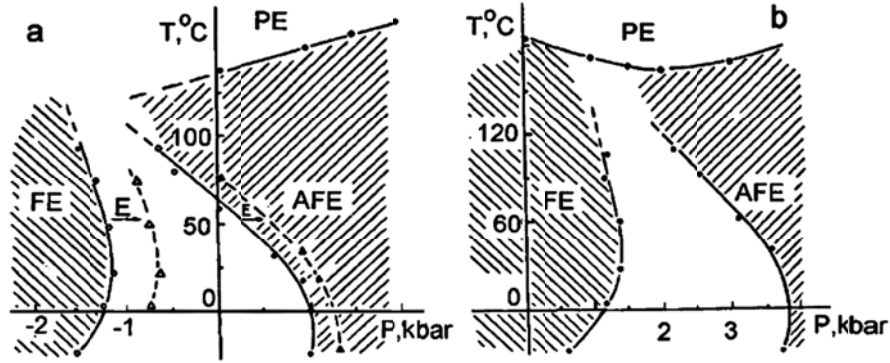

**F**ig.3.3. P-T phase diagrams for PLZT solid solutions with 6% content of La, Zr/Ti: a - 80/20; b - 65/35 [6, 18, 21]. Arrows show the shift of the phase boundaries by electric field 500 *V/cm*.

The form of experimental *P-T* diagrams conditions the specific form of the *T-E* ones discussed above. The model for the FE-AFE phase transformation was also used to predict the evolution of the *T-E* phase diagrams for substances subjected to the action of hydrostatic pressure (Fig.2.2i and the commentary to it). The experimental *T-E* diagrams for the 6/80/20 series of solid solutions (which belongs to the boundary hysteresis region of the Y-*T* diagram) corresponding to the case when the sample is under hydrostatic pressure are shown in Fig.3.4. They are completely similar to those predicted by the model from Ch.2 (Fig.2.2i).

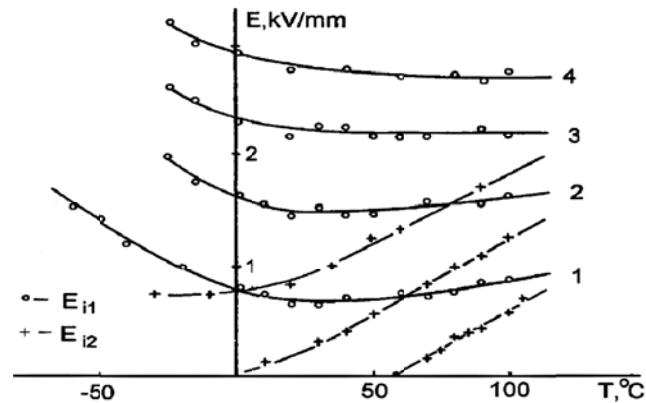

Fig.3.4. $E_{i1}$ and $E_{i2}$ critical fields as functions of temperature for 6/80/20 PLZT solid solution for different pressures (MPa): 1 - 0.1; 2 - 100; 3 - 200; 4 – 300 [18, 21].

The phase state in any part of each of the diagrams may be easily determined by comparing E-*T* diagrams with the Y-*T* and *P-T* diagrams. The phases participating in transformations that take place while crossing any phase boundary in these diagrams may be easily found. In particular, the dependence of phase transformations on the history of the sample may be readily interpreted. The authors would like to emphasize that the analysis of the whole set of the diagrams testifies that only three states - FE, AFE and PE - participate in the phase transformations. Other states are not present. (More complete sets of diagrams can be found in [6]). Nowadays the *T-E* diagrams of PLZT solid solutions may often be found in the literature. In our opinion, it is impossible to build the completely clear picture of the properties of investigated substances on the base of only one type of measurements.

The properties of PLZT solid solutions in the literature are most often discussed on the base of the ones of the X/65/35 solid solutions. The X-*T* diagram for this series is shown in Fig.3.1b. The X-*T* diagram for X/65/35 solid solutions in which praseodymium is substituted for lead one can find in [6, 18]. The way by which such a diagram may be obtained from the set of diagrams shown in Fig.3.1a of the present paper (the said procedure is realized and shown in Fig 3.1b for X/75/25 and X/65/35 PLZT) is described in Ch.2 (Fig.2.3). It is experimental confirmation of the fact that the X-T diagram contains the states observed in the Y-*T* diagrams only. It is the most important circumstance for our consideration and the other states are absent and cannot take place at all. The latter circumstance is to be specially emphasized. In the samples of the X/65/35 series the temperature changes may give rise only to those phase transitions which occur in any of the X/100-Y/Y series sample. This means that in the samples of the X/65/35 series only the transitions between FE, AFE and PE states may take place. Naturally, these PT will have peculiarities defined by a weak difference between the energies of FE and AFE. Moreover, in the praseodymium series these effects are to be revealed at lower contents of the substituting element, since the ionic radius of $Pr^{3+}$ is smaller than that of $La^{3+}$. Such a statement has been completely confirmed by experiments.

## 3.2. Coexistence of the FE and AFE phase domains

Direct observation of the coexisting FE and AFE phase domains by the method of transmission electron microscopy (TEM) was realized in our experiments performed in early eighties [9, 10, 23] on the samples with the composition 7/65/35 and later on the 8/65/35 solid solutions. These substances belong to the intermediate region which separates the states with the FE and AFE ordering.

TEM images for two orientations of crystallite spalling plane with respect to electron beam are shown in Fig.3.5.

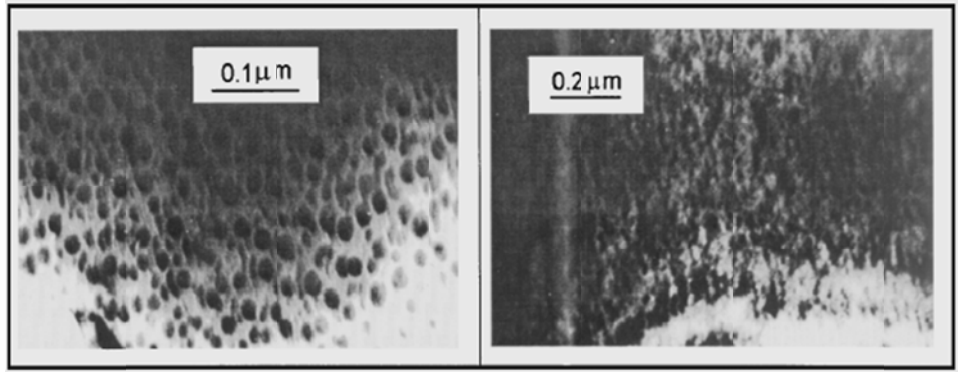

Fig.3.5. TEM bright-field images for 7/65/35 PLZT. On the left—for (110) the foil plane, on the right—for (120) the foil plane [9, 10, 23].

The TEM image on the left is the bright-field image at (110) orientation of foil plane (the spalling plane is (110)). Under such conditions a clear two-phase structure is observed inside the crystallite. The oval shape second-phase inclusions have dimensions of the order of $(2-4) \cdot 10^{-6}$ *cm* and the density of these inclusions is of the order of $10^{11}$ *cm*$^{-2}$. The right part of the Fig.3.5 presents the image of the same part of the crystallite obtained at another position of foil plane (120). In this case a "fibrous" two-phase structure manifests itself. Analysis of the images obtained at different orientations of crystallite with respect to electron beam shows that the second-phase inclusions have cylindrical shape and grow through the whole depth of the thin crystallite in the case of (110) spalling. In bulk samples such inclusions have ellipsoidal shape close to spherical one and approximately similar dimensions [24-26].

The "fibrous" structure seen on the right in Fig.3.5 is a consequence of slanting passage of electron beam through a thin crystallite at an angle with the axis of cylindrical inclusions.

It is known that the FE and AFE phases differ in configuration volume; therefore, their coexistence may be accompanied by formation of dislocations along the interphase boundaries or by the emergence of elastic stresses. The images of the single crystals in Fig.3.5 with a thickness not exceeding 0.2 *μm* (obtained by cleaving the crystallites of the coarse-grained PLZT ceramics with the grain size about 10 *μm*) have shown the AFE matrix containing inclusions (domains) of the FE phase. No dislocations along the FE-AFE interphase boundaries have been observed. This signifies that the inclusions of the metastable FE phase have a coherent character. Photographs did not shown typical manifestations of matrix distortions caused by elastic stresses.

The TEM pictures for PLZT available in the literature are obtained for samples with different prehistory (as a rule, they are of the type shown in the right part of Fig. 3.5). In [27, 28] these images are presented for annealed samples and for the same samples subjected to the action of AC electric field at room temperature. As noted in the present paper, such a prehistory influences the samples' microstructure. In our opinion the phenomena observed in [27, 28] is a consequence of ion segregation at interphase boundaries and of the long-duration diffusion process of their formation.

Now we shall dwell on another effect caused by the inhomogeneous two-phase state of the samples. As an example, consider the 7.5/100-Y/Y PLZT series. A section of the phase diagram for these substances corresponding to the vicinity of the boundary between the regions of FE and AFE ordering is shown in Fig.3.6a.

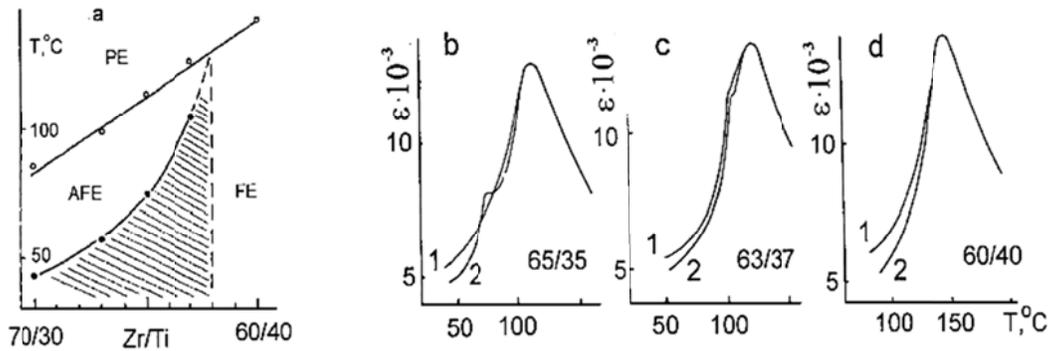

Fig.3.6. The Y-$T$ phase diagram (a) and temperature dependencies of dielectric constant (b, c, and d) for 7.5/100-Y/Y PLZT solid solutions: 1- before the electric field action; 2 - after the electric field action.

It is analogous to those presented in Fig.3.1a. The $\varepsilon$ ($T$) dependences for the solid solutions from the said series located in three different regions of the Y-$T$ diagram are shown in Fig.3.6.b-d. These dependences were measured before (on annealed samples) and after the action of DC electric field with an intensity of 2.5 $kV/mm$. For the 7.5/60/40 solid solution the dependences have no anomalies except a maximum at the point $T_m$. There exists only a maximum at $T = T_m$ associated with PE PT in annealed samples with 7.5/65/35 composition. An additional anomaly is observed at the temperature $T_0$ in polarized samples. It manifests itself only after the action of high-intensity electric field in the samples in which FE phase has been induced. However, in the 7.5/63/37 solid solution the anomaly at $T_0$ is observed in both annealed and polarized samples.

As seen from the Y-$T$ diagram of the 7.5/100-Y/Y series, the equality of the free energies takes place in the solid solution which contains zirconium and titanium in the ratio 62/38. The 7.5/63/37 solid solution is located in the vicinity of the boundary composition (dashed line in Fig.3.6.a) therefore its FE and AFE state free energies have close values. Moreover, small values of the intensity of the field inducing FE phase point to low values of the height of the potential barrier which separates the free energy minima corresponding to FE and AFE states. In view of the above-mentioned, approximately equal quantities of FE and AFE phases are formed in the volume of the sample in the process of cooling below the temperature $T_m$ (for instance, for the 7.5/65/35 solid solution a larger share of FE phase is observed only after the action of electric field). Therefore, manifestation of FE phase at subsequent heating takes place in all the cases, even if it has not been specially induced by electric field. The same picture is observed in all the solid PLZT solutions which have approximately equal stability of FE and AFE states. However, such behavior may be explained only on the base of the corresponding diagrams of phase state.

The described behavior is called spontaneous transition from the relaxor state to the normal FE state (though in fact such a transition is caused by electric field and manifests itself in experiments performed in FC regime) in the literature. For a better insight into the phenomenon in consideration, one should refer to the expression which describes the interaction of coexisting FE and AFE phase domains (Sect.1). The condition $T_{c,af} \geq T_{c,f}$ is fulfilled for the 7.5/100-Y/Y PLZT. The spontaneous polarization $P_s$ in FE domains essentially changes (decreases) under heating in the vicinity of the temperature $T_{c,f}$, whereas the properties (order parameters) in AFE domains weakly depend on temperature (in the first approximation they may be considered constant). In this case the action exerted by the interphase interaction on the FE subsystem may be described by the effective field $E_s$ such that $W_{int} = -\vec{E}_s \cdot \vec{P}_s$. As discussed in Ch.1, $T_{c,f}$ is not a true point of PT in this situation, and above it (for $T > T_{c,f}$) $P_s \neq 0$ (though for these temperatures $P_s$ has small values). Whereas, a sharp decrease of spontaneous polarization and the change of FE ordering at the $T_{c,f}$- point itself.

Let us now consider the influence of the domain of FE and AFE phases on dielectric properties, using PLLZT solid solutions as example. The phase diagrams for the said solid solutions are shown in Fig.3.1a after the action of the electric field. The boundary regions (shown by dashes in figures) are the hysteresis regions for the FE–AFE transformation (see Ch.2). X-ray studies of PLLZT with compositions from these boundary regions of the phase diagrams [13] have shown that the domains of FE and AFE phases coexist in the sample volume. It is necessary to note that the legible boundaries between the regions of the single phase (FE and AFE) states and the region of coexisting phases are absent in the phase diagrams before the application of the electric field and that induces the FE phase.

The upper curve in the Fig.3.7 contains the dependence $\varepsilon_m$ ($\varepsilon_m$ is the maximum of the $\varepsilon(T)$ dependence) on the Ti content in the 10/100-Y/Y PLLZT series. The dependences of $\varepsilon(Y)$ at 20 $^oC$ are presented in the lower part of the Fig. 3.7. Curve 1 shows the dependence, obtained on the samples which have not been affected by an electric field. In this case the maximum of $\varepsilon(Y)$ corresponds to ~ 19 % of Ti content. Curve 2 shows the dependence $\varepsilon(Y)$ after the application of an electric field. Now, the maximum of $\varepsilon(Y)$-curve located near ~13 % of Ti content.

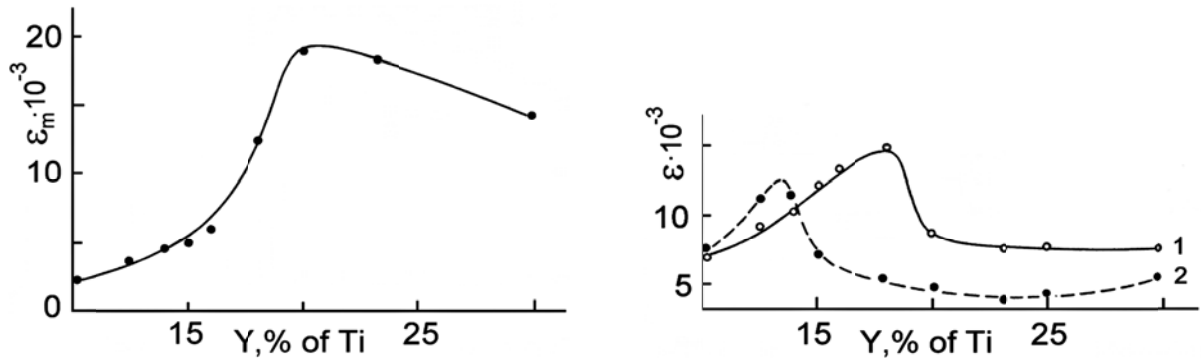

Fig.3.7. Dependence of $\varepsilon_m$ on Y (on the left) and $\varepsilon(20\ ^oC)$ on Y for annealed (1) and polarized (2) samples (on the right) for 10/100-Y/Y PLLZT solid solutions [13, 29].

The discussed dependencies of dielectric constant on the solid solution composition can be consistently explained taking into account the coexistence of domains of the FE and AFE phases in the bulk of those solid solutions which compositions fall into the dashed regions in diagrams of the PLLZT phase states. Low-frequency measurements (at frequencies less than $10^8$-$10^9$ Hz) reveal the essential contribution of the domain wall oscillations to dielectric constant of ferroelectrics. In the system of coexisting FE and AFE domains the oscillations of the interphase boundaries must have a noticeable influence on the value of $\varepsilon$, because they have high mobility in an electric field. Therefore the behavior of the dielectric constant near the FE-AFE-PE triple point gives essential information about the system of coexisting phases.

The ε(Y) dependence at 20°C obtained on samples which have not been subjected to the action of an electric field is shown in the lower part of Fig.3.7 (curve 1). In this case the boundary between FE and AFE states in phase diagram is at ~ 19% of Ti (and it location is nearly independent on temperature), and the maximum of ε(Y) is located at the same composition. After the application of DC electric field (and induction of FE state in the dashed region of the Y-T diagram in Fig. 3.1a) the position of the boundary separating the FE and AFE regions is ~ 13% of Ti concentration. This value corresponds to the maximum of the ε(Y) dependence for the polarized samples. This behavior of the ε(Y) dependence is confirmed by the results, presented in upper part of Fig.3.7, where the $\varepsilon_m(Y)$ dependence is shown. In this case the maximum is also located near the FE-AFE boundary.

### 3.3. Difficulties in identification of phase states in PLZT system of solid solutions

The PLZT solid solutions are being referred to the class of the so-called relaxor FE during recent 10-12 years. This period was preceded by 18 year-long studies of physical processes in PLZT (since the synthesis of these materials and the opening of their unique electrooptical properties in 1969). A huge amount of experimental results was accumulated (a detailed review of results of studies of properties of the PLZT solid solutions and phase transitions in this substance as of 1985 one can find in [6]). In particular, the crystalline structure was found to be not cubic (the degree of distortion being low) at the temperatures below that of $\varepsilon'(T)$ maximum, $T'_m$. This structure was characterized as pseudo-cubic. However, later PLZT crystalline structure was considered cubic at $T < T'_m$ (before the action of electric field). Therefore, we will show that the crystalline structure of the PLZT solid solution system is not cubic at $T < T'_m$, as it must take place for ferroelectrics and antiferroelectrics.

The evolution of the $Y - T$ phase diagram in the process of successive increase of the La substitution in $Pb_{1-3x/2}La(Zr_{1-y}Ti_y)O_3$ after the action of the strong electric field and induction of FE phase (see ch.2) is presented in Fig.3.1a. The said phase diagrams before the action of the field are shown in Fig.3.8.

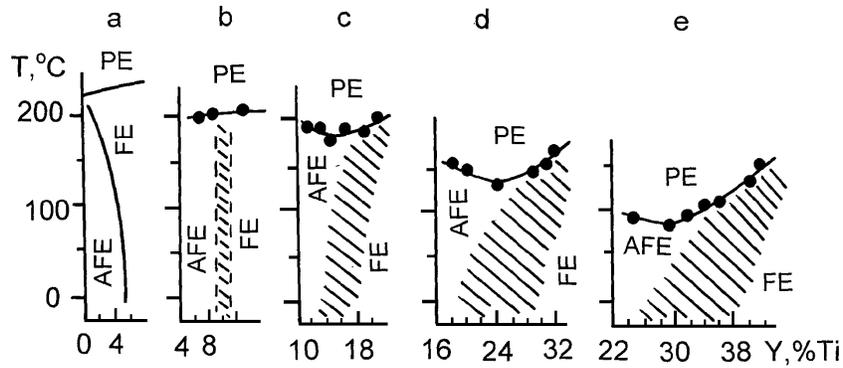

Fig.3.8. "Ti-composition–temperature" phase diagrams for $Pb_{1-3x/2} La_x(Zr_{1-y}Ti_y)O_3$ solid solutions with different La contents before the exposure to the electric field : (a) $x = 0$; (b) $x = 0.02$.; (c) $x = 0.04$; (d) $x = 0.06$; (e) $x = 0.08$.

The boundary regions (shown by dashes in figures) are the hysteresis regions for the FE–AFE transformation. X-ray and TEM studies of PLZT with compositions from this boundary region of the phase diagram [3, 23, 30] have shown that the domains of the FE and AFE phases coexist in the bulk of the sample (the transition electron microscopy images of coexisting domains of FE and AFE phases is shown in Fig.3.5). It has to be reminded that clearly manifested boundaries between the single phase FE and AFE states in the diagrams (Fig.3.8) are absent before the exposure of samples to the electric field and induction of the FE phase.

PLZT samples were obtained by the co-precipitation of components from the mixture of the aqueous solutions of lead and lanthanum nitrates and zirconium and titanium chlorides. After washing and drying the precipitates were calcinated at 550°C and 850°C. Ceramic samples were sintered at the temperatures 1320–1340 ◦C in a controlled PbO atmosphere. The grain size was from 5 to 7 $\mu$m.

As seen in Fig.3.1 and 3.8, the phase $Y-T$ diagrams for all the series with the La content higher than 4% are equivalent from the viewpoint of physics. Therefore, the largest part of our study in the scope of the present work was carried out on PLZT with 6% of La. We investigated the profiles of the X-ray diffraction lines in the solid solutions in question. Our attention was focused on (200) and (222) x-ray diffraction lines, which are the most characteristic lines for studying the crystalline structure of perovskite type. The (200) line is a doublet in the presence of tetragonal distortions of the elementary cell, whereas the (222) line is a singlet in this case. In the presence of rhombohedral distortions the (200) line is a singlet and the (222) line is a doublet.

These lines are shown in Fig.3.9. The variation of profiles of these lines with the change of Ti content in solid solutions is clearly seen in Fig.3.9.

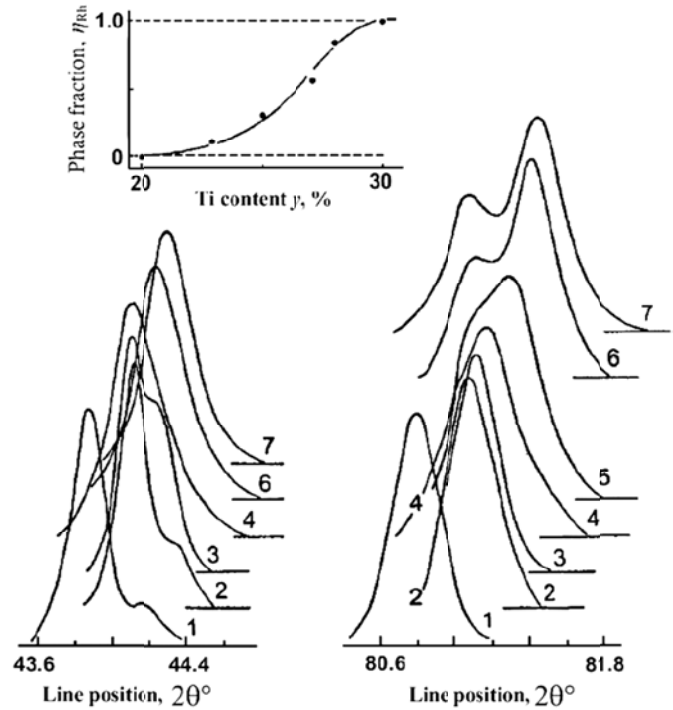

Fig.3.9. Profiles of the (200) (at the left) and (222) (at the right) X-ray lines
for the PLZT solid solutions with 6% of La [30].

The complex character of the profiles of the X-ray lines undoubtedly proves that the crystal structure is not cubic at temperatures below the Curie point. The crystal structure in the AFE state is characterized by tetragonal distortions of the elementary crystal cell, whereas the one in the FE state undergoes rhombohedral distortions. In the case of tetragonal distortions, the X-ray patterns contain weak superstructure reflexes that substantiate the AFE ordering in these solid solutions. The splitting of the X-ray lines decreases when the composition reaches the boundary region separating the FE and AFE states in the $Y - T$ diagram. The shape of the X-ray lines is also dependent upon the location of the solid solution in the $Y - T$ phase diagram. In particular, the intensity of components of the composite lines redistributes as the substance reaches the boundary region. Analysis of the intensities of components of the (200) and (222) lines allows us to obtain the dependence of the fraction of the rhombohedral (FE) phase in the sample volume as a function of Ti content. This dependence is given in the inset in Fig.3.9. The position of the peak of each component of the X-ray diffraction line depends on the Ti content in the solid solution. The dependences of these positions on Ti content for the (200) and (222) x-ray lines are presented in Fig.3.10.

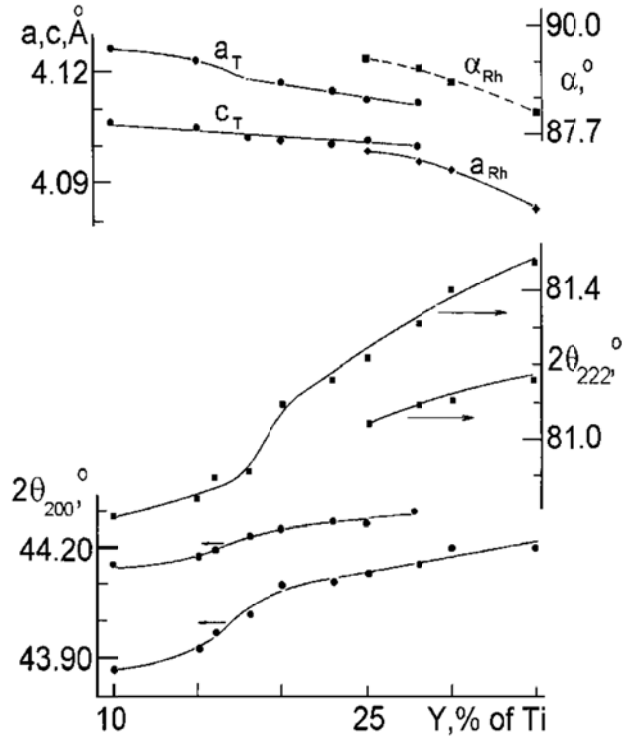

Fig.3.10. Dependences of the crystal cell parameters on Ti content in PLZT with 6% of La are shown at the top of the figure [30]. The parameters $a_T$ and $c_T$ correspond to the tetragonal (AFE) phase and the crystal cell parameter $a_{Rh}$ and the rhombohedral angle $α_{Rh}$ are for the rhombohedral (FE) phase. Dependences of the positions (the $2θ_{002}$ and $2θ_{222}$ angles) of the peaks of components for the (200) and (222) X-ray lines on Ti content in PLZT with 6% of La are presented at the bottom of the figure.

The dependence of the crystal cell parameters $a_T$ and $c_T$ for the tetragonal (AFE) phase and crystal cell parameter $a_{Rh}$ and rhombohedral angle $α_{Rh}$ for the rhombohedral (FE) phase on the Ti content is shown in the upper part of the Fig.3.9 to highlight the change of the solid solution crystal structure with the variation of composition.

The crystal structure of the solid solutions that belong to the boundary region of the $Y – T$ diagram is two phase. The phases with tetragonal and rhombohedral distortions of the perovskite elementary cell coexist. The simultaneous finite non-zero splitting of both (200) and (222) X-ray lines (see Fig.3.9 and Fig.3.10) and the widening and asymmetry of the lines (in particular, for PLZT with $x = 6\%$ and Zr/Ti = 75/25 these lines are quasi-singlet) is evidence of the two-phase structure of the solid solution with this composition. Thus, the crystal structure of PLZT with 6% of La is unambiguously non-cubic at the temperatures below the Curie point. These results also demonstrate that the identification of the structure of the PLZT with 6% of La and the Ti concentrations from the boundary region of the $Y - T$ phase diagram can be hindered by the following circumstance. The domains of the FE and AFE phases coexist in the samples with compositions that belong to the boundary region. So, one can easily make a mistake while identifying the structure of PLZT from the boundary region, if the examination the structure of the solid solutions from all the regions of the $Y - T$ diagram has not been done. It also has to be noted that the intensity of the X-ray lines superstructure decreases when the composition of PLZT approaches the boundary region moving from the region of zirconium-rich solid solutions.

The change of the character of the (200) and (222) X-ray lines that reflects a change of the crystalline structure is caused by the interaction between the coexisting FE and AFE phases (Ch.1, [3, 20]. The contribution of this interaction increases as the PLZT composition reaches the value that corresponds to equal free energies of the FE and AFE phases. At the same time, the deviations from properties of ordinary FE or AFE phases become more pronounced in solid solutions that belong to the interval of compositions where the free energies of the FE and AFE phases are close or equal. These deviations reveal themselves, for example, in measurements of the dielectric or electro-optical hysteresis loops, in the dispersion of the dielectric permittivity in the vicinity of the point of paraelectric phase transition, and in the smearing of this phase transition.

In the PLZT with 8% of La the observed picture is even more complicated. The increase of lanthanum amount diminishes the energy barrier that separates the free energy minima corresponding to the FE and AFE states. Therefore, the interaction between these phases manifests itself greatly. The boundary region in the $Y - T$ diagram becomes wider and the degree of crystal lattice distortions decreases, so the X-ray splitting is less pronounced. One more circumstance that is important has to be noted. The morphotropic phase boundary that is located at the point that corresponds approximately to the composition Zr/Ti = 53/47 [31] in the $Y - T$ diagram of $PbZ_{1-y}Ti_yO_3$ is displaced towards the solid solutions with higher percentages of Zr as the La concentration increases. At 8–9% of La, the $FR_T$-$FE_{Rh}$ morphotropic boundary is observed in the vicinity of the 65/35 Zr/Ti composition [32]. This composition corresponds to solid solutions, which are called 'relaxor ferroelectrics' and are the object of active discussions in the literature. The substance with this composition contains three phases, and it is manifested in the observed structure of the X-ray lines. Each of these lines is a superposition of the lines of FE rhombohedral, FE tetragonal and AFE tetragonal phases. As far as we know, such a fact has not been considered at the identification of the crystalline structure of PLZT and consequently, at the decomposition of the X-ray diffraction lines into simple components. Here we would like to mention that the transmission electron microscopy investigations [9, 23, 33, 34] of PLZT show that the size of the domains of the coexisting FE and AFE phases is of the order of 20–30 nm.

Complexity of identification of the crystal structure of PLZT is furthermore complemented by the specifics of the hot pressing method used for preparation of samples for which the data of crystal lattice investigations are available in the literature. According to [36] the hot pressing prevents the achievement of a high degree of homogeneity due to such factors as violation of the stoichiometry resulting from hot pressing; the "underannealing" effects caused by low temperatures applied in the process of hot pressing; and the presence of residual mechanical stresses arising during hot pressing. Even in the case when the hot-pressed PLZT samples have a high optical quality, some nanometer scale regions containing chemical elements that have not reacted completely are present in the samples' volume. In particular, this was confirmed in [36] by means of transmission electron microscopy. It was demonstrated in [37–39] that the hot pressed PLZT samples of high optical quality also contain nanodomains with the composition close to that of pure $PbZr_{1-y}Ti_yO_3$.

The results presented here explicitly demonstrate that the crystal structure of series of PLZT solid solutions with 6% of La is non-cubic at the temperatures below the Curie temperature. One can identify it undoubtedly only if the solid solutions from different regions of the $Y-T$ phase diagram are investigated simultaneously. There is a considerable probability of error in the structure identification when only one solid solution belonging to the boundary region of the $Y-T$ phase diagram (dashed regions in Fig.3.9 and Fig.3.10) is studied. It is also worth mentioning that the coexistence of FE and AFE phases or the even more complicated three-phase structure of PLZT with La content 6% could be the reason for relaxor behavior at the phase transition from the paraelectric to the dipole ordered phase [40, 41].

# 4. TWO-PHASE FE-AFE SYSTEM IN DC ELECTRIC FIELD. INTERMEDIATE STATE IN FERROELECTRICS AND ANTIFERROELECTRICS

## 4.1. Model theoretical consideration

Phase transition via the intermediate state (IS) is one of the most interesting phenomena in physics of magnetism and superconductivity. Such IS represents a thermodynamically stable state of coexisting phases between which the phase transition of the first order takes place under the action of external magnetic field. This was demonstrated by L. Landau for superconductors and by V. Bar'yakhtar and co-workers for magnetic crystals. The existence of the IS in above-mentioned substances is caused by action of the demagnetizing field resulting from the finite size of the samples. IS cannot exist in the infinite superconductor or magnetic material. Thermodynamic potentials used for phenomenological description of phase transitions in magnetic substances and superconductors on the one hand and ferroelectrics and antiferroelectrics on the other hand are akin to each other. That is why the suggestions were made that the IS analogous in its nature has to be present in ferroelectrics and antiferroelectrics. However, such IS has not been experimentally observed. It was shown in [42] that IS similar to the ones in magnetic substances and superconductors cannot exist in ferroelectrics and antiferroelectrics in principle.

However, it was shown that the IS that represents the thermodynamically stable structure of domains of the FE and AFE phases can exist. The nature of the intermediate state is vastly different from the one in magnetic substances and superconductors. The main difference of the FE and AFE substances form, for example, magnetic substances consist in the way how the field is applied to the sample [42] (form the physics point of view this difference is in the caused by the different form of one of the Maxwell equations applied to magnetic substances and dielectrics).

Let us consider behavior of the system with coexisting FE and AFE phases placed into a DC electric field, and described by the thermodynamic potential (1.1). In this case the eq.(1.4) is replaced by expression :

$$\varphi_\lambda = \varphi - \eta_{\alpha,i} E_{\alpha,i;ext} - \lambda\left(\sum_\alpha \xi_\alpha - 1\right), \qquad (4.1)$$

where $E_{\alpha,i;ext}$ is the external field conjugate to the order parameter $\eta_{\alpha,i}$. The conventional procedure of minimization of the nonequilibrium thermodynamic potential leads to the following system of equations for the determination of the equilibrium values for the order parameters

$$\xi_\alpha \left( \partial \varphi_\alpha / \partial \eta_{\alpha,i} + E_{\eta_{\alpha',i}} - E_{\alpha,i;ext} \right) = 0, \qquad (\xi_\alpha \neq 0) \tag{4.2}$$

$$\varphi_\alpha + \eta_{\alpha,i} E_{\alpha,i;int} = \lambda = const, \tag{4.3}$$

$$E_{\alpha,i;int} = E_{\alpha,i;ext} - E_{\eta_{\alpha',i}}. \tag{4.4}$$

As seen from the equation (4.3), the condition for the coexistence of thermodynamically balanced structure of the coexisting domains of the phases is the equality of their thermodynamic potentials, taking into account the external and internal effective fields.

As may see from (4.3) and (4.4.), the fields $E_{\eta_{\alpha,i}}$ are spatially varied. Therefore, at the same value of the external field $E_{\alpha,i;ext}$ AFE $\to$ FE transition may take place only in certain local regions of the sample (but not in the whole of the volume of the sample). This means that, within a certain interval of the external electric field intensity, the domains of the phases, participating in the transition, coexist in the volume of the sample. By analogy with magnetic materials or superconductors, such a state should be called the intermediate state (IS) in AFE. In view of this, let us obtain a relation between the external field and the field of the phase transition for the case of thermodynamically balanced structure of coexisting FE and AFE phases, and define the external field intensity corresponding to the onset and completion of the PT into the IS. The condition of PT is the equality of the internal field within local regions of the sample to the PT field ($E_{\alpha,i;int} = E_{pt}$). From this condition and from (4.4) we have the following relation:

$$E_{pt} + E_{\eta_{\alpha',i}} = E_{\alpha,i;ext}. \tag{4.5}$$

Now let us express $E_{\eta_{\alpha',i}}$ in terms of the share of the volume of the sample which has undergone the transition, and put down the order parameters in accordance with (4.3). Thus we obtain the dependence of the share of the phase, which has undergone the transition into FE state, on the external field intensity within the limits of the IS existence.

To write down the thermodynamic potential for each of the phases take into account that the coordinates of FE and AFE minima now depend on electric field intensity. The nonequilibrium potentials for each of the phases may be written up to the quadratic terms as (by analogy with eq.(1.8) ):

$$\varphi_1 = \varphi_{1,0}^E + U_1\left(P_1 - P_{1,0}^E\right)^2 + V_1\left(\eta_1 - \eta_{1,0}^E\right)^2.$$

(4.6)

$$\varphi_2 = \varphi_{2,o}^E + U_2\left(\eta_2 - \eta_{2,0}^E\right)^2 + V_2\left(P_2 - P_{2,0}^E\right)^2$$

The interphase interaction energy has the form (1.14):

Let us now refer to eq.(4.4). For the considered system with AFE $\rightarrow$ FE PT induced by electric field, we have the expression for the intrinsic field for the domains of the AFE phase:

$$E_{2;\text{int}} = E_{ext} - \xi_1 C_1 P_1.$$

(4.7a)

Since the equilibrium solution of $P_1$ for the thermodynamic potential of the two phase system (taking into account (2.6)) is $P_1 \cong P_{1,0}^E$ (see expressions (1.11)), we obtain:

$$E_{2;\text{int}} \cong E_{ext} - \xi_1 C_1 P_{1,0}^E.$$

(4.7b)

For the intermediate state $E_{\text{int}} = E_{pt}$. Therefore, by analogy with magnetic materials we obtain:

$$\xi_1 \cong \frac{E_{ext} - E_{pt}}{C_1 P_{1,0}^E}.$$

(4.8)

And by analogy with magnetic systems, for antiferroelectrics we have the boundaries of IS:

$$E_1 \cong E_{pt}; \quad E_2 \cong E_{pt} + C_1 P_{1,0}(E_{pt}).$$

(4.9)

(These boundaries are determined by the conditions $\xi_1 = 0$ and $\xi_1 = 1$, respectively).

The comparison of the expressions (4.8) and (4.9) with analogous ones for IS in magnetic materials [43, 44] shows that the equations which describe the IS in AFE are identical to those for magnetic substances. However, the fields $E_{\eta_{\alpha,1}}$ and $H_{diam}$ (the latter is the demagnetization field and it defines the nature and range of existence of IS in magnetic materials) differ in their physical nature. This results in some distinctions in the conditions of observation of the IS in magnetic materials and ferroelectrics or antiferroelectrics. First of all, the interval of the existence of the IS in AFE does not depend on the geometric characteristics of the sample, whereas in magnetic substances this dependence is pronounced. Moreover, in ferroelectrics the appearance of the IS is also possible when the FE $\rightarrow$ AFE PT is induced by hydrostatic pressure. In fact, with the increase of pressure the FE phase stability diminishes, but that of the AFE state increases. At some critical pressure $P_{pt}$ the thermodynamic potentials of these phases acquire the same value, and this leads to the emergence of nuclei of a new AFE phase. They induce the fields $E_{\eta_{\alpha,i}}(x,y,z)$, due to which different local regions of the sample are under different conditions. This is just the fact that gives rise to the coexistence of the domains of the phases participating in the PT.

Thus, in this section it has been shown that at PT induced in AFE (or FE) by electric field (or hydrostatic pressure), the existence of the IS is possible. Such a state represents the coexistence of the phases participating in the PT within a rather wide interval of the thermodynamic parameters initiating the transition.

### 4.2. Experimental results

4.2.1. PLLZT solid solutions

The *E-T* phase diagrams for the 10/100-Y/Y series of the PLLZT solid solutions are presented in Fig.4.1.

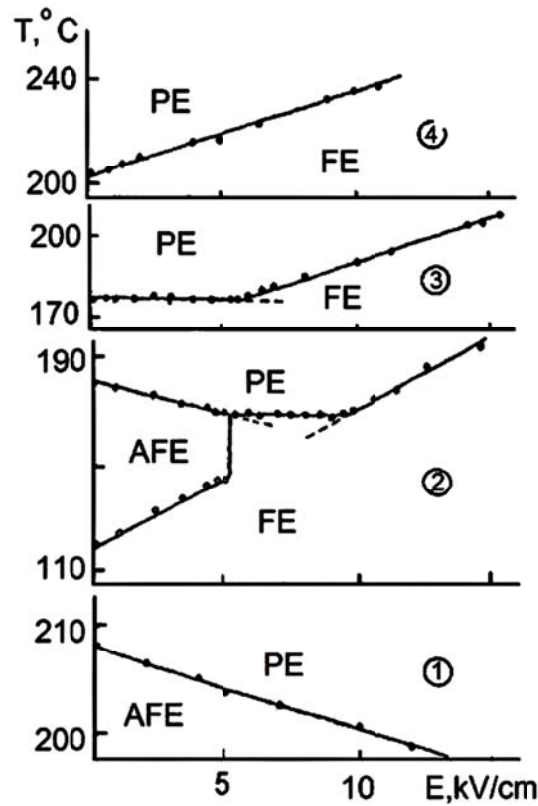

Fig.4.1. "Electric field-temperature" phase diagrams for 10/100-Y/Y PLLZT series.
Ti-content Y,%: 1 – 10; 2– 16; 3 – 18; 4 – 25.

Measurements performed in bias DC electric field show that for 10/90/10 and 10/87.5/12.5 PLLZT. The transition temperature decreases with the increase of the field, $dT/dE = -0.7 \cdot 10^{-3}$ deg·cm·$V^{-1}$. For 10/75/25 PLLZT from the FE region of the Y-T phase diagram $dT/dE = 3.0 \cdot 10^{-3}$ deg·cm·$V^{-1}$. The dependences of $T_c(E)$ on the electric field intensity for solid solutions from both the FE and AFE regions of the Y-T phase diagram are described on the base of the ordinary thermodynamic point of view.

The solid solutions from the borderland region of Y-T phase diagram possess an unusual dependence of the PE PT temperature on the electric field intensity. Fig.4.1 shows the phase E-T diagrams for 10/84/16 and 10/82/18 PLLZT solid solutions. A salient feature of these diagrams is the presence of a wide interval of electric field intensity where the PE PT temperature is constant. As seen from the E-T and Y-T diagrams, such a peculiarity manifests itself only in the border region, which separates the regions of FE and AFE states. This feature of the E-T phase diagrams cannot be described on the base of the ordinary thermodynamic point of view.

However, all the mentioned facts are easily explained based on the model concepts, discussed in Ch.1 and pt. 4.1 [42, 45]. For the solid solutions from the borderland region on the Y-T and E-T phase diagrams, FE and AFE states have approximately equal stabilities. Therefore, the domains of these phases coexist in the volume of the sample. As seen from the expression (4.8), the change in the electric field intensity leads to the redistribution of the shares of the FE and AFE phases – the intermediate state exists. The thermodynamic potentials of these phases remain unchanged (constant), till the field intensity lies within the limits of the existence of the IS. A consequence of these factors is the independence of the PT temperature on the field. If the DC field intensity exceeds the upper critical field $E_2$, the dependence $T_c(E)$ becomes a typical linear one. For 10/82/18 and 10/80/20 PLLZT $E_1$ is zero. For 10/84/16 PLLZT, the low-temperature state is antiferroelectric in the field interval up to 5 kV/cm. At higher fields there appears the IS region which then is superseded by the region of FE state.

4.4.2. PLZT solid solutions

PLZT solid solutions are characterized by somewhat unusual type of dielectric hysteresis loops, in particular, by the so-called narrow hysteresis loops which result in the appearance of quadratic ones of electrooptical hysteresis. As shown in Ch.3, they must manifest themselves in those substances where the domains of FE and AFE phases coexist and AFE-FE PT induced by electric field is realized. Such an effect was experimentally investigated in 4 series of PLZT solid solutions: 6/100-Y/Y, 7.5/100-Y/Y, 8.25/100-Y/Y and 8.75/100-Y/Y [46]. Since the Y-$T$ diagrams of all these series are physically equivalent, below we present only the results for the 8.25/100-Y/Y series. The Y-$T$ diagram of the said series is shown in Fig.4.2a.

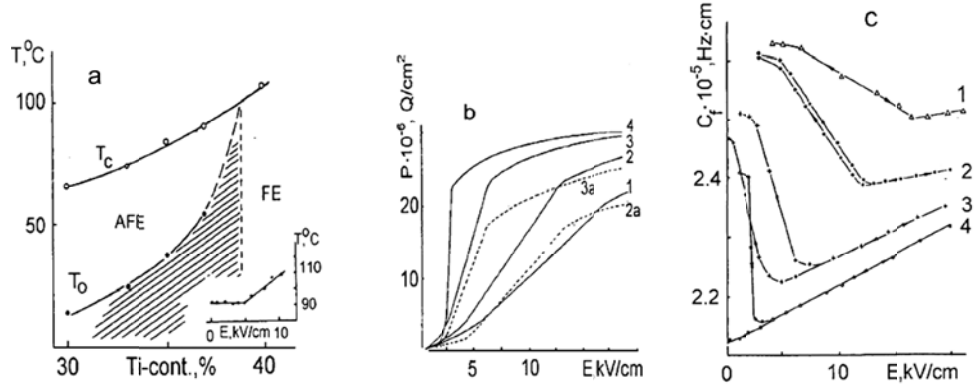

Fig.4.2. Phase Y-T diagram (a), polarization vs electric field (b), and frequency constant vs field (c) for 8.25/100-Y/Y PLZT solid solutions. Content of Zr/Ti: 1 - 72/28; 2 - 70/30; 3 - 67/33; 4 - 65/35.

First of all, it should be noted that there exists an interval of fields for the dependences $T_c(E)$ of the solid solutions located near the FE-AFE boundary in which $T_c$ does not depend on the field intensity. The mentioned effect has been discussed in the present paper for PLLZT solid solutions (sect.4.2.1).

Fig.4.2b presents the polarization dependence on electric field intensity (the samples were previously annealed at 600°C). The dependences of the resonance frequency (strictly speaking, the frequency constant $C_f = f_r\, r$, here $f$ is the frequency of the first radial resonance, and $r$ is the radius of the sample) of the first harmonic for radial oscillations for these samples on electric field are shown in Fig.4.2c. The study of the resonance characteristics allows to obtain an important information on the elastic properties, however, for understanding the problems considered in this paper the data presented in Fig.4.2 are sufficient. In contrast to $P(E)$, the dependences $C_f(E)$ are nonmonotonic ,therefore, they allow to fix the critical fields of PT more precisely.

As one can see from the curves 1 in Fig.4.2b and 4.2c, there are three intervals of external electric field, where the properties of the samples noticeably differ when the field intensity increases. In the absence of the electric field this solid solution is in the AFE state and it is preserved within the field interval $0 < E < E_{cr,1} = 4.0$ *kV/cm*. At high fields ($E > E_{cr,2} = 12.5$ *kV/cm*) the sample is in FE state. When the field intensity is in the interval of values between 4.0 *kV/cm* and 12.5 *kV/cm* the polarization linearly increases. This interval is the region of the intermediate state in the studied solid solutions. Increase of the Zr content in solid solution leads to the widening of the intermediate state region.

The results of measurements discussed above allowed us to build the "composition – electric field" phase diagrams that represent the dependencies $E_{cr,1}(Y)$ and $E_{cr,2}(Y)$. These diagrams containing the intermediate state region are presented in Fig.4.3 for 8.25/100-Y/Y and 8.75/100-Y/Y solid solutions.

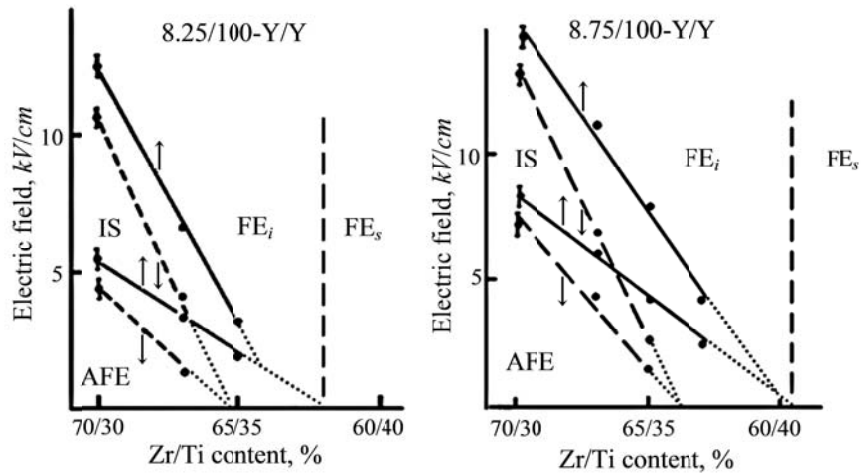

Fig.4.3. "Composition – electric field" phase diagrams for 8.25/100-Y/Y (on the left) and 8.75/100-Y/Y (on the right) PLZT solid solutions.

In this figure the notations IS, $FE_i$, and $FE_s$ are used for the intermediate state, induced FE phase, and spontaneous FE phase, respectively. As one can see, the region of the AFE phase stability is shifted toward the solid solutions with higher Ti content when the La concentration increases. The intermediate state region is also shifted in the same direction when the La content increases (see Fig. 4.3).

It should be noted that in high-intensity fields (up to 20 *kV/cm* and more) the dependence $P(E)$ is not saturated. This is bound up with the fact that the AFE domains are still preserved in the volume of the sample after inducing the FE phase. Such a situation was observed in experiments on light scattering in the solid solutions with the composition close to that of 8.25/67/33 PLZT and belonging to the same region of the *Y-T* diagram [23, 47]. The increase of polarization at $E > 12.5\ kV/cm$ is connected with the change of the internal state of the preserved AFE phase domains. The processes caused by the field-induced polarization having the form $P_i = \varepsilon E$ do not seem to be complete, since there are still no saturation on the curves 1a and 2a in Fig.4.2, which present the sample polarization after the subtraction of $P_i$. On completing the discussion of the behaviour of the 8.25/63/37 PLZT solid solution, note that the processes caused by the electric field are characterized by a weak hysteresis.

As can see from Figs 4.2b and 4.2c, when the solid solution composition moves away from the value corresponding to the equilibrium between the AFE and FE phases (that is, when the Zr concentration increases) the value of the electric field at which the induction of the FE phase takes place increases. It is not something unusual because the stability of the AFE phase increases with the increase of Zr content and one needs stronger fields for the FE phase induction.

It follows form the phenomenological consideration [47-49] that the AFE →FE phase transition has to occur in a sharp abrupt manner. However, the different behavior is observed in our experiments. The transition takes place in a finite interval of electric fields. The width of the transition increases when the solid solution composition moves away from the region of FE states in the "composition-temperature" phase diagram.

The width of the intermediate state region from (4.9) $E_{end} - E_{onc} = CP_{1,0}(E_{pt})$ is determined by the value of induced polarization inside the FE phase domain. Since the stability of the AFE state in solid solutions in question increases with increase of the Zr content, the value of the field of transition $E_{pt}$ will also increase. As follows form the phenomenological theory of the AFE→FE phase transition [30, 47-49] the $P_{1,0}(E_{pt})$ value noticeably increases when the induction of the FE state field takes place at higher values of the electric field intensity. It means that the value of $P_{1,0}(E_{pt})$ increases with the increase of the Zr content in solid solutions under investigation. As a consequence of this the interval of electric field values within which the intermediate state takes place also increases with increase of Zr content in solid solution.

The share of FE phase increases linearly with the increase of the external DC field in the region of intermediate state (in compliance with (4.8)). The increase of the share of the polar phase should lead to changes in the piezoelectric characteristics of the samples. The experimental results confirm this conclusion. Dependencies of piezoresonance parameters on electric field are presented in Fig.4.4 and Fig.4.5 for some solid solutions from the series 8.25/100-Y/Y.

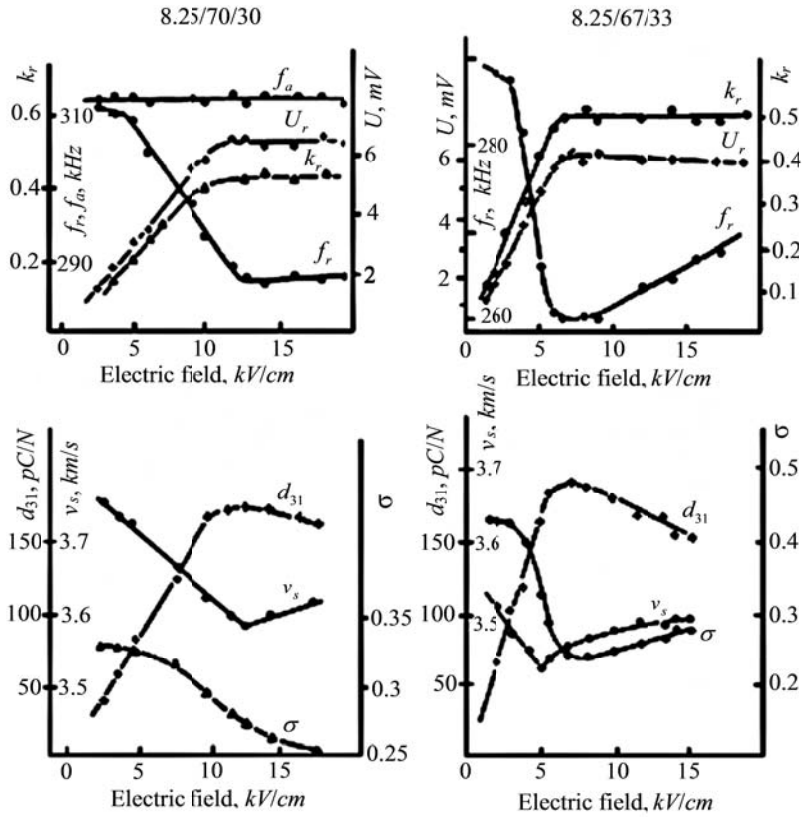

Fig.4.4. Influence of external electric field on the parameters
of the 8.25/70/30 PLZT and 8.25/67/33 PLZT (radial resonance).

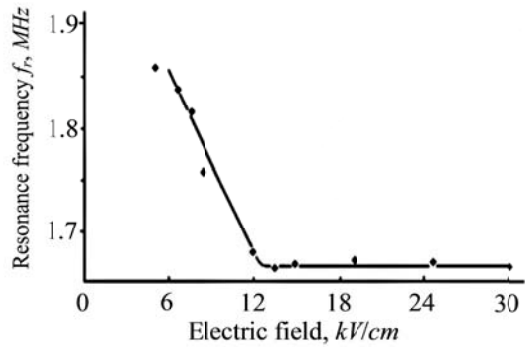

Fig.4.5. Dependence of the resonance frequency for the thickness resonance
on the external electric field intensity in 8.25/70/30 PLZT.

The resonance curves for 8.25/70/30 PLZT solid solution at different intensities of the external DC electric field are shown in Fig.4.6 as an example.

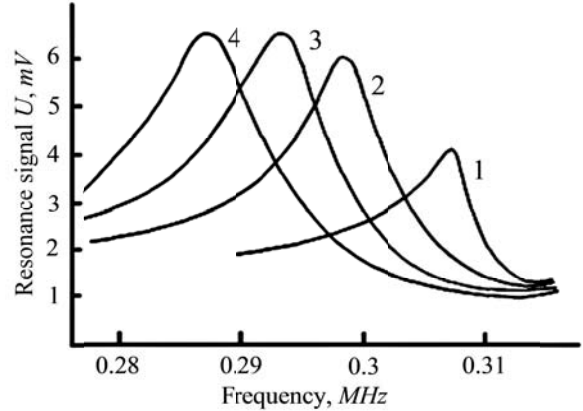

Fig.4.6.Resonance curves of 8.25/70/30 PLZT at different intensities of external electric field
($E, kV/cm$) : 1- 5.0,  2 – 8.0,  3 – 10.0,  4 – 13.0.

Above presented results show that AFE → FE phase transition via intermediate state leads to increasing of the piezoactivity of materials. This circumstance allows using this phase transition for effective control of piezoelectric parameters of material by external electric field (see Fig.4.4 - Fig.4.6). Significant change of the resonance frequency, which exceeds similar characteristics of known materials by several orders of magnitude, has engaged out attention. It should be particularly emphasized here that this change of the resonance frequency is not linked to the change of linear sizes of resonators in an electric field. The dependencies of the sample deformations on the intensity of an applied electric field given in [24] for the solid solution compositions close to the ones studied in this paper. The relative deformations less than $10^{-3}$ observed in [24] in the fields with intensities of the order of 10 $kV/cm$ cannot be compared with the relative changes of resonance frequency observed in our experiments. Moreover the relative deformations in the direction along the applied electric field and in the perpendicular direction have different signs. The dependence of the frequency of the longitudinal resonance (it is a thickness resonance for the geometry of our samples) on the intensity of the applied field at the phase transition via intermediate state is given in the Fig.4.5 for comparison. As one can see, the sign of the effect for both longitudinal and transverse (radial) resonances is the same. These data allow to neglect the changes of linear sizes of resonators in applied electric field during the analysis of the resonators behavior in external electric field.

As indicated above, the phase transition via an intermediate state accompanied by the displacement of the interphase boundaries without change of the internal state of domains of both coexisting phases ends when the share of the induced FE phase is close to unity. However, the complete transition to the homogeneous FE state does not take place in the values of the electric field intensities achieved in our measurements. The domains of the AFE phase are still present in the volume of the samples (in the FE matrix). When the intensity of the electric field is higher than the second critical field the further increase of the field intensity leads to the change of the internal state of the AFE and the displacement of the interphase boundaries is not a major factor any more. Such process causes another mechanism of resonance frequency changes. As it is seen in the Fig.4.2c, the sign of the change of the resonance frequency becomes opposite to the sign that was observed when the values of the applied field were within the interval $E_{cr,1} < E < E_{cr,2}$. The reason for this change is that the effective rigidity of the system increases (on the expense of decrease of contribution caused by the mobile interphase boundaries) when the phase state of the system becomes increasingly homogeneous. The controllability of the resonance frequency becomes higher in the fields $E > E_{cr,2}$. The hysteresis is completely absent and the changes in the piezoelectric parameters are weak.

## 5. PHASE COEXISTENCE AND IONS SEGREGATION. LONG-TIME RELAXATION.

### 5.1. Model consideration

The problem of stability for the inhomogeneous state of the coexisting FE and AFE phases has been considered in Ch.1. The negative value of $\Delta W$ in (1.13) or in (1.16) shows that the considered interphase boundary possesses a negative surface energy. It seems that the negative value of this boundary's energy should have stimulated the division of the sample volume into unlimited number of very small domains. However, it was only a simplified consideration of the problem. In particular, the effects connected with the condition of continuity for elastic medium at the interphase boundary were not taken into account.

The crossing of the IDW (from one phase to the other) goes on simultaneously with continuous conjugation of the atomic planes (free of breaks and dislocations) (see Ch.2). This coherent IDW structure leads to an increase of the elastic energy. Such increase is the more essential; the larger is the difference in the configuration volumes of the FE and AFE phases. Just this effect defines the positive value of the surface energy density for the boundaries separating ordinary domains in ferroelectrics [17, 18]. This elastic energy restricts the increase of the area of interphase boundaries and, consequently, the reduction in size for the domains of the coexisting phases. These stresses weaken the condition of the inhomogeneous state existence.

In the substances under consideration, that is the substances where the FE and the AFE states are possible, equivalent crystallographic sites are occupied by ions that are different either in size or in the value of charge or in both. In a single-phase state (inside the domains of each of the coexisting phases), the ions, forming the crystal lattice, are not subjected to the action of forces in the absence of external factors (more correctly, the resultant force affecting each ion is equal to zero). Opposite situation is observed for the ions located near the "bare" IDW. The balance of forces affecting each of these ions is upset. "Large" ions are pushed out into those domains, which have a larger configuration volume and, consequently, a larger distance between crystal planes. "Small" ions are pushed out into the domains with a smaller configuration volume and a smaller interplanar distance. Such process is accompanied with both a decrease of the elastic energy concentrated along the "bare" IDW and an increase of the energy bound up with the segregation of the substance. The considered process of ion segregation will be completed when the new-formed IDW structure provides the energy minimum. Such a "clothed" IDW will be further called real IDW or simply IDW.

Thus, the formation of the heterophase structure of the coexisting domains of the FE and AFE phases is followed by the emergence of chemical inhomogeneity of the substance. The said process is realized owing the ion diffusion at relatively low temperatures ($T < T_c$). In this case, the diffusion coefficients are too small, and the process is the long-time one. For some PZT-based solid solutions characteristic times of this process exceed 100 h at 20°C. In more details this question will be considered in experimental part.

## 5.2. Experimental results

The above-discussed mesoscopic segregations were observed in [56, 57]. However, at that time there was no complete understanding of their formation mechanism even though the dependence of the segregation parameters on the solid solutions composition (and, consequently, on the location of the solid solution on the phase diagrams), as well as the influence of DC electric field on formation of segregations were studied [57].

Now we will consider the kinetics of the process of segregation of the solid solution near IDW separating the domains of FE and AFE phases [58, 60]. The investigation was performed on two series of PZT-based solid solutions with coexisting domains of the FE and AFE phases. The substitution of the lead ions in PZT by the isovalent complex $(La_{1/2}Li_{1/2})^{2+}$ (PLLZT) or by $La^{3+}$ ions (PLZT) allows to achieve wider regions of solid solution compositions where the phase coexistence takes place [6, 20, 47, 59]. The studies were carried out on PLZT with 6% content of lanthanum, i.e. on the 6/100-Y/Y PLZT series and on PLLZT with 15 % of $(La_{1/2}Li_{1/2})^{2+}$, i.e. on the 15/100-Y/Y PLLZT series. The phase diagrams for the said solid solutions (before they were subjected to an electric field action) are presented in Fig.3.8 and Fig.3.10 for PLZT and in Fig.5.1 for PLLZT.

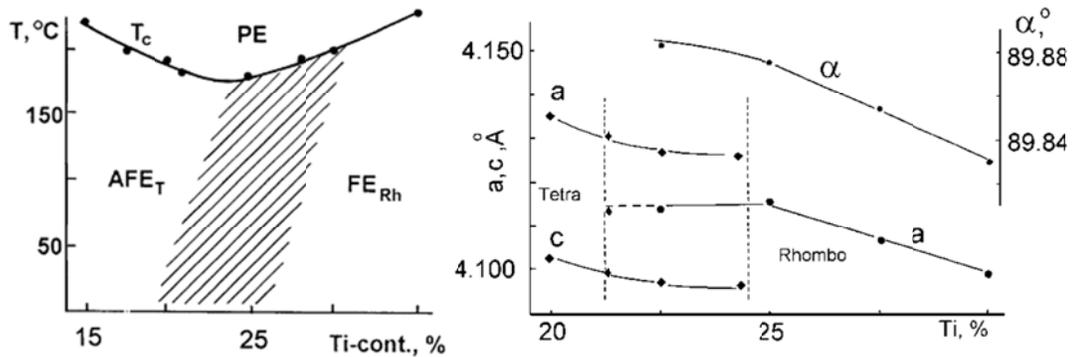

Fig.5.1. The Y - $T$ phase diagram for 15/100-Y/Y PLLZT series (on the left) and the dependence of lattice parameters on solid solution composition (on the right).

The decomposition of the solid solution is manifested in the appearance of weak diffusion lines (halos 1 and 2 in Fig.5.2, pattern 2) near the basic diffraction lines of the X-ray diffraction pattern, which are characteristic for the perovskite structure of the solid solutions [56, 58, 60].

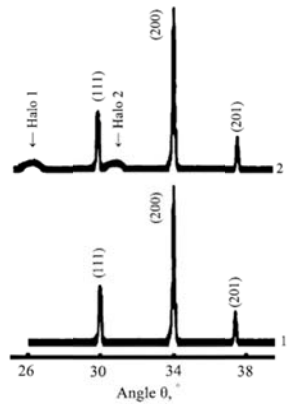

Fig.5.2. X-ray Debye–Scherrer diffraction patterns of 6/73/27 PLZT solid solution
obtained at the room temperature right after the quenching (1) and after ageing during 30 days (2).

Therefore, there arises the problem of recording of x-ray diffraction patterns in a wide range of angles during a short time interval while investigating the kinetics of the long-duration relaxation and the formation of the mesoscopic structure of segregations. For this purpose, the method of registration of scattered X-rays from the samples placed in the Debye X-ray chamber with subsequent photometry of X-ray diffraction patterns, is the most suitable (the Debye-Scherrer method). In our studies we used a chamber with a diameter of 57 mm. The registration was carried out by the method of "plane section" with filtered $CuK_\alpha$ radiation. The vanadium oxide selective absorber was used for X-ray filtration. The layer thickness was chosen experimentally. The process of registration lasted for 20 minutes (10 minutes for each of the two positions of the sample plane symmetric about the X-rays direction). The sample rotation speed was 1 r.p.s. The described method was repeatedly used in earlier investigations while studying the processes of atomic ordering in oxide ferrite substances [61, 62], as well as at examining clustered structures in PZT-based solid solutions [56, 58, 60].

X-ray diffraction measurements by the Debye-Scherrer method were carried out according to the following scheme. The samples were annealed during 22 hours at 650°C. After the annealing they were quenched at room temperature. Then the samples were aged at room temperature during the period of time τ, and the X-ray diffraction patterns were obtained afterwards.

The solid solutions belonging to the shaded region of the phase Y-*T* diagram, i.e. 15/77/23 PLLZT and 6/73/27 PLZT have been selected for the studies of the long-time relaxation. At room temperature, the domains of FE phase (with rhombohedral type of distortions of the elementary perovskite cell) and AFE phase (with tetragonal type of distortions of the elementary perovskite cell) coexisted in the bulk of the samples.

As a first step, the samples of the PLZT solid solutions were annealed at 600°C for 22 h. The X-ray patterns that were obtained at 600°C on the annealed samples and they contained only strong singlet X-ray lines caused by the coherent scattering from crystal planes of cubic perovskite lattice. It is important to note that the ferroelectric Curie temperatures of the solid solution under investigation are well below the annealing temperature, but well above the ageing temperature (room temperature).

After the high-temperature X-ray studies the samples were quenched to room temperature and then left to age at 22°C during the time interval τ. At the end of each time interval τ the X-ray patterns were recorded by the Debye-Scherrer method. Right after quenching (τ ≈ 0) the X-ray patterns contained only strong singlet diffraction lines as in high-temperature case (Fig.5.2 (pattern 1). The structure of the X-ray patterns becomes more complicated during ageing. A splitting of the singlet lines takes place. Broadened diffuse lines (halos) with a significantly lower scattering intensity appear in addition to the diffraction lines (Fig.5.2, pattern 2). These new lines are the result of the incoherent scattering from chaotically oriented segregates at the interphase ferroelectric–antiferroelectric boundaries [58, 60, 61]. We studied the behavior of the halos located in the two angle intervals $\theta = 25°–27°$ (Halo 1) and $\theta = 29°–32°$ (Halo 2). The intensity, location, and shape of the halos changed with time. The shape and location of the diffraction lines, which characterize the crystal structure of the solid solution under investigation, also changed with time.

The dependences of the elementary cell volume (the calculations were made based on the position of the (200) x-ray diffraction peak in the pseudo-cubic approximation) on the ageing time for the solid solutions under investigation are shown in Fig.5.3.

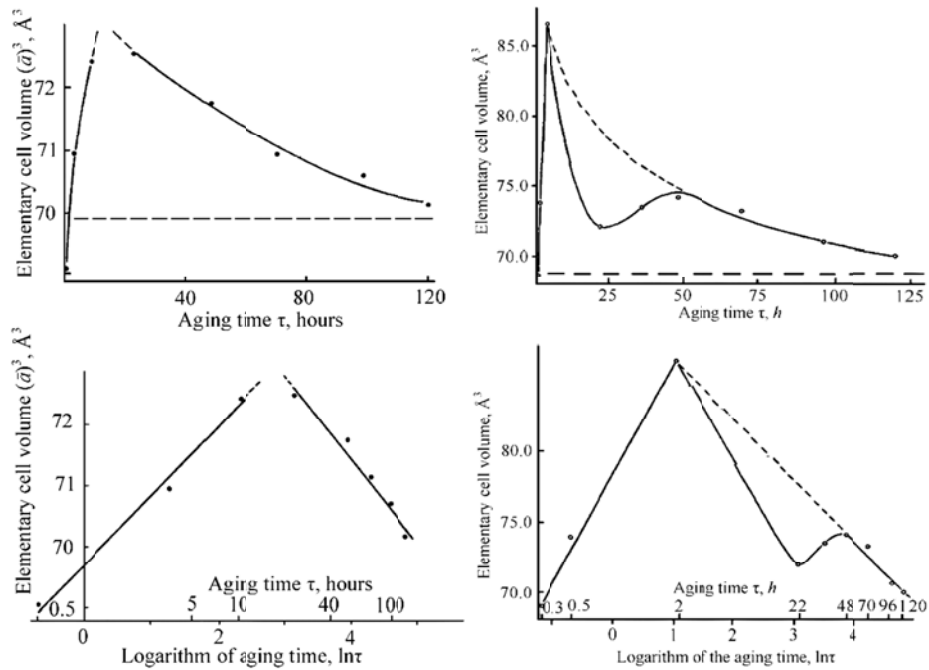

Fig.5.3. Dependence of the elementary cell volume on the ageing time for the 6/73/27 PLZT solid solution (on the left) and for the 15/77/23 PLLZT solid solution (on the right)

In the first stage of the process of ageing (of about 15–20 h for PLZT and about 3 h for PLLZT), the volume of the elementary cell of the perovskite crystalline structure of the solid solution increases. At longer ageing times, the volume decreases. As one can see, the change of the volume that occurs virtually follows an exponential behavior for both the first and the second stage of ageing. There is a peculiarity in behavior of the $V(\tau)$ dependence near $\tau \approx 22$ h which we will discuss later.

The changes of shape, intensity and position of the halos with ageing time are presented in Fig.5.4 and 5.5. The following peculiarities of these dependences attract attention. Both the profile and the angular position of the diffuse scattering lines change during the aging process. It should be mentioned that there is a clear correlation between the change of the positions of the X-ray diffraction lines (Fig.5.6) and the change of the location of the diffuse scattering lines. The dependence of the intensity of the diffuse lines on the ageing time reaches saturation in approximately 25 h.

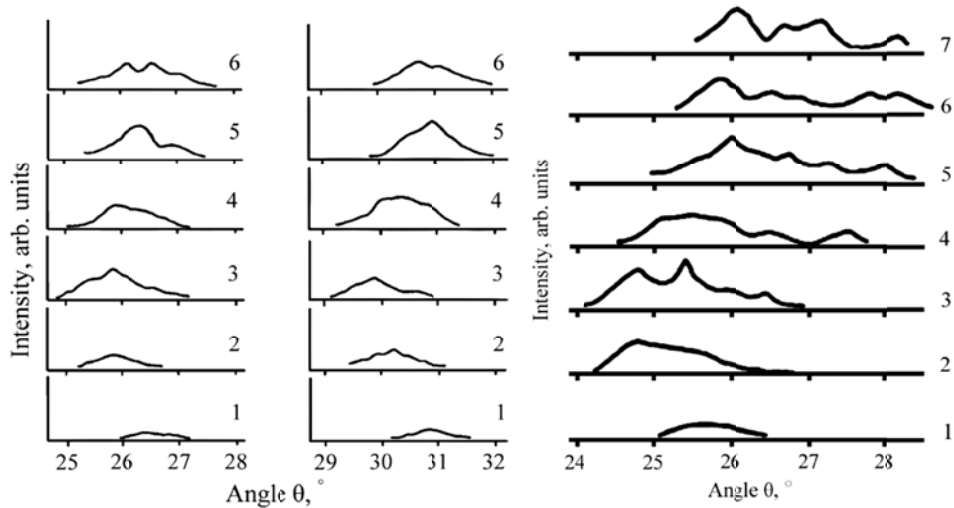

Fig.5.4. Changes of profile and position of the two halos in the process of ageing after the quenching of the 6/73/27 PLZT solid solution (on the left). Ageing time $\tau$ (hours): 1–0.5, 2–3.5, 3–23, 4–48, 5–72, 6–120.

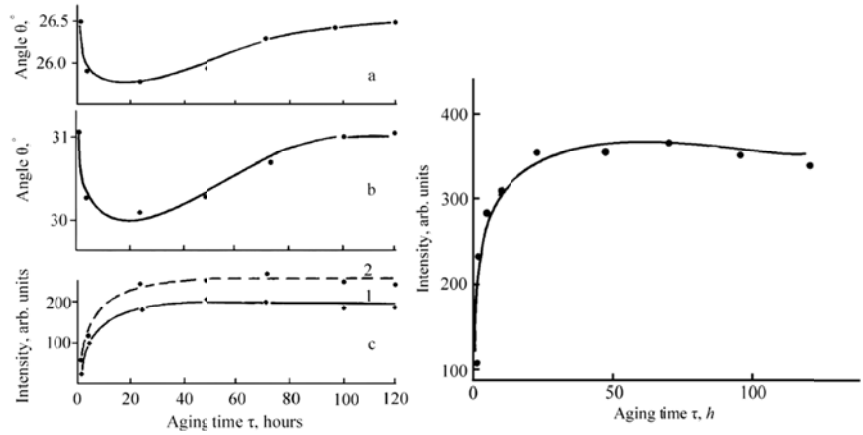

Fig.5.5. The ageing time dependencies of positions of two diffuse lines (on the left): halo 1 (a), and halo 2 (b) for the 6/73/27 PLZT solid solution. Dependencies of the intensity of diffuse lines on the ageing time $\tau$ (c): 1–halo 1, and 2–halo 2. The ageing time dependence of intensity of diffuse line (on the right) for the 15/77/23 PLLZT solid solution.

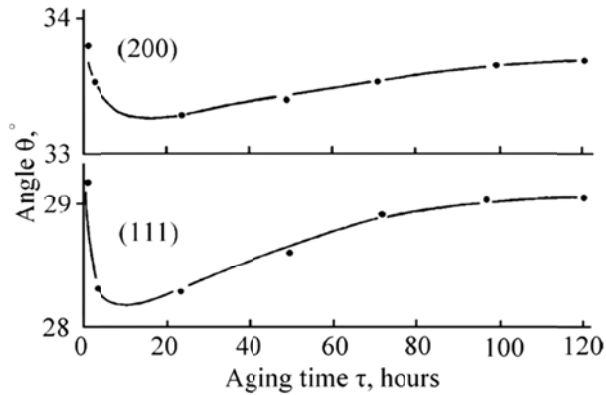

Fig.5.6. Dependencies of positions of the (111) and (200) X-ray diffraction lines for the bulk sample on ageing time for the 6/73/27 PLZT solid solution.

Changes in the shape and position of X-ray diffraction lines during the samples' ageing process are given in Fig.5.7. We analyzed the behavior of the (111) and (200) X-ray lines, which are the most typical for the perovskite structure of the solid solutions investigated. The former line is a singlet in the case of tetragonal lattice distortions, and it is a doublet line in the case of rhombohedral distortions. On the contrary, the latter line is a doublet in the first case and a singlet in the second case.

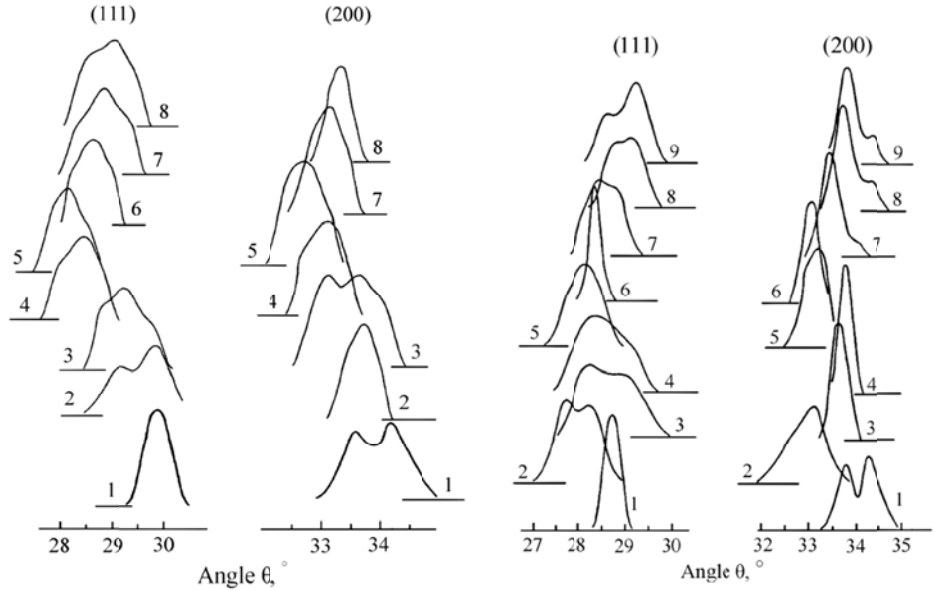

Fig.5.7. On the left: variations in the shape and position of the (111) and (200) X-ray diffraction lines in the process of the samples' ageing after quenching of the 6/73/27 PLZT solid solution. Ageing time $\tau$ (hours): 1–0.25, 2–0.5, 3–3.5, 4–23, 5–48, 6–72, 7–96, 8–120. On the right: variations in the shape and position of the (111) and (200) X-ray diffraction lines in the process of the samples' ageing after quenching of the 15/77/23 PLLZT solid solution. Ageing time $\tau$ (hours): 1–0.3, 2–0.5, 3–3.0, 4–22, 5–48, 6–72, 7–120.

Analysis of the profile of the said diffraction lines allows us to deduce a structural relationship between the low-temperature phases during the process of ageing. Immediately after quenching, the phase with the tetragonal type of distortions of the perovskite crystal structure predominates in the bulk of the sample. Then the phase with the rhombohedral type of distortions grows to dominate, with some ageing. Only with further ageing the low-temperature phases, namely, the phase with the tetragonal type of lattice distortions and the phase with the rhombohedral type of lattice distortions coexisting as the equilibrium two-phase structure are established in the bulk of the sample.

The domains of the FE and AFE phases coexist in the sample volume of the investigated solid solutions. It is known, that the characteristic time of establishment of the equilibrium low-temperature state at the structural phase transitions is of the order of $10^{-4}$–$10^{-6}$ s [63]. The paraelectric–ferroelectric or paraelectric–antiferroelectric phase transitions are also structural ones. Therefore, all the processes that we investigated in this study (after speed cooling from 600°C) take place in a system which is essentially a two-phase system containing domains of coexisting ferroelectric and antiferroelectric phases. Thus, one has to take into account the two-phase nature of the system and the presence of the interphase boundaries while interpreting the results.

The time dependences of the shape and the intensity of diffuse X-ray lines (halos), as well as the absence of the said lines in the X-ray patterns obtained at $600^\circ$C, confirms the connection between the long-time relaxation and the formation of segregates in the vicinity of the interphase ferroelectric to antiferroelectric phase boundaries. The establishment of an equilibrium state is a long-time process in solid solutions in which the state of coexisting ferroelectric and antiferroelectric phases is possible. As one can see from the X-ray data, it continues for not less than 120 h. However, taking into account the limited sensitivity of this method one can assert that this process takes even longer.

The segregation processes are multistage. It is clearly seen from the results given in figures 2–6 that there are different relaxation times caused by different mechanisms. In addition to segregation, one has to note that the mechanisms responsible for establishment of equilibrium values of the structural order parameters occur on the time scale of $10^{-4}$ s (such time intervals are beyond the abilities of our experimental methods). Without elucidation of particular mechanisms responsible for attainment of the equilibrium state, one can assume that the long time constant of this process is connected with the diffusion processes associated with the local decomposition of the solid solutions along the interphase domain boundaries. The estimation of the size of the segregates (using the shape of the diffused X-ray lines) gives values of 8–15 nm [56, 57] (similar approach to estimations of the average size of nanoregions one can be found in [64-66]).

Long-time relaxation processes are non-monotonic processes due to the condition of 'strong deviation from equilibrium' in the initial stages following quenching. In the case of 'weak deviation from equilibrium' (at the final stage) the relaxation process is monotonic and is described by an exponential law. Nonetheless, the PLZT and PLLZT solid solutions do differ by the presence of vacancies in the A-positions of the crystal lattice of the PLZT system of solid solutions. There is a peculiarity of the ageing process for the PLLZT system of solid solutions occurring at ageing times 20–30 h, which was attributed to accumulation of elastic stress and the subsequent drop in strain [60, 58], not present in the PLZT system. Otherwise, the compositional (and structural) relaxation process follows similar patterns for both the PLZT and PLLZT solid solutions systems.

Now let us dwell on the mechanisms defining the kinetics of the processes in question at different periods of time (stages). There are different mechanisms, which contribute to this long-time relaxation. Two of these mechanisms should be pointed out among the others. The contribution of the crystal lattice defects, in particular oxygen vacancies and the diffusion of the cations in the vicinity of the interphase boundaries, caused by local mechanical stresses, command the greatest attention. Under the conditions of our experiments the concentration of the vacancies in the lead sites remained practically constant because the volatilization of lead in the PLZT solid solutions starts only at temperatures $T > 800°C$. The difference in size (and, consequently, in mobility) and charge of the ions, which are located in the equivalent sites of the crystal lattice, should be taken into account. At the same time, a permanent rearrangement of the multiphase domain structure also takes place. This domain structure rearrangement is due to the change of the local composition of the solid solution and, as a consequence, to the change of the local phase stability. Complete analysis is still beyond our grasp as the influence of the oxygen sublattice defects on the crystalline structure of these solid solutions is investigated insufficiently at present, but by drawing from the experimental results obtained for related oxide materials with perovskite or perovskite-type structures, some additional insights are possible.

Annealing of the samples at $600°C$ leads to the growth of the concentration of oxygen vacancies; the equilibrium concentration of the oxygen vacancies grows rapidly as the temperature rises. Quenching down to room temperature leads to freezing of the nonequilibrium elevated concentration of the vacancies in the bulk. Oxygen vacancies in ionic–covalent compositions, to which the solid solutions with perovskite structure belong, lead to an increase of the crystal lattice parameters [67-70]. Alongside the increase in lattice parameter, vacancies in perovskite and perovskite-like compositions favor the increase of the stability of the phases with tetragonal type of crystal lattice distortions, caused by a static $T_2 \times e$ Janh-Teller effect [70, 71] (for example, on $Ti^{3+}$ ions in PZT, $BaTiO_3$ et. al.).

During the ageing process the oxygen vacancies move towards the sample surface and leave the sample (actually, the diffusion of oxygen into the bulk of the sample across the surface takes place) and, therefore, the crystal lattice parameters decrease. Since at room temperature, the diffusion coefficient is comparatively low, and the surface maintains a steady state concentration of vacancies, the said process is the long-time one.

Let us consider the ion diffusion in the vicinity of the domain boundaries separating domains of the ferroelectric and antiferroelectric phases. Mechanical stresses arise at these interphase boundaries after the quenching and formation of domains of the coexisting ferroelectric and antiferroelectric phases inside the sample's volume. These stresses are caused by the difference of the interplanar distances in the neighboring domains. There is an associated increase in the elastic energy. This increase in strain is reduced by the redistribution of the ions in the vicinity of the interphase boundaries and, as a result, by the local decomposition of the solid solution and the formation of segregates. Since mechanical stresses are now the motive force of ion diffusion, this process must have a higher rate than the process of establishment of the equilibrium concentration of the oxygen vacancies. As is seen from Fig.5.5(c), the dependence of the intensity of the diffuse X-ray lines, connected with the formation of the segregates, on the ageing time reaches saturation already in 20–25 h.

Ions that essentially differ in ionic radii (and charge) participate strongly in the diffusion processes, which define the formation of the segregates and the local decomposition. These are lead ions $Pb^{2+}$ and lanthanum ions $La^{3+}$, which occupy the A-sites of perovskite crystal lattice, and zirconium $Zr^{4+}$ and titanium $Ti^{4+}$ ions, which occupy the B-sites of the crystal lattice. Obviously, the rates of their diffusion differ as well. This leads to the change of the chemical composition both of the segregates and the solid solutions inside the domains with time. As the ions with smaller ionic radii from the domains of one of the coexisting phases reach the interphase boundary, the solid solution inside the domains becomes enriched with 'larger' ions. Consequently, the position of the solid solution in the phase $Y-T$ diagram changes in time, the crystal lattice parameters increase and the type of the crystal lattice distortion changes. This is clearly seen in Fig.5.5. When at the first stage of the local decomposition "small" lanthanum and titanium ions reach the interphase boundaries, the composition of the solid solution inside the domains correspond to PZT with an elevated content of zirconium. Such solid solutions are characterized by the rhombohedral type of crystal lattice distortion. Therefore, as the intensity of the diffuse lines grows the profile of the X-ray diffraction lines changes. This circumstance points to the fact that the predominating amount of the tetragonal phase is replaced by that of the rhombohedral phase. That is, in the bulk of the sample the share of the phase with rhombohedral distortions increases.

At this stage of the process the change of the elementary cell volume is defined by the competition of two processes – the reduction of the volume owing to the decrease of the concentration of the oxygen vacancies and the increase of this volume due to the enrichment of the composition inside the domains with the 'larger' ions. The diffusion of oxygen vacancies is the slower process. As a result, in the first stage of ageing the volume increases and a maximum value is achieved in approximately 20 h.

As further ageing takes place, the 'larger' lead and zirconium ions reach the interphase boundaries and the composition inside the domains approaches its nominal formula composition; the solid solution regains its stable location on the composition versus temperature phase diagram (that is, a return to the two-phase region of the state diagram takes place). In the profiles of the X-ray lines this fact manifests itself (after approximately 50–60 h of ageing) by gradual establishment of precisely such shape that is characteristic of the domains after sintering and ageing on a very long timescale (the line shape that is established after one year).

The changes in the diffusive scattering profile also take place during the above-described ageing process. These changes confirm the fact that the concentration of different ions in the segregates varies at the interphase boundaries. The profile of the diffusive lines is defined by the resulting enveloping curve obtained after summation of the X-ray scattering from the crystal planes in the segregates. Since the local chemical composition of the segregates constantly changes in the process of ageing after quenching, the profile of the diffusive lines changes as well.

As we have indicated in Ch.4 (pt.4.3), many difficulties in the identification of the crystal structure are associated with the presence of coexisting domains of the FE and AFE phases. Local decomposition of the solid solutions in the vicinity of the boundaries between the coexisting phases and the formation of the mesoscopic structure of segregates at these boundaries in the bulk of the samples poses a particular problem. In the X-ray diffraction patterns, it manifests itself in the appearance of supplementary diffuse lines that accompany the diffraction lines, on which basis the PLZT crystal structure is actually identified. Often these diffuse lines may appear as satellites of the main Bragg peaks. In this case, in the process of mathematical treatment of the experimental X-ray diffraction patterns according to the existing programs a third phase is brought into consideration. We have faced such a phenomenon while treating the experimental results. This is connected only with the fact that in the development of the corresponding programs the phenomenon of local decomposition of the solid solution has not been taken into account. As far as we know, the effects associated with the coexistence of domains of the ferroelectric and antiferroelectric phases and the local decomposition of the solid solution in the vicinity of the interphase boundaries have not been discussed in the literature up to now.

In conclusion, let us consider the peculiarity of the time dependence of the elementary cell volume in PLLZT solid solutions which manifests itself at the adding time intervals of 22 h. The X-ray patterns obtained for the 10/77/23 PLLZT solid solution after different aging time intervals are presented in the Fig 5.7. A pronounced complex structure of the (200) X-ray line is clearly seen for the aging time interval of 22 h. We carried out the estimation of sizes of the regions of the coherent scattering after different aging time intervals, namely, for $\tau = 22$ h and $\tau = 48$ h for comparison. In the first case, the size of the above regions was $3 \cdot 10^{-4}$ cm, and in the second case, it was $1.5 \cdot 10^{-4}$ cm. It has to be noted that the regions of the coherent scattering are nonuniform and this circumstance leads to the oblong shape of reflexes in diffraction pattern obtained after the aging time of 22 h. The ratio of the long $a$ and short $b$ axes of the coherent scattering region is $a/b \approx 2.25$. In the case of $\tau = 48$ h the shape of the coherent scattering regions is close to equiaxial.

Complex structure of the (200) X-ray line is the evidence of the presence of texture in the samples after the aging time interval $\tau = 22$ h. This texture was absent at the shorter intervals of aging time. The said texture disappears again at $\tau = 48$ h and longer intervals of aging time. It can be attributed to the presence of internal mechanical stress and to the increase of the elastic energy caused by this stress. Such processes are well known and studied for the process of alloy aging (accompanied by the process of local decomposition) [72]. The relaxation toward the equilibrium stress-free state (with smaller value of elastic energy) takes place by means of ordering in the system of segregates-precipitates including appearance of the texture in the samples. Hence, we attribute the peculiarity in the $V(\tau)$ dependence at $\tau = 22$ h to the appearance of internal mechanical stresses, which are a consequence of the above-considered diffusion processes accompanied by the local changes of chemical composition of the solid solution. This peculiarity in the $V(\tau)$ dependence disappears in the process of further aging because of the relaxation of mechanical stresses. The absence of such peculiarity in the $V(\tau)$ dependence for the PLZT solid solution is explained by the fact that mechanical stresses relax by means of redistribution of vacancies in the regions where mechanical stress appears. Let us remind that in the PLZT solid solutions (also investigated in this paper) each two lanthanum ions lead to appearance of one additional vacancy in A-cites of the crystal lattice.

Thus, in spite of similarity of the processes of local decomposition in PLZT and PLLZT solid solutions some differences are connected with essentially different number of vacancies in the crystal lattice in these systems of solid solutions and therefore with the presence of additional mechanical stresses in the PLLZT solid solutions.

## 6. TWO-PHASE NUCLEATION AND DIFFUSE PHASE TRANSITIONS

### 6.1. Model consideration

In Ch.1 and 3 we considered stability of the two phase state at the temperatures lower then Curie point. Some physically correct conclusions regarding the peculiarities of the behavior of the considered system may be also made for temperatures close to $T_{c,f}$ ($T_{c,f} > T_{c,af}$) and higher than one. Within the interval of temperature close to PE transition the thermodynamic potential of the system has only one minimum. Therefore, the substance described by such a potential is ferroelectric, and the appearance of AFE phase domains at temperatures near $T_{c,f}$ is possible only in the form of fluctuations. The said domains, arising in the volume of the substance, interact with the FE matrix. Such the interaction changes the density of the thermodynamic potential in the volume of the sample within which it takes place. For the other parts of the volume the density of FE state potential remains unchanged.

Thus, for the volume, in which the interphase interaction takes place, the density of the nonequilibrium thermodynamic potential may be written as [3, 20]:

$$\varphi = \frac{\alpha_1}{2} P_1^2 + \frac{\alpha_2}{4} P_1^4 + \frac{\beta_1}{2} \eta_1^2 + \frac{A}{2} P_1^2 \eta_1^2 + \frac{m}{2} P_2^2 + \frac{n}{2} \eta_2^2 + C P_1 P_2 + D \eta_1 \eta_2 \qquad (6.1)$$

In this expression the first four terms correspond to the density of the nonequilibrium thermodynamic potential of the FE phase for $T > T_{c,af}$ ($\beta_1 > 0$, so we assume that $\beta_2 = 0$), the next two terms correspond to the density of the nonequilibrium potential of the fluctuational domains of the AFE phase ($m > 0$, $n > 0$), and the last two terms represent the interphase interaction.

The minimization of (6.1) with respect to $P_2$ and $\eta_2$ yields:

$$P_2 = -(C/m)P_1; \qquad \eta_2 = -(D/n)\eta_1. \tag{6.2}$$

Substituting the obtained expression into (6.1) we have:

$$\varphi = \frac{1}{2}\left(\alpha_1 - \frac{C^2}{m}\right)P_1^2 + \frac{\alpha_2}{4}P_1^4 + \frac{1}{2}\left(\beta_1 - \frac{D^2}{n}\right)\eta_1^2 + \frac{A}{2}P_1^2\eta_1^2. \tag{6.3}$$

As one can see from the last expression, the interphase interaction leads to the renormalization and increase of the Curie point (see the coefficient at $P_1^2$). Moreover, under certain conditions this interaction may also stabilize the AFE state in some part of the sample's volume (see the coefficient at $\eta_2^2$). This means that there is a possibility of existence of the two-phase (FE+AFE) domains at temperatures exceeding $T_{c,f}$.

More consistent approach to obtaining the change of the Curie temperature in the region of the crystal where the interphase interaction is present can be achieved using equations (1.13) and (1.7). It is necessary to substitute eq. (1.13) into eq. (1.7) and take into account that $P_{1,0}^2$ and $\eta_{2,0}^2$ are the values of these parameters that do not take into account the interphase interaction. Coordinate dependencies of $C_1(x, y, z)$ and $D_2(x, y, z)$ give the distribution of Curie temperature values in the sample volume and as a consequence the inhomogeneous state of domains of the FE and AFE phase in the paraelectric matrix of the substance.

As one can see from (1.13), (1.16) and (6.3), the borderland interphase regions have a lower free energy in comparison with the neighbouring FE and AFE domains. Therefore, at lowering the temperature from high ones, the space regions of the phase with $P \neq 0$ and $\eta \neq 0$ are the first to nucleate in the PE matrix of the sample. Then, with further lowering the temperature, the domains of the FE and AFE phases are added to these regions, and for $T < T_{c,f} \cong T_{c,af}$ they will turn out to become the boundary regions separating the domains of the coexisting FE and AFE phases.

Our simplified approach does not take into account the effects connected with the presence of elastic stresses caused by the phase coexistence and the interphase boundaries. Therefore some additional remarks have to be done. It is a general knowledge that the configuration volume increases in the course of the PE-FE phase transition, whereas at the transition into the AFE state this parameter decreases. Therefore the existence of the domains of each phase separately at temperatures exceeding the Curie points is accompanied by the appearance of essential elastic stresses, and is not advantageous from the energy point of view.

However, the increase of the energy does not take place for the complex two-phase (FE + AFE) domain. In this case the configuration volume does not vary for the two-phase domain as a whole. The presence of domains of such structure allows to remove the elastic stresses and to raise the dipole disordering temperature due to the interaction (1.9) as much as possible (i.e. to decrease the free energy to the utmost). The ratio of the phase volumes in the two-phase domain which is present in the PE matrix of the crystal is defined by both the relative stability of the FE and AFE phases (i.e. by the relative difference of their free energies) and the changes in the elastic energy, which may be brought about by each of the phases. A characteristic property of such a domain is the fact that it should easily match the varying external conditions: electric field, pressure (including those of the PE matrix). An insignificant polarization of this domain, almost complete absence of elastic stresses and the absence of distortions in the PE matrix, where the domain arises, provide its existence at temperatures exceeding both $T_{c,f}$ and $T_{c,af}$.

For simplicity the latter phenomenon may be thought of as a complex two-phase nucleation. It is realized in the form of a spatial domain which lies within the PE matrix of the crystal.

Now let us proceed to considering the PE-FE transition itself. The diffusion of paraelectric PT usually is bound up with the "entrapment" of the low-temperature (LT) phase remnants into the high-temperature region with $T > T_c$. Thus, the ordered low-temperature state is the starting-point for the process in question. On the contrary, in PT physics it is the symmetry and properties of the high-temperature (HT) phase that define all possible LT phases and PT under any temperature changes. This idea should be a base while considering diffusive phase transition (DPT).

Typical temperature dependence for the inverse dielectric constant (curve 2) in the vicinity of the FE-PE PT point is shown in Fig.6.1 [73].

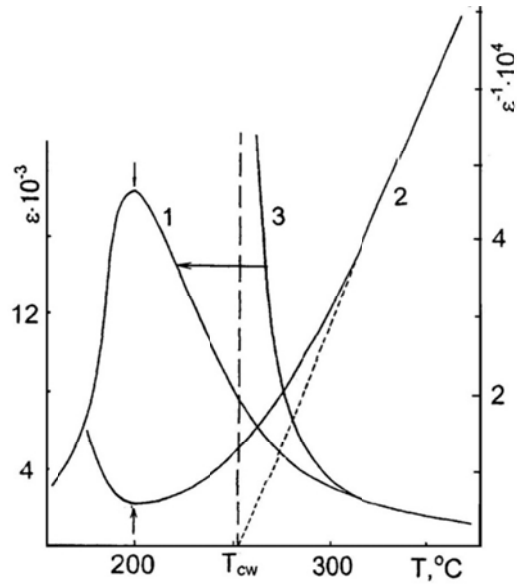

Fig.6.1. Temperature dependences of ε (1) and 1/ε (2) for ferroelectric with diffuse phase transition and ε(T) calculated from Curie-Weiss law dependence (3). Arrow shows the shift of ε(T) dependence under effect of ferroelectric-nonactive impurity [73].

Hear a special attention has to be paid to the value of the Curie-Weiss temperature $T_{cw}$, since it is above the Curie point (to the right of ε(T) dependence maximum). Up to now this fact has been neither explained nor even discussed in the literature. However, from the viewpoint of PT physics it is just at $T_{cw}$ point that the transition into the ordered state must be realized. (The transition from PE into FE state is due to the condensation of the polar mode with the wave vector in the center of the Brillouin zone. This mode is responsible for the increase of ε while approaching the Curie point.). Curve 3 built in agreement with the Curie-Weiss law presents the temperature dependence of ε which reflects the properties of the HT phase and which must describe the behaviour of dielectric constant when approaching the Curie point from above. Nevertheless, in the vicinity of the transition the experimental curves ε(T) (curve 1) and, consequently, 1/ε(T) are shifted towards lower temperatures. In other words, at lowering the temperature the PT is blocked, displaced to the region of lower temperatures. Moreover, due to the diffuseness of the transition the said blocking should differ in intensity in different local regions of the crystal.

In our opinion, the main problem in diffusive phase transition (DPT) physics is to clarify the mechanism (or mechanisms) of "pressing out" the ordered phase towards the low-temperature region. Our recent results including those presented in this paper allow proposing a mechanism of "pressing out" paraelectric PT towards the low-temperature region with its simultaneous diffuseness. The decisive factor which defines the said PT behaviour is the above-considered mechanism of ions separation at the interphase FE-AFE boundary into the two-phase domains.

The dynamics of DPT may be presented as follows. At the first step, in the process of cooling from high temperatures, the behaviour of such a substance corresponds to the classical theory of phase transformations. The temperature dependence of dielectric constant obeys the Curie-Weiss law. Then the discussed two-phase domains appear at cooling. At the interphase FE-AFE boundary, "separation" of the elements, which form the crystalline lattice of the given substance, is initiated due to the difference in their ionic radii. There appear local segregations with "foreign" composition and mesoscopic dimensions; in the general case they are not FE- or AFE-active. It is just at this step that the deviation from the Curie-Weiss law manifests itself on the $1/\varepsilon(T)$ and, consequently, $\varepsilon(T)$ dependences. In the process of cooling the share of such segregations in the volume of the substance increases and the deviation from the Curie-Weiss law becomes more noticeable.

However, the passive dielectric segregations not only reduce both the share of the FE component in the substance and the value of dielectric constant. Resent achievements of the physics of inhomogeneous solids testify that the introduction of non-magnetic impurities into magnetic materials or of passive dielectric ones into ferroelectrics and antiferroelectrics suppresses the formation of low-temperature ordered phase. The literature on this problem is ample enough, but here the author would like only to dwell on the well-investigated case of percolation systems. At present it is shown [74] that when approaching the percolation threshold from the ordered phase (this case corresponds to the increase of the content of passive impurities in the system), the PT temperature decreases and tends to zero at $x \to x_c$. At present this problem is investigated in details not only on point defects of crystal lattice, but also on extended (mesoscopic-scale) inhomogeneities [75]. Similar mechanisms work in the systems under discussion. The segregations which are developing in the process of cooling "press out" the LT phase and, consequently, PT, towards lower temperatures and lead to PT diffusion. The latter is explained by the fact that in the places of segregation development it is impossible to find out pronounced boundaries separating the active FE and non-active parts of the sample. Moreover, there exist local mechanical stresses which are not compensated due to the segregation of the chemical elements.

In this connection it should be noted that for the majority of investigated solid solution systems (where the change in the content of one of the components results in the substitution of FE ordering by AFE one), the concentration dependences $T_c(x)$ are characterized by a fall in the vicinity of the phase boundary between the regions of the said states in the *x-T* diagrams. As seen from the results presented here, it is another convincing reason to substantiate our ideas of the blocking near the triple FE-AFE-PE point.

The above-said is valid for AFE PT, too. The distinction from FE PT consists only in the fact that the transition from PE into AFE state is caused by the condensation of the non-polar mode with the wave vector at the boundary of the Brillouin zone, whereas the increase of dielectric constant at $T \rightarrow T_c$ results from softening (but not condensation) of the polar mode with the wave vector in the center of the Brillouin zone. The letter mode interacts with the soft non-polar mode. Therefore, in the case of AFE the extrapolation of the straight-line part of the dependence $1/\varepsilon(T)$ does not yield the point of AFE PT (i.e. the point of stability loss for the high-symmetry phase with respect to relatively small anti-polar displacements of the crystal lattice ions).

Thus, in the proposed model of DPT the basic role belongs to chemical segregations caused by the difference in the configuration volumes of the FE and AFE phases. Such segregations were revealed experimentally and reported in [56]. At that time the mechanism of their formation was not clear. Detailed studies of dependence of properties of segregations on the solid solution's composition (the position of the solid solution in the phase *x-T* diagram) carried out in [57]. The influence of DC electric field on properties of segregations was also studied in [57]. At present we have experimental results concerning the development of the process of chemical segregation at the interphase boundaries in time.

The diffusion processes bound up with inhomogeneous deformations of solids have already been considered in the literature. As an example, see the Gorsky effect ([76], sec.11.2).

We would like to note that the considered "stratification" must also take place at the FE(AFE)-PE interphase boundaries. However, in this case the change in the interplanar distances while crossing such a boundary is essentially lesser in comparison with that for the FE-AFE boundary. Therefore, the driving force of the process will be essentially lesser, too. In our opinion, the decomposition of solid solution near the latter boundary should are decisive.

In the upper proposed is a mechanism of DPT into paraelectric state for substances characterized by a faint difference between the energies of FE and AFE ordering. Now we shall show that the ideas in consideration are applicable to the understanding of the nature of DPT in a vast class of FE and AFE

The results of experimental investigations of oxygen-containing octahedral substances obtained in the course of many years show that for all real FE and AFE the distinction in the energy for different types of dipole ordering is weak. From the viewpoint of free energy this means that for most oxides with the perovskite structure the free energy in the space of the order parameters of FE and AFE phases has two minima at rather low temperatures. The expansion of the free energy (described in terms of the Landau theory) is to be realized with respect to the powers of two order parameters. However, since the depths of the minima are different, for some substances the metastable minimum does not become stable. Under such conditions only one phase transition (FE-PE or AFE-PE) will be observed in simple experiments, and it is sufficient to take into account only one order parameter in the Landau expansion. But this by no means signifies that the presence of the second minimum cannot be revealed in experiments.

Now let us refer to Fig.6.2. The lower part of this figure contains the generalized phase diagram. Here the temperature is shown on the ordinate and the external parameter (e.g. hydrostatic pressure P or chemical composition $x$) are plotted on the abscissa. There is the triple FE-AFE-PE point in this diagram. Near the FE-AFE boundary low-temperature phases coexist. As has been discussed above, the diffuseness of PE phase transition greatly depends on the type of LT state (see the upper part of Fig.6.2). For the regions of FE or AFE states remote from the triple point diffuseness parameter $\delta$ is small. As the triple point is approached when the external pressure $P$ or concentration $x$ are changed, the diffuseness of PT increases.

In the generalized diagram of states the majority of known FE or AFE are located within the interval 1-2 or 3-4. (The particular location may be determined only from the complete set of phase diagrams of the substance in consideration *x-T, P-T, x-E, x-P, E-P, E-T*, etc.).

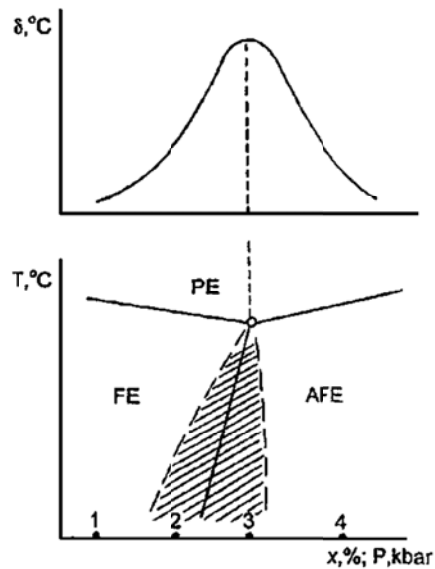

Fig.6.2. Generalised diagram of phase states (below) and the dependence of diffuseness parameter on external factors (above) near the triple point FE-AFE-PE [73].

If the location of the substance corresponds to the points 1 or 4 (or approaches them), then the free energy is characterized by the presence of only one minimum, and the substance manifests the properties of an ordinary FE or AFE. The experimental data may be easily interpreted. But when the location of the given substance approaches the point 2 or 3 in the diagram of states, there appears the second (metastable) free energy maximum, and its depth approaches that of the minimum corresponding to the stable state while reaching the phase boundary. In this case it is difficult to interpret the experimental results. Many physical properties of these compounds are defined by the coexistence of domains with FE and AFE dipole ordering in the volume of the substance. Though real FE-AFE transition is not observed in most experiments, such coexistence takes place indeed. What new states have been devised to account for this fact! But the explanation is quite simple: for this purpose one must only have the complete set of phase diagrams.

In view of the above-said it is clear that the experiments must be performed in a way allowing to realize the transition from the point 1 to the point 4 in the diagram of states. This can be achieved either by changing the solid solution composition or by applying external factors. In the latter case the experiments using hydrostatic pressure will have the most unambiguous interpretation since then it is easy to reveal the origin of "unusual" properties. It seems strange that for several decades experiments aimed at building complete sets of phase diagrams have not been carried out while studying many substances

Thus, the authors adhere to the viewpoint that the developed here ideas about the DPT in the vicinity of the triple FE-AFE-PE point are suitable for the description of paraelectric PT in a wide class of FE and AFE. This viewpoint will be consistently confirmed further using a set of experimental phase diagrams as a background.

**6.2. Experimental results**

6.2.1. Diffusive phase transitions near FE-AFE-PE triple point

Now let us consider the phase transitions into PE state for the PZT-based solid solutions located near the boundary between the regions of FE and AFE ordering of the phase Y-*T* diagram. Shown in Fig.6.3a and 6.3b are the dependences of the parameter of paraelectric PT diffusion, $\delta$, on the composition for two series of PLLZT solid solutions with 10% and 15% contents of the substituting complex. The main peculiarity of the presented dependences is the increase of the diffusion in the solid solutions which belong to the boundary region separating those of FE and AFE ordering in the Y-T diagram (see Fig3.1). The displacement of the boundary region towards higher contents of titanium at increasing the concentration of $(Li_{1/2}La_{1/2})$-complex in the solid solution results in the shift of the maximum of $\delta(Y)$ dependence.

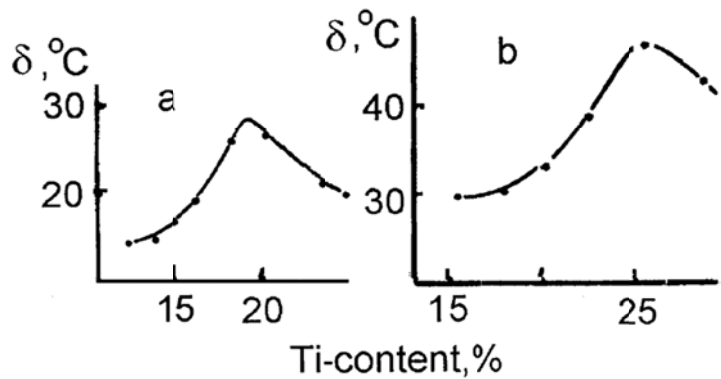

Fig.6.3. Diffuseness parameter $\delta$ vs Ti-content for PLLZT 10/100-Y/Y (a) and 15/100-Y/Y (b) solid solutions [13].

The dependences of the diffuseness parameter $\delta$ on the Ti-content in PLZT solid solutions with sequentially increased contents of lanthanum are presented in Fig.6.4. Under such a substitution the phase boundary, separating the FE and AFE regions in Y-T diagrams (Fig.3a and 3.8) is shifted towards the increasing of Ti content. As see from Fig.6.4, δ(Y) dependences are shifted in the same way, and their maximum cincides with the location of the borderland between the regions of spontaneous FE and AFE states in the Y-T diagrams. Thus, it is obvious that in PLZT the maximum diffuseness of paraelectric PT takes place in the vicinity of the equilibrium of FE and AFE states.

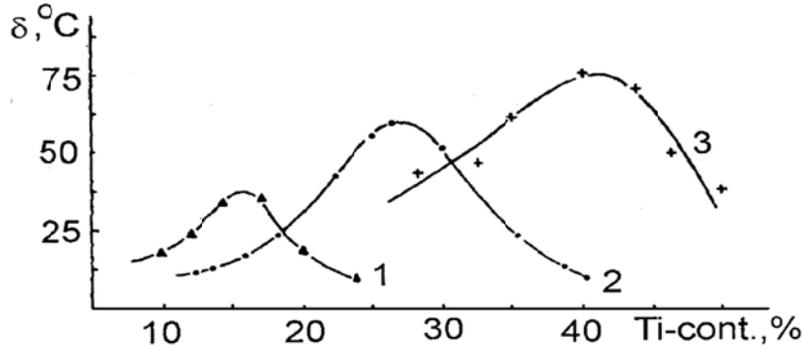

Fig.6.4. Diffuseness parameter δ vs Ti-content for PLZT solid solutions [47, 77].
Content of La, %: 1 – 4; 2 – 6; 3 – 8 .

Now consider the results of investigation of the effect of hydrostatic pressure on diffusive PT in PLZT. Fig.6.5 shows the diffuseness parameter as a function of hydrostatic pressure for PLZT solid solutions with the compositions 6/65/35 and 6/80/20. Under normal conditions the former solid solution is in FE state and undergoes FE-PE PT at temperature of about $180°C$ (Fig.3.1a and 3.3). The increase of pressure leads to transition of the solid solution into the state of coexisting FE and AFE phases, and then into the single-phase AFE state. The latter solid solution is in the state of coexisting phases under normal conditions; the increase of the pressure induced the transition into single phase AFE state.

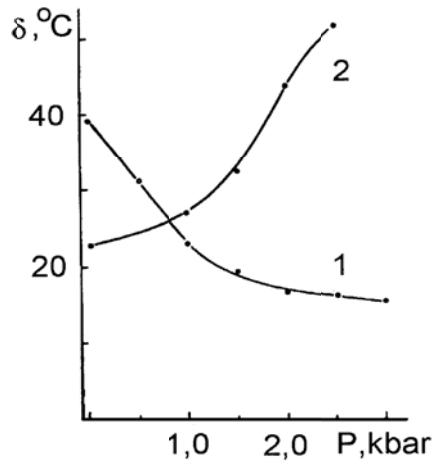

Fig.6.5. Diffuseness parameter *δ* vs pressure for 6/100-Y/Y PLZT solid solutions [47, 77].
Content of Zr/Ti: 1 – 6/80/20; 2 – 6/65/35.

As it may be seen from Fig.6.5, the increase of the pressure up to 3 kbar leads to the growth of the diffuseness parameter in PLZT 6/65/36 and to decrease of this parameter in the 6/80/20 solid solution.

Thus, the present experimental results show that in the vicinity of the FE and AFE phases equilibrium the transition into PE state are diffusive, and diffuseness degree diminishes when moving from the phase equilibrium. We have also stated that such feature does not depend on the origin of the phase equilibrium (a variation of the content of solid solution components or a variation of the hydrostatic pressure). As stated in [77], in the case when the phases are redistributed by an electric field the PT diffuseness degree changes too. In particular, if the field favours the emergence of the state of the coexisting FE and AFE phases, then the diffuseness parameter increases.

To better understand the problem under discussion, let us consider the results of our earlier papers [6, 45] devoted to studying diffusive phase transitions (DPT) in PZT solid solutions 0.97 Pb($Zr_{1-y}Ti_y$)$O_3$ + 0.03 Cd($Ta_{2/5}W_{1/3}$)$O_3$ using hydrostatic pressure. The addition is introduced with the aim to obtain nearly vertical FE-AFE phase boundary in the "pressure-temperature" (*P-T*) phase diagram (Fig.6.6). As is known, for small values of *y* the AFE state is stable at the temperatures below the Curie point. On increasing *y* FE state becomes stable at low temperatures. Due to the choice of the solid solution composition (*y*) the change of the temperature results in the emergence of either FE-PE or AFE-PE phase transition at the Curie point (for atmospheric pressure). We have investigated the solid solutions with *y* = 0.02, 0.03 and 0.04. Shown in Fig.5 is the phase *P-T* diagram of the solid solution with *x* = 0.02. As we see, the pressure $P_0 \cong 3$ kbar corresponds to the equilibrium of FE and AFE states within a wide temperature range.

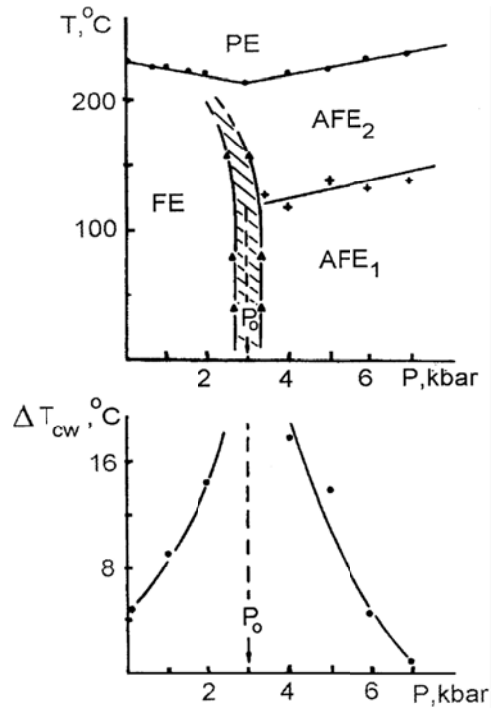

Fig.6.6. "Pressure-temperature" phase diagram for PZT solid solution (at the top) and
the temperature interval of deviation from Curie-Weiss laws pressure (at the dawn) [6, 45. 47].

Experiments show that the diffuseness of PE transition essentially depends on the pressure, and the maximum diffuseness degree takes place at 3 kbar. For $P \gg P_0$ and $P \ll P_0$ the reverse dielectric constant obeys the Curie-Weiss law irrespective of the type of LT state (FE or AFE). The interval $\Delta T_{cw}$ of the deviation from this law in the vicinity of $T_c$ is narrow. The interval $\Delta T_{cw}$ sharply increases when $P \to P_0$ from the side of both FE and AFE phase. The dependence $\Delta T_{cw}(P)$ for $y = 0.02$ is also shown in Fig.6.6. While increasing the value of $x$ in the solid solution the boundary between FE and AFE states on the $P$-$T$ diagram shifts to the right by ~ 0.9 kbar per 1% of Ti. The anomaly of the dependence $\Delta T_{cw}(P)$ is displaced in the same manner.

The present results show that the diffuseness of paraelectric PT essentially depends on the proximity to equilibrium line of the FE and AFE states in the $P$-$T$ phase diagram. The pressure $P_0$ is the value at which the thermodynamic potentials of FE and AFR states become equal, and, therefore, the probability of the coexistence of these phases is the highest.

In more detailed the mechanism of paraelectric PT diffuseness in the vicinity of the triple FE-AFE-PE point and its expansion for a large number of ferro- and antiferroelectrics is discussed in [6, 20, 73].

### 6.2.2. X-ray investigation of two-phase (FE+AFE) nucleation in PE phase

We investigated the dependence of the shape of X-ray diffraction lines on temperature at $T > T_c$ (into PE phase) for 10/100-Y/Y PLLZT solid solutions [13]. For this purpose (222) and (400) lines, the most typical for perovskite crystal structure, were chosen. The former line is singlet in the case of tetragonal crystal lattice distortion; in the case of rhombohedral type of lattice a distortion it is doublet. The latter line is doublet when the distortion is of tetragonal type, and it is singlet when the distortion is rhombohedral. The analysis of the shape of these X-ray diffraction lines allows to conclude, which of the low-temperature phases is extended to the high-temperature (H-$T$) region, disposed above the point of the PE PT for the main part of the samples' volume.

Fig.6.7 illustrates the temperature dependences of the asymmetry of the X-ray lines ($\gamma = S_l/S_r$) and the crystal cell parameters for PLLZT solid solutions belonging to different regions of the Y-T diagram. Here $S_l$ and $S_r$ are the areas below the contour of the X-ray diffraction line located from the left and from the right of the vertical line drawn from the vertex of the X-ray pattern, $S = S_l + S_r$, where S is the total area of the X-ray diffraction line.

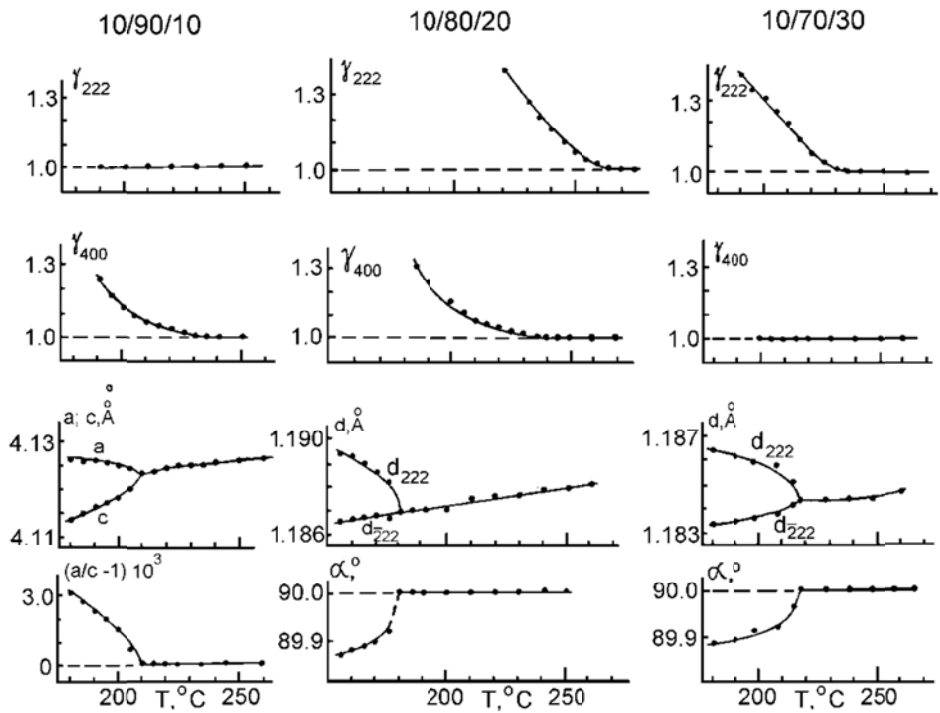

Fig.6.7. Temperature dependences of X-ray parameters for 10/90/10 (to the left), 10/80/20 (in the middle) and 10/70/30 (to the right) PLLZT solid solutions [13,47].

According to the data of calorimetric, dielectric and X-ray diffraction measurements, 10/90/10 PLLZT undergoes PE PT at 209°C. The temperature dependences of the crystal lattice parameters *a*, *c* and (*a/c* - 1) near the point of PT are shown in the left column in Fig.6.7. The asymmetry of the (400) line is preserved at temperatures up to 230°C. The (222) X-ray diffraction line is symmetric for temperatures above 209°C. This fact directly testifies that the diffuseness of PT in this solid solution is connected with the extension of the domains of the tetragonally distorted AFE phase to the high-temperature range above the transition point.

10/70/30 PLLZT undergoes PE PT at 218°C. The temperature dependences of the crystal lattice parameters near the point of PT are shown in the right column in Fig.6.7. The asymmetry of the (222) line is preserved at temperatures up to 235°C. The (400) X-ray diffraction line is symmetric within the whole of the interval of temperatures above 218°C. This fact shows that the diffuseness of PT in this solid solution is connected with the extension of the domains of rhombohedrally distorted FE phase to H-*T* range above the transition point.

For the 10/80/20 PLLZT, located closely to the boundary between the regions of FE and AFE ordering in the Y-*T* diagram the observed picture differs from that described above. In this case both the considered (222) and (400) X-ray diffraction lines are asymmetric in the temperature region above the transition point (these data are shown in the middle column in Fig.6.7). The temperature interval, within which the lines' asymmetry is preserved, is much wider than those for 10/90/10 and 10/70/30 PLLZT. The main part of the samples' volume undergoes PT at 180°C, whereas the complete symmetry of the X-ray diffraction patterns becomes obvious only at temperatures above 260°C.

Analysis of the profile of X-ray diffraction lines, at the temperatures higher than the Curie point revealed that the samples have a complex phase composition. In the first approximation, it may be concluded that the PE matrix of the sample as a whole contains two-phase domains which, in their turn, consist of adjoining domains with FE and AFE ordering.

# 7. RELAXATION DYNAMICS OF INTERPHASE (FE-AFE) DOMAIN WALLS.

## 7.1. Model consideration

The influence of DC electric field for system, in which the free energies of FE and AFE states slightly differ and the domains of these phases coexist, is considered in Ch.4. It is shown that under the action of the field the state of the substance is changed due to the displacement of IDW, the internal state within the domains being unchanged. Such a motion of IDW is inertial and accompanied with relaxation processes. The relaxation dynamics of IDW should manifest itself most vividly while investigating the said substances in AC electric field.

The main properties of IDW dynamics under action of AC electric field can be obtained while examining the forces which act on IDW. The basic equation derived from the condition of balance of forces affecting IDW, has the form:

$$P_{E(t)} + R(u) + P(u,t) = 0, \qquad (7.1)$$

where $u = u(t)$ is the time-dependent displacement of the IDW, $E(t)$ is the electric field intensity; $P_{E(t)}$ is the pressure acting on IDW associated with the expansion of FE domain volume in external field. The forces which counteract the displacement of the IPW under the effect of the field can be presented as a sum of two terms: here $R(u)$ is the pressure of the forces caused by the interaction of the IDW with immobile crystal structure defects, and $P(u,t)$ is the pressure of the so-called after-effect forces. The latter forces may be caused by different factors: the interaction of the IDW with mobile crystal lattice defects, as well as nonzero duration of the thermal processes connected with phase transitions in those regions of the crystal through which IDW passes.

Eq. (7.1) describes the equilibrium of the IDW under the action of different kinds of physical forces. While considering this relation as a motion equation (Newton equation), one can easily see that here the terms $m_w \ddot{u}(t)$ ($m_w$ is the effective mass of the IDW) and $\gamma m_w \dot{u}(t)$ are not taken into account. These terms may be not considered in the analysis of motion if the effective mass $m_w$ is small and if the frequency of measuring AC electric field is essentially lower than the frequency of the IDW oscillation in the potential well $\int du R(u)$ created by immobile crystal lattice defects.

The effective pressure exerted by external electric field on IDW can be presented in the form

$$P_{E(t)} = E(t)P_s. \tag{7.2}$$

For simplicity we shall consider a uniaxial ferroelectric, and the direction of external field coinciding with the polar axis.

The pressure of the force affecting IPW from static defects can be presented by the first term of the power series expansion in displacements (i.e. by a quasi-elastic force):

$$R(u) = -\alpha u, \quad (\alpha > 0). \tag{7.3}$$

The pressure $P(u,t)$ is defined by the whole of the sample's history. Therefore it must depend on the values of $u$ at any time $t'$ preceding the present measurement time $t$ ($t' < t$). Analytically such dependence can be chosen at different degree of complexity. Here we choose it in the simplest form. Proceeding from the most general consideration, $P(u,t)$ can be expressed as

$$P(u,t) = -w\int_0^t F[u(t),u(t')]g(t-t')dt', \tag{7.4}$$

where $F[u(t),u(t')] = F[u(t)-u(t')]$, $g(t-t')$ is a function of aftereffect which describes the contribution of relaxation effects existing at the time $t'$ ($0 < t' < t$), to the resultant relaxation at the time moment $t$. For a system with one relaxation time the function $g(t-t')$ has a simple form:

$$g(t-t') = \frac{1}{\tau}\exp\left(\frac{t-t'}{\tau}\right). \tag{7.5}$$

Further, by analogy with (7.3), for small displacements of IDW the function $F[u(t) - u(t')]$ can be presented in a linearized form:

$$F[u(t)-u(t')] = -k[u(t)-u(t')]. \tag{7.6}$$

Thus, the expression which describes the pressure of aftereffect acquires the form:

$$P(u,t) = kw\int_0^t [u(t)-u(t')]\cdot\exp[-(t-t')/\tau]\frac{dt'}{\tau}. \tag{7.7}$$

The motion equation for IDW will have the following form in the above-mentioned approximation:

$$-\alpha u(t) + kw\int_0^t dt' \left[u(t) - u(t')\right]\exp\left(-(t-t')/\tau\right) + P_s E_0 \exp(i\omega t) = 0$$

or  (7.8)

$$[1+\eta G(t)]u(t) - \eta\exp\left(-\frac{t}{\tau}\right)\int_0^t u(t')\exp(t'/\tau)\frac{dt'}{\tau} = u_0 \exp(i\omega t),$$

where $\eta = kw/\alpha$, $u_0 = P_s E_0/\alpha$ is the IDW displacement under the action of DC electric field with the intensity $E_0$ and $G(t) = (1-\exp(-t/\tau))$ is the time function. This equation should be solved either by the method of successive approximations (in the first approximation) or by transforming it into a differential equation which can be easily solved in the stationary case $t \to \infty$ ($G(t) = 1$). The steady-state solution is sought for in the form $u(t) = \tilde{u}\exp(i\omega t)$ and $\tilde{u}$ is obtained in the form [78]:

$$\tilde{u} = u_0 \frac{(1+i\omega\tau)}{[1+i\omega\tau(1+G(t)\eta)]} \to u_0 \frac{(1+i\omega\tau)}{[1+i\omega\tau(1+\eta)]}. \tag{7.9}$$

Since the IDW displacements under the action of electric field lead to a change of the dipole moment in the volume of the substance involved by this displacement, then it is easy to determine the dielectric susceptibility $\chi$ connected with this process:

$$\chi(t) = P_s S u(t)/E(t), \tag{7.10}$$

where $S$ is the area of IDW oscillation. Let us put $\chi_0 = P_s^2 S/\alpha$ is the static susceptibility. Thus, from (7.10) and (7.9) we shall obtain:

$$\chi = \chi' - i\chi'' = \chi_0 \frac{1+\omega^2\tau^2(1+\eta) - i\omega\tau\eta}{1+\omega^2\tau^2(1+\eta)^2}. \tag{7.11}$$

From here it follows that:

$$\chi'(\omega,T) = \chi_0 \frac{1+\omega^2\tau^2(1+\eta)}{1+\omega^2\tau^2(1+\eta)^2} \equiv \chi_0(T)F_1(\omega,T), \tag{7.12a}$$

$$\chi''(\omega,T) = \chi_0 \frac{\omega\tau\eta}{1+\omega^2\tau^2(1+\eta)^2} \equiv \chi_0(T)F_2(\omega,T). \tag{7.12b}$$

As noted earlier, the relaxation dynamics of the IDW is influenced by the aftereffect of the first-order PT between FE and AFE states, taking place in that part of the sample where the IDW motion is realized. In this case the said states are separated by a potential barrier, and thermoactivation processes take place. Therefore it is natural to assume that the temperature dependence of the relaxation time has the following form:

$$\tau = \tau_0 \exp(\Delta/kT). \tag{7.13}$$

Here $\Delta$ is the activation energy, $\tau_0$ is the value reciprocal of the frequency of the thermal activation (its order of magnitude equals $10^{-11}$ - $10^{-13}$ s).

As shown in Ch.4, the coexistence of FE and AFE phase domains is provided by the equality of their phases thermodynamic potentials within a wide interval of field intensity. The change in the external field intensity results in the displacement of interphase boundary, the energy of each of the coexisting phases remains unchanged ($\Delta = 0$). However, the presence of the effect of "IDW lag", caused by the aftereffect forces (7.4), changes this simple picture. Now the condition of field compensation inside the domains is not fulfilled, and $\Delta \neq 0$ (though being a small value). Therefore, under the condition of week fields one may put $\omega\tau \ll 1$.

Under the said condition the functions $F_1(\omega,T)$ and $F_2(\omega,T)$ acquire the following form:

$$F_1(\omega,T) \approx 1 - \omega^2\tau^2\eta(1+\eta) \approx 1 - 2\omega^2\tau_0^2\eta(1+\eta)\Delta/kT, \tag{7.14a}$$

$$F_2(\omega,T) \approx \eta\omega\tau \approx \eta\omega\tau_0(1+\Delta/kT). \tag{7.14b}$$

The dependence $\chi_0(T)$, as well as the typical dependences $F_1(T)$ and $F_2(T)$ for different frequencies near the temperature of the $\chi_0(T)$ maximum are given in Fig.7.1.

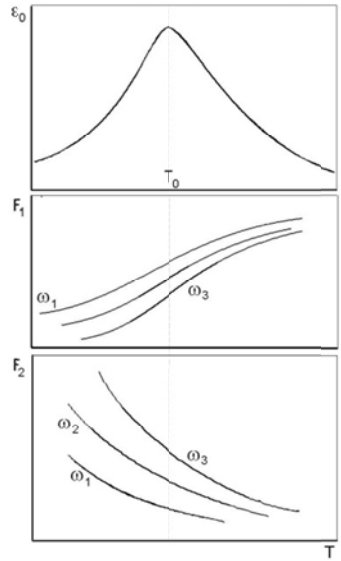

Fig.7.1. Schematic dependences of $F_1$ and $F_2$ on temperature near the maximum of $\varepsilon(T)$ for different frequencies $\omega_1 < \omega_2 < \omega_3$.

Taking into account the fact that the expression for $\chi'(T)$ and $\chi''(T)$ are the product of $\chi_0(T)$ into the functions $F_1(T)$ and $F_2(T)$, respectively, one can make the following conclusions:

- with the increase of the measuring field frequency the dependences $\varepsilon'(T)$ and $\varepsilon''(T)$ are shifted towards higher temperatures;

- the maximum of the real component of the permittivity, $\varepsilon'(T'_m) \equiv \varepsilon'_m$, decreases with the increase of the frequency, whereas the maximum of the imaginary component of the permittivity, $\varepsilon''(T''_m) \equiv \varepsilon''_m$, increases; the rate of $\varepsilon''_m$ rise is higher in comparison with the decrease of $\varepsilon'_m$.

- $T'_m$ is always higher than $T''_m$.

We can obtain expressions which describe the shift of $T'_m$ and $T''_m$, caused by the change of the field frequency. In particular, taking into account (7.12) and (7.14) and taking a derivative with respect to $(1/T)$ and equating it to zero. The following expression for $T'_m$ will be obtained:

$$\left(\frac{1}{T'_m}\right) = -\frac{k}{\Delta}\ln\omega - \frac{k}{\Delta}\ln\tau_0 + \frac{k}{\Delta}V(1/T'_m). \tag{7.15}$$

Here $V(1/T)$ is a slow varying function. The introduction of the effective temperature $T_f$, i.e. the linearization of (7.15) using the method of Pade approximation [79], yields another form of this expression:

$$\frac{1}{T'_m - T_f} = -\frac{k}{\Delta}\ln\omega - \frac{k}{\Delta}\ln\tau_0, \tag{7.16}$$

which can be also written down in the form of the Vogel-Fulcher relation:

$$\omega = \left(\frac{1}{\tau_0}\right)\exp\left[-\frac{\Delta}{k(T'_m - T_f)}\right]. \tag{7.16a}$$

The expressions for $T''_m(\omega)$ analogous to (7.16) and (7.16a) are obtained in the same way.

From (7.16) a mathematical expression which relates the location of the $\varepsilon'(T)$ maximum with the electric field amplitude, can be easily derived. At first, we will consider DC electric field. When the electric field is applied to the sample, the eq. (7.13) acquires the following form:

$$\tau = \tau_0 \exp\left[(\Delta - \alpha P_s E)/kT\right], \tag{7.17}$$

where the second term in the exponent describes the changing of FE minimum of the thermodynamic potential, caused by the field. For obtaining the dependence $T'_m(E)$ the substitution $\Delta \to \Delta - AE$ is to be made in the exp. (7.16). It is easily seen that the increase of the field intensity leads to a near linear decrease of $T_m'$.

$$\frac{1}{T_m' - T_f} = -\frac{k}{\Delta - AE}\ln\omega - \frac{k}{\Delta - AE}\ln\tau_0. \qquad (7.18)$$

Expression (7.18) can be used for consideration of the AC field influence on the location of $T_m'$. The applicability of (7.17) for the description of IDW motion in AC electric field in the scope of this model is caused by the fact that the process of order parameters relaxation to the steady state under the influence of the field and the potential barrier height change can by considered simultaneously.

The relaxational dynamics of IDW is determined by the term $P(u,t)$ в (7.1). Let us remind that this term may be caused by different factors: the interaction of the IDW with mobile crystal lattice defects, as well as nonzero duration of the thermal processes connected with phase transitions in those regions of the crystal through which IDW passes. Characteristic time of the above-mentioned processes is rather long. AT the same time the processes of establishment of the order parameter during the structural phase transitions (to which FE and AFE phase transitions belong) are the fast ones. Characteristic times of these processes are several orders of magnitude shorter. Based on the assumptions used to write eq. (7.1), one can consider the establishment of the potential barrier value $\Delta$ between two minima corresponding to FE and AFE phases in eq. (7.13) as instantaneous when the AC electric field is applied to the sample(that is the value $\Delta$ follows the instantaneous value of the field). That is why the dependence $T_m'$ on the amplitude of the AC field will be the same as the dependence on the intensity of the DC field, namely, increase of the field amplitude will lead to a near linear decrease of $T_m'$.

We have considered the model of oscillation for IDW separating FE and AFE phase domains which are in equilibrium. Though the studied version of the model is the most simplified, the obtained results allow us to predict a wide range of experimental phenomena.

One of the model simplifications consists in the choice of the relaxation function $g(t-t')$ in the form (7.5). This case corresponds to a single relaxation time. Actually, while considering a real physical system, one should take into account the existence of a spectrum of relaxation times. However, even such slight complication of the model will lead to certain mathematical (but non physical) problems, while investigating Eq.(7.8). For instance, in [80-82] the solutions of equations analogous to (7.8) were studied, and purely mathematical difficulties were observed. The numerical methods are used to solve these equations, and the physical nature of the obtained solutions became not evident. Moreover, at the final stage of the equations solutions authors made simplifying assumptions. Therefore, we suppose that at the present stage of studying of relaxation dynamics of IDW it is sufficient to use the single-time relaxation function $g$. The complication of the function $g(t-t')$ may be justified at subsequent interpretation and comparison of experimental results obtained for physical objects of different nature of the same substances (for instance, crystalline, ceramics and film samples).

The most important result of the present chapter consists in the fact that the considered model allows us to describe the temperature dependences of real and imaginary part of permittivity and theirs evolution under varying AC electric field frequency and amplitude. Equations (7.12) and (7.14) correspond to an isolated IDW. To obtain the dependences $\varepsilon'(T)$ and $\varepsilon''(T)$ for the whole investigated sample containing a large number of IDW, the changes in the quantity and in area of such interphase boundaries caused by temperature changing of the sample, are to be taken into account. At $T < T_0$ the number and dimensions of IDW weakly depend on the temperature; therefore the permittivity of the sample is described by the expressions analogous to those for the isolated boundary. The most essential change in the IDW number and area take place at temperatures close to $T_0$ and higher. Moreover, complicated FE+AFE domains exist in the PE matrix of the substance at temperatures higher than $T_0$ (Ch.6 [5, 6, 10]). That is, at $T > T_0$ the IDW exist, and theirs oscillations contribute to the permittivity of the substance. However, as the temperature rises, the number and volume of FE+AFE domains and, consequently, the number and area of IDW diminish. There exist the temperature $T_t$ such that for all temperatures $T > T_t$, no FE+AFE domains are present in the PE matrix of the substance [10- Pec-5], and, consequently, the IDW are also absent. So, there is no contribution of the IDW oscillations into the permittivity. Therefore, at $T > T_t$ the functions $\varepsilon'(T)$ and $\varepsilon''(T)$ do not depend on the AC field frequency and intensity, and $\varepsilon'(T)$ coincides with the function $\varepsilon_0(T)$. Such a situation is shown in Fig.7.2.

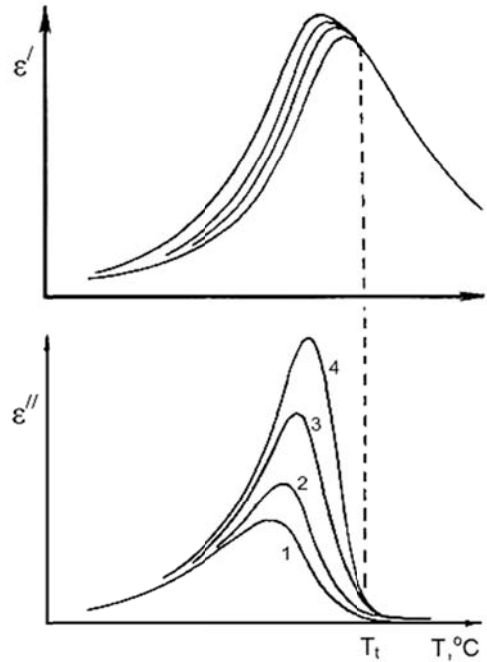

Fig.7.2. Schematic dependences of $\varepsilon'(T)$ and $\varepsilon''(T)$ on temperature at phase transition into PE state for different frequencies $\omega_1 < \omega_2 < \omega_3 < \omega_4$.

## 7.2. Experimental results

Now let us discuss the experimental results on dielectric properties of those substances that have a weak difference between the energies of FE and AFE states. As an example, we consider the 10/100-Y/Y PLLZT and 6/100-Y/Y PLZT solid solutions. Phase diagrams of the said solid solutions are shown in Fig.3.1(e), and Fig.3.8(d) (see also Fig. 3.10), respectively.

7.2.1. PLLZT solid solutions

Fig.7.3 shows the dielectric response obtained using AC field at 1 kHz, with the amplitude varying between 1 and 150 *V/mm*, for PLLZT solid solutions located in the three different regions of the Y-*T* phase diagram. First of all, it should be noted that for the said values of electric field intensity the dependences *P(E)* are linear. The effects bound up either with the reorientation of FE phase domains or with FE phase induction manifest themselves in fields which noticeably exceed the used ones.

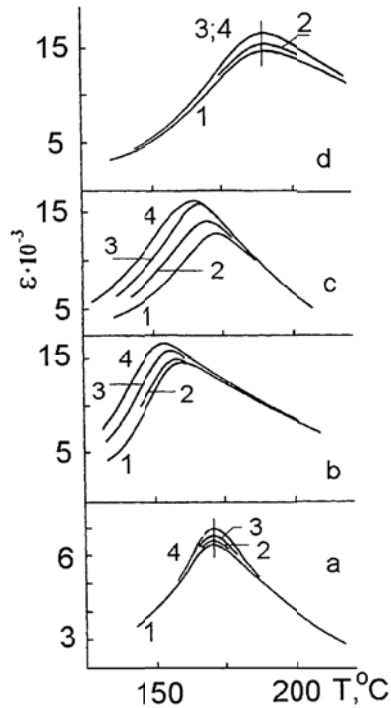

Fig.7.3. Temperature dependences of $\varepsilon'$ for 10/100-Y/Y PLLZT [47, 83]. Ti-content Y, %: a - 15; b - 20; c - 23; d – 30; $E_\sim, V/mm$: 1 - 1; 2 - 50; 3 - 100; 4 - 150.

For the 10/90/10 PLLZT from AFE region of Y-T diagram all $\varepsilon'(T)$ dependences at varying AC field practically fall on the same curve. For the 10/15/85 PLLZT the dependences $\varepsilon'(T)$ for four values of AC field are presented in Fig.7.3a. As is seen, the temperatures $T'_m$ coincide, but $\varepsilon'_m$ slightly rises itself at increasing the field. For 10/30/70 PLLZT solid solution from FE region of Y-$T$ diagram the behaviour of $\varepsilon'(T)$ is similar to that of 10/15/85 PLLZT: $T'_m$ does not depend on the AC field amplitude, and the dielectric constant somewhat increases in the vicinity of $T'_m$ (Fig.7.3d).

Different situation is observed in the solid solutions located near the boundary between the stability regions of FE and AFE states in the Y-$T$ diagram that is for 10/20/80 and 10/23/77 PLLZT.

For fields up to 20 V/mm the observed $\varepsilon'(T)$ behaviour is analogous to that described above: $T'_m$ does not depend on the field amplitude and $\varepsilon'(T)$ measured in fields with different intensities are practically the same. For AC fields higher than 20 V/mm the behavior of $\varepsilon(T)$ changes.

The $\varepsilon'(T)$-dependences for the 10/20/80 and 10/23/77 PLLZT for AC field intensities higher then 20 $V/mm$ are presented in Fig.7.3b and 7.3c. As is obvious, in these solid solutions both $\varepsilon'_m$ and $T'_m$ noticeably depend on the AC signal intensity. On increasing the AC field amplitude $T'_m$ diminishes and $\varepsilon'_m$ rises. The dielectric constant increases at all $T < T'_m$ with increasing AC field amplitude, whereas at $T > T'_m$ $\varepsilon'$ is practically independent on the AC field.

Dispersion of dielectric constant was investigated within $10^2 - 10^5$ Hz frequency range. For the series 10/90/10 and 10/70/30 of PLLZT solid solutions located in the Y-T phase diagram within the region of uniform AFE and FE states, respectively, $\varepsilon'(T)$ and $\varepsilon''(T)$ ($T'_m$ and $T''_m$) do not depend on frequency. However, for the 10/80/20 and 10/77/23 PLLZT the considered changes in the behaviour of $\varepsilon'(T)$ turn out to be cardinal (Fig.7.4). With the growth of the field frequency the $\varepsilon'(T)$ dependences are being displaced towards higher temperatures and $\varepsilon'_m$ noticeably decreases.

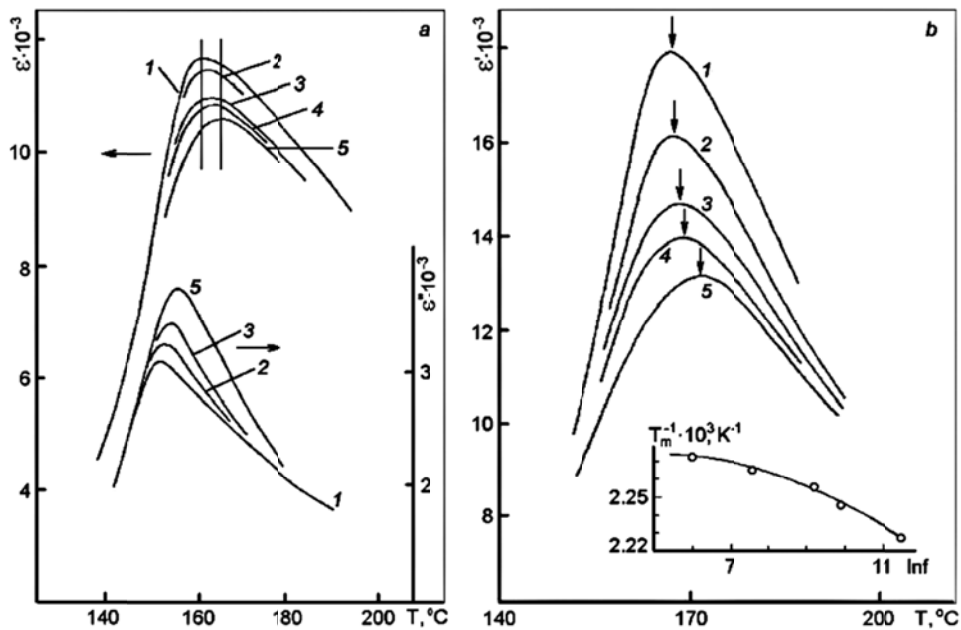

Fig.7.4. Temperature dependencies of $\varepsilon'$ and $\varepsilon''$ for 10/80/20 PLLZT (left) and temperature dependence of $\varepsilon'$ for 10/77/23 PLLZT (right) [47, 83, 84]. Frequency kHZ: 1 – 0.4; 2 – 2.0; 3 – 10.0; 4 – 20.0; 5 – 100.0.

The $\varepsilon''(T)$ dependences also essentially change: $\varepsilon''(T)$ curves are shifted towards higher temperatures and $\varepsilon''_m$ increases. For example, for 10/77/23 PLLZT the Vogel-Fulcher relation is fulfilled in the following form: $\omega = (1/\tau_0)\exp\left[-\Delta/k(T_m - T_f)\right]$, with the effective temperature $T_f = (425 \pm 2)\ K$. In more detailed these results are discussed in [83, 84].

The investigations performed with DC bias applied to the samples, were carried out on PLLZT solid solution with 15 % of Ti. The *E-T* phase diagram for 10/85/15 PLLZT is presented in Fig.7.5a. For low-intensity DC bias only one AFE-PE phase transition takes place on the line $T_c(E)$ if the temperature is varying (in accordance with *Y-T* phase diagram). For DC bias $E > E_{tr} \cong 15\ kV/cm$, FE-PE transition is observed.

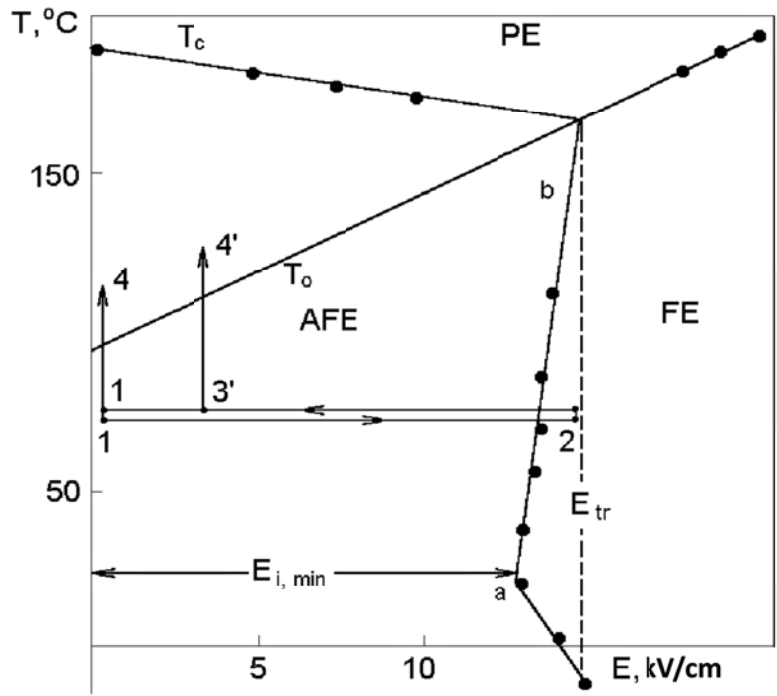

Fig.7.5a. Phase *E-T* diagram for 10/85/15 PLLZT solid solution [47, 83].

The temperature dependences of the dielectric constant $\varepsilon'(T)$ for different DC bias and measuring AC fields are presented in Fig.7.5b. For zero DC bias (Fig.7.5b (a)) the location of $\varepsilon'(T)$ maximum is independent of the AC field amplitude, and $\varepsilon'_m$ somewhat increases with the rise of AC field. Similar behavior is observed for the DC bias intensity $E = 22\ kV/cm$ (Fig.7.5b (c)). The $\varepsilon'(T)$ dependences obtained for different AC field amplitudes are practically coincident in the whole of the temperature range studied.

Different dielectric behavior is observed for the DC bias $E = 12\ kV/cm$ (Fig.7.5b (b)). The maximum value of dielectric constant $\varepsilon'_m$ increases and the temperature $T'_m$ decreases when the AC field amplitude rises. Simultaneously, the large increase of the dielectric constant takes place for $T < T_m$.

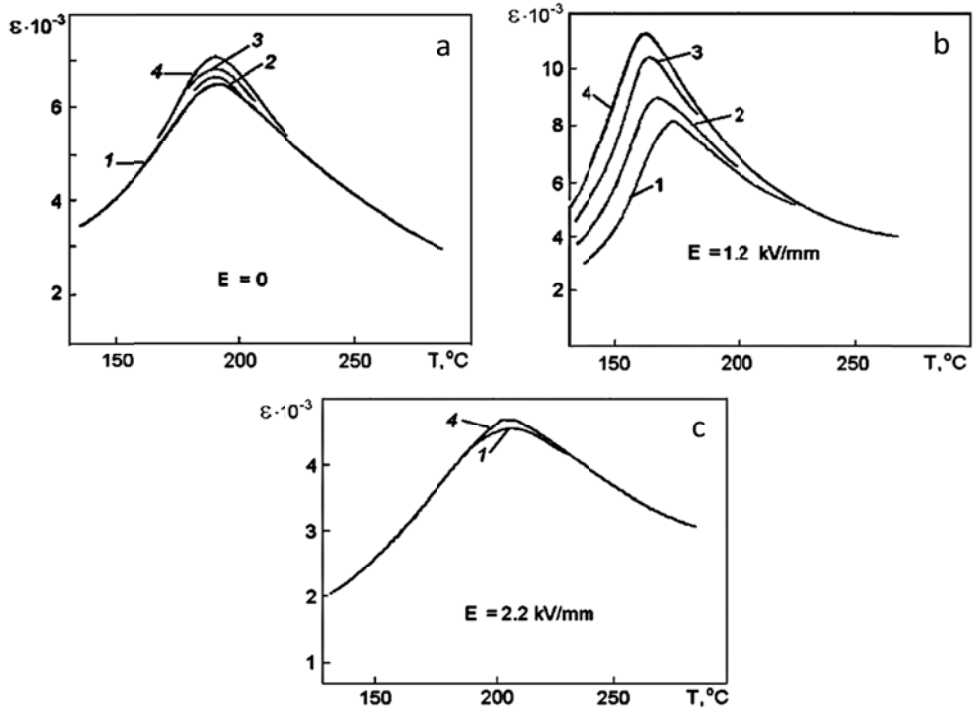

Fig.7.5b. Temperature dependences of $\varepsilon'$ for 10/85/15 PLLZT solid solution at different AC and DC field intensities. AC filed amplitude (*V/mm*): 1 – 1; 2 – 50; 3 – 100; 4 – 150

As seen from Fig.7.3, 7.4 and 7.5, peculiarities in the behavior of the permittivity caused by the increase of the measuring AC field amplitude are revealed only in the solid solutions located near the boundary separating the regions of FE and AFE states in the Y-*T* or *E-T* phase diagrams. These solid solutions have close values of free energy for FE and AFE states, and this leads to the coexistence of the domains with FE and AFE ordering in the volume of the samples. In this case one should take into consideration the contribution of interphase boundary oscillations. As seen from the present results, the deviations of the permittivity from the classic behaviour take place only in the case when the contribution of the said boundaries has the form predicted by the model [20, 85] for FE-AFE phase transformations. At the same time, it is not essential which of the external factors – change of the DC bias intensity or change of the solid solution composition – results in the appearance of the equilibrium two-phase state.

The difference in the behaviour of the dielectric constant under changing intensity of the measuring AC field should be specially considered for the case of weak and strong fields (with intensities lower and higher than 20 *V*/*mm*, respectively). In our opinion, IDW pinning on crystal structure defects manifests itself in weak fields. In more detail this problem is discussed in [83].

7.2.2. PLZT solid solutions

The dielectric properties of 6/100-Y/Y PLZT solid solutions and their dependence on the location of the substance in the phase Y-*T* diagram (Fig.3.8, 3.10) are identical to those of PLLZT solid solutions. For the 6/90/10 and 6/52/48 solid solutions located far from the AFE/FE boundary region of the Y-*T* diagram (and, thus, belonging to the region of homogeneous AFE and FE ordering, respectively) the curves $\varepsilon'(T)$ are independent of the frequency and the intensity of AC field. For the 6/72/28 solid solution located near the above-said boundary there is a considerable dispersion of both components of dielectric constant at the transition into PE state (Fig.7.6), the Fogel-Fulcher law is fulfilled. In the literature such an effect is called relaxor behavior.

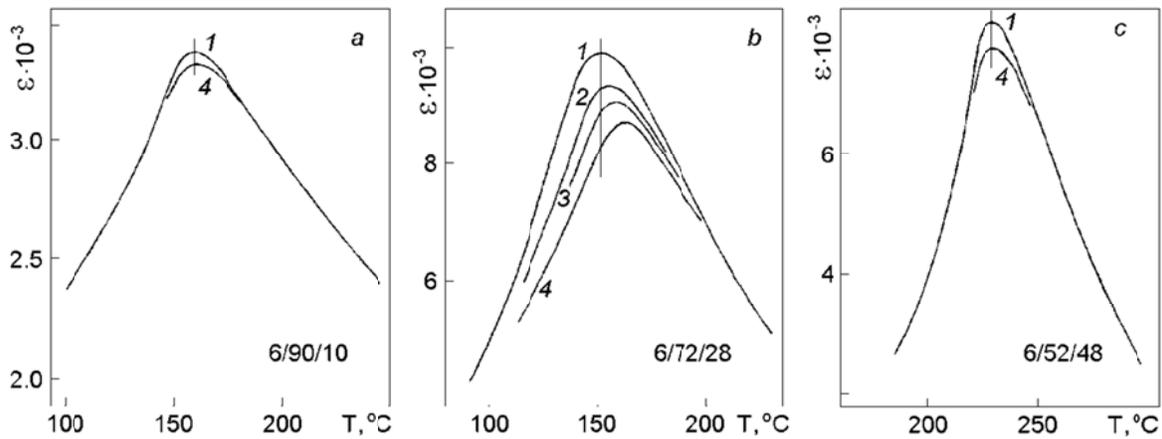

Fig.7.6. Dependence $\varepsilon'(T)$ for 6/100-Y/Y PLZT solid solutions [47, 83].
Ti-content Y,%: a - 10; b - 28; c - 48. Frequency, kHz: 1 - 1; 2 - 10; 3 - 20; 4 - 100.

The presence of a noticeable dispersion of permittivity in the vicinity of the temperature $T_m$ and the fulfilment of the Fogel-Fulcher law were already discussed for the 6/65/35 solid solution (located not far from the boundary region of the Y-*T* diagram of 6/100-Y/Y PLZT and characterized by equal values of free energy for FE and AFE states) [27, 28]. However, the authors of the mentioned papers investigated only one solid solution and therefore could not bound up the studied dielectric properties and their dispersion with the position of the given solid solution in the phase diagram. Naturally, they could not interpret the observed phenomenon correctly. As seen from the presented experimental results, those properties of permittivity which form a base for attributing PLZT solid solutions to the so-called relaxor ferroelectrics are caused by the coexistence of FE and AFE phase domains and by the contribution of the interphase boundary separating these domains, to the value of permittivity (in detail these problems are discussed in [20], Sect.6 and in [85]). In the solid solutions where FE-AFE interphase boundaries are absent, no dispersion of permittivity (or the so-called relaxor behavior) is observed.

## 8. FERRO-ANTIFERROELECTRIC PHASE TRANSITIONS AND THE CO-CALLED "RELAXOR FERROELECTRICS"

Lead magnesium niobate ($PbMg_{1/3}Nb_{2/3}O_3$) is being investigated since the middle fifties; however, physical nature of phenomena observed in this compound is still unclear. We shall not dwell on different models used for interpretation of the behavior of this substance during such a long period. All these models are based on the idea of "chemically inhomogeneous substance", the inhomogeneity being predetermined in the process of obtaining the substance itself. The mentioned concept of "inhomogeneous substance" underlies the model of the so-called relaxor ferroelectrics [86] often used at present while interpreting the properties of complex-composition oxides with perovskite crystalline structure.

The whole set of phenomena in relaxor ferroelectrics have not been explained up to now. One of the main results achieved in the frame of relaxor FE approach is the expression which describes the dispersion of permittivity in the vicinity of the temperature corresponding to the maximum of the dependence $\varepsilon(T)$. As shown in our paper [20], the explanations of physical phenomena observed in such substances as $(Pb,Li_{1/2}-La_{1/2})(Zr,Ti)O_3$ (PLLZT), $(Pb,La)(Zr,Ti)O_3$ (PLZT), $Pb(Mg_{1/3}Nb_{2/3})O_3$ (PMN), $Pb(In_{1/2}Nb_{1/2})O_3$ (PIN) self-consistently follows from the ideas of FE-AFE transitions. The applicability of the developed model concept for analysis of the behavior of real substances is demonstrated in [47] while analyzing the phase diagrams of PZT-based solid solutions and the experimental data on the behavior of the said substances under the action of external effects (temperature, hydrostatic pressure, electric field or any combination of these factors). The goal of the present chapter is to demonstrate the suitability of the ideas of the FE-AFE PT for description of the whole complex of peculiarities in the behavior of PMN, PIN and their related substances.

## 8.1. Phase states and phase transitions in PMN

The scheme used while discussing PLZT and PLLZT solid solutions [47] is rather hard for consideration of peculiarities in the behaviour of PMN, inasmuch as for this substance the number of phase diagrams of different type available in the literature is very limited. Taking into account that PMN is being studied since middle fifties such a situation should be surprising.

We have managed to find only two types of phase diagrams for PMN: the Y-$T$ and $T$-$E$ ones for the system of PbTiO$_3$-PMN (PT-PMN) solid solutions. Therefore, in the course of our discussion they will be often compared with diagrams of other substances for which the nature and properties of phase transitions have been established reliably. First of all we shall analyse the phase diagram of lead titanate in which titanium ions are being sequentially substituted by the ionic complex (Mg$_{1/3}$Nb$_{2/3}$).

However, before considering the influence of PMN on the stability of phase states in lead titanate let us consider the changes in the phase states of other substances with perovskite structure taking place at the formation of solid solutions on the base of lead titanate. For lead-containing oxides with perovskite structure the substitution of A-ion by those with smaller ionic radii increases the energy stability of AFE state with respect to that of FE state. The same picture is observed when B-ion is substituted by ion with greater radii [17, 87]. Most well-known is the diagram of the solid solutions obtained at the substitution of titanium by zirconium, i.e. the diagram of PZT solid solutions. Below the Curie point, when the content of zirconium reaches 95%, FE ordering is superseded by AFE one. The same picture is observed at the substitution of titanium by hafnium.

The "composition – temperature" phase diagrams (both Y-$T$ and X-$T$) for different compounds are presented in in Fig.8.1. The former of them are obtained under the conditions when titanium is substituted not by an individual ion but by (Zr$_{0.6}$Sn$_{0.4}$) or (Zr$_{0.4}$Sn$_{0.6}$) complex [18].

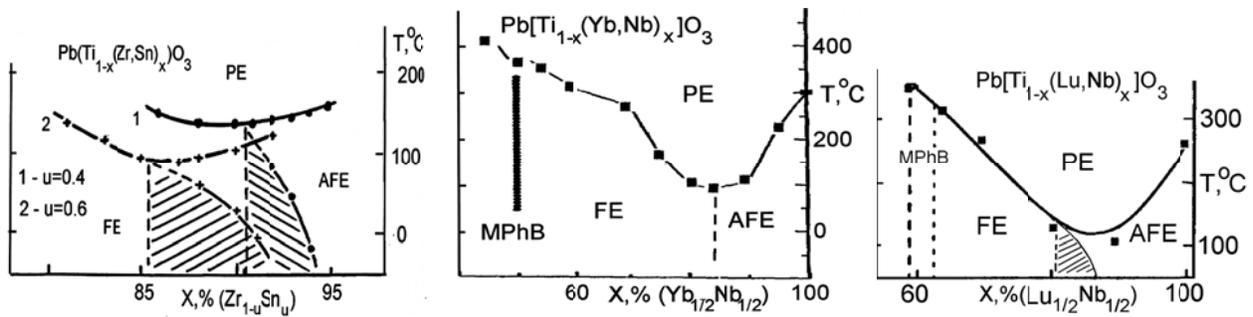

Fig.8.1. Phase diagrams "composition-temperature" for the solid solutions on the base of the lead titanate with B-site substitution: $PbTiO_3 - Pb(Zr_{1-u}Sn_u)O_3$ [18], $PbTiO_3 - Pb(Yb_{1/2}Nb_{1/2})O_3$ [88], $PbTiO_3 - Pb(Lu_{1/2}Nb_{1/2})O_3$ [89, 90].

As one can see, in this case the described above regularity also manifests itself: FE state is substituted by AFE one. The only difference from the situation observed for PZT solid solution system consists in the fact that now a more noticeable contribution to the system's free energy belongs to the effects connected with the anharmonicity of the elastic crystalline potential (this is caused by the presence of $Sr^{4+}$, $Sn^{4+}$ and $Ti^{4+}$ ions with different ionic radii in the equivalent crystal sites. The diagrams have a broad region (dashed in the Fig.8.1) which actually is the hysteresis region of FE-AFE transitions. Analogous phase diagrams take place at the substitution of titanium by the complexes ($Lu_{1/2}Nb_{1/2}$) or ($Yb_{1/2}Nb_{1/2}$) [88-90]. They are presented in Fig.8.1 also.

The following examples are connected with the substitution in A-site. The X-$T$ diagrams for PZT with the 80/20 composition when lead is substituted by strontium $Sr^{2+}$ [6, 18] or ($La_{1/2}Li_{1/2}$)$^{2+}$-complex are shown in Fig.3.1b. The 80/20 PZT is a ferroelectric at $T < T_c$. The low-temperature AFE state becomes stable at the increase of strontium or ($La_{1/2}Li_{1/2}$)-complex content. As in the previous case, there appears an intermediate hysteresis region

The X-$T$ diagrams obtained at the substitution of lanthanum for lead have already been discussed in Ch.3. They are completely identical to the ones for strontium $Sr^{2+}$ [6, 18] or ($La_{1/2}Li_{1/2}$)$^{2+}$-complex substitutions (Fig.3.1b).

Now let us pass to PT-PMN solid solution system. From the viewpoint of physics and taking into account the above-considered examples it seems natural to expect that the substitution of titanium by the complex ($Mg_{1/3}Nb_{2/3}$) should lead to the change from low-temperature FE ordering to AFE ordering, and the Y-$T$ diagram of the system PT-PMN should be analogous to those discussed already in this section and presented in Fig.8.1 and 3.1b. The hysteresis region should manifest itself, too. This supposition is confirmed by experiments. The Y-*T* diagram of the system PT-PMN shown in Fig.8.2 (obtained based on the results of the papers [91-98]) is analogous to other diagrams Fig.8.1 and 3.1 and is analogous to the model diagram in Fig.2.3 (see also [20]) for systems with FE-AFE phase transition. (We do not consider phase diagram of PT-PMN system from morphotropic region of phase diagram).

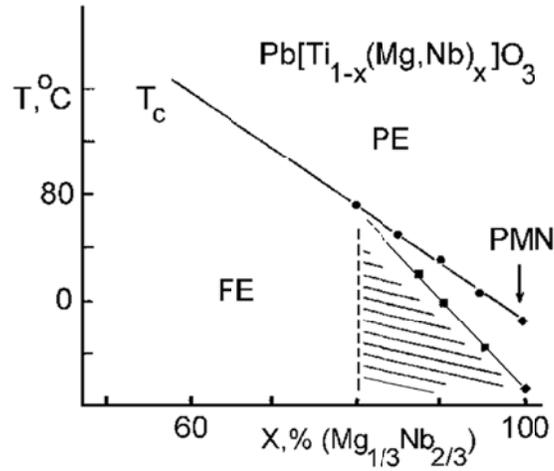

Fig.8.2. Phase diagram "composition-temperature" for solid solutions on the base of the lead titanate with substitution of titanium by complex $(Mg_{1/3}Nb_{2/3})$ [91-98]

Therefore, one may state that at room temperature PNM belongs to the AFE region of Y-*T* phase diagram and at temperatures below 220 *K* it belongs to the hysteresis region of FE-AFE states. If electric field has not been applied, the macroscopic state of the samples is AFE for temperatures $T < T_c$. After the action of the field with an intensity higher than a critical value, FE state is induced in the sample's volume at temperatures below $T_0 \approx 220$ *K*. Subsequent heating leads at first to FE→AFE phase transition at the temperature $T_0$ and then to AFE→PE transition. It should be noted that FE and AFE states in PMN are not entirely identical to the classic ones realized in the systems for which the interaction of the domains of these phases is negligibly small. For PZT-based solid solution the said problem has been discussed in [20, 47] and above in the present paper; for PMN it will be considered below.

The understanding of the origin of phase states and of their FE-AFE nature in PMN allows to predict a number of its basic properties and to search for their experimental verification. The analysis of the Y-*T* diagram (Fig.8.2) of solid PT-PMN solutions allows us to conclude that PMN should have the *P-E* and *T-E* diagrams analogous to ones presented in Fig.2.2c and 2.2i, respectively. The authors have not succeeded to find the *P-T* diagram for PMN in the literature, but the *T-E* diagrams for this substance completely confirm this statement.

Phase T-E diagrams for different PMN samples are presented in Fig. 8.3. The diagram in Fig. 8.3a is taken from [99] for ceramic samples (this is generalized diagram based on experimental data [100-102]). The diagrams in Fig. 8.3b and Fig. 8.3c are for the crystals for different directions of applied electric field. These diagrams are completely identical to the model diagram Fig. 2.2i for substances undergoing FE-AFE PT.

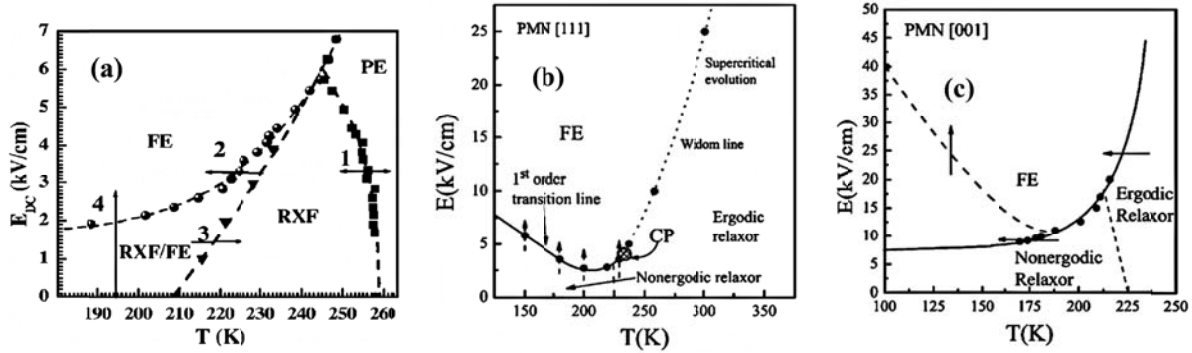

Fig. 8.3. "Temperature-electric field" phase diagrams for PMN. (a) The standard empirical history-dependent phase diagram is presented in [100-102]. "FE" represents a ferroelectric state, "PE" a paraelectric state with nanodomains, and "RXF" a glassy relaxor state in terms of [99]. (b) Phase diagram of PMN crystal for the bias field in the [111] direction [103]. (c) Phase diagram of PMN crystal for the bias field in the [001] direction [103]. Here, arrows show the path of the approach to the FE state for different transition lines.

The minimal critical field $E_{i,min} \equiv \Delta_i$ exists in T-$E$ diagrams of PMN. Moreover, PMN satisfies the requirements to the temperature behavior for FC and ZFC states which follow from the model T-E diagram shown in Fig.2.2i (see the commentary to Fig.2.2i in the present paper or to Fig.2 in [18]). The difference of FC and ZFC behavior of PMN is caused by the existence of minimal field for induction of FE phase. Such situation has been discussed in Ch.2 and has been confirmed experimentally for PZT-based solid solutions in Ch.3 of the present paper.

As we discussed in Ch.2 the application of hydrostatic pressure leads to increase of $E_{i,min} \equiv \Delta_i$ (see Fig. 3.3 and Fig. 3.4). On the other hand the increase of the lead titanate content in (1-x)Pb($Mg_{1/3}Nb_{2/3}$)$O_3$ – $xPbTiO_3$ has increase the stability of the AFE phase and decrease of $E_{i,min} \equiv \Delta_i$. The AFE state appears spontaneously at decrease of the temperature below the Curie point when the equality $E_{i,min} = 0$ is achieved. Based on the "composition-temperature phase diagram for the (1-x)Pb($Mg_{1/3}Nb_{2/3}$)$O_3$ – $xPbTiO_3$ solid solutions presented in Fig. 8.2 the vanishing of $E_{i,min}$ in the diagrams "temperature - electric field" has to take place in the vicinity of $x \approx 0.20$. This conclusion is confirmed by the experimental results given in the Fig. 8.4 [104].

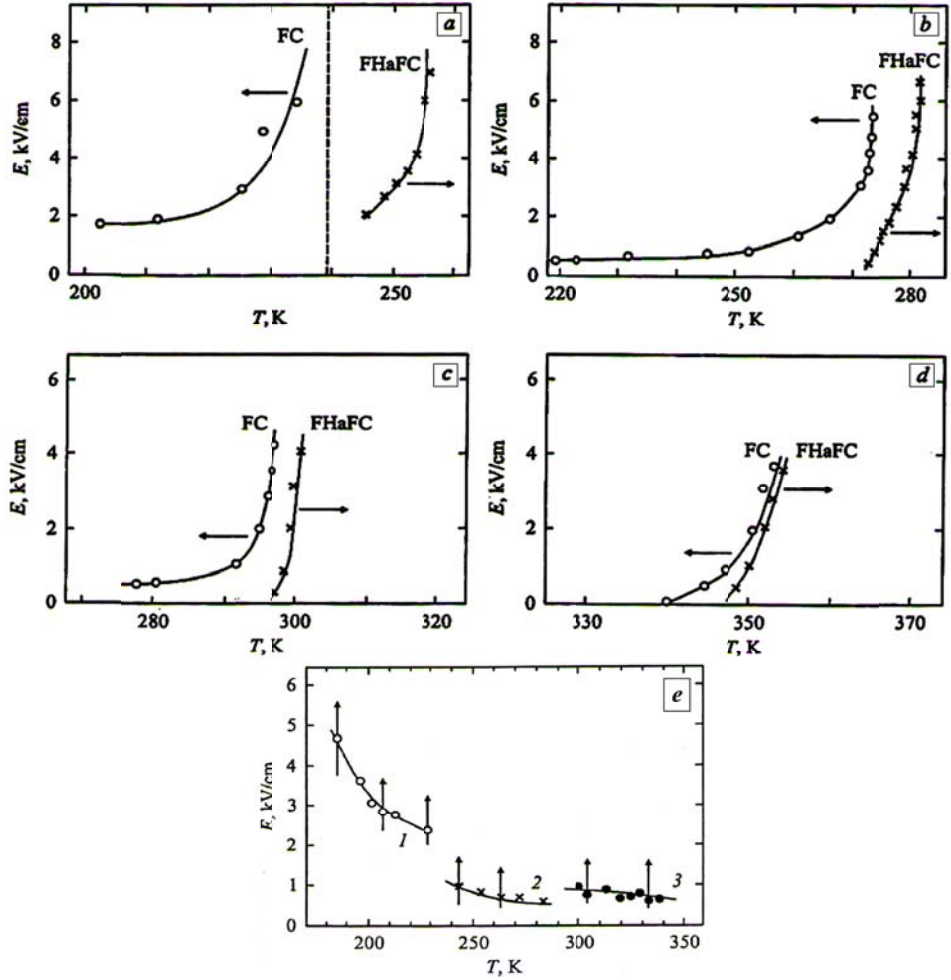

Fig.8.4. The *T-E* phase diagrams for single crystals with compositions (1-x)Pb(Mg$_{1/3}$Nb$_{2/3}$)O$_3$ – xPbTiO$_3$. DC electric field was applied along [001] direction [104]. Arrows show the directions of change of external parameters (temperature or DC field). Content of PbTiO$_3$, x: a - 0.06, b - 0.10, c – 0.13, d – 0.20. In the diagram (e) PbTiO$_3$ content x: 1 – 0.06, 2 – 0.13, 3 – 0.20. Based on the results, presented in the work, "the phase boundary" for composition 3 in diagram (e) (*x* = 0.20) corresponds to the usual process of domain ordering in DC electric field.

As one can see from the results obtained for single crystalline samples, the intensity if electric field necessary for induction of the FE phase decreases with the increase of the PbTiO$_3$ content *x* in the solid solution. The value of the parameter $E_{i,min} \equiv \Delta_i$ also decreases and at the compositions in the vicinity of 20% of lead titanate this value is equal to zero.

Thus, the phase diagrams for the PMN-PT system of solid solutions available in the literature can be completely described in the frames of view point of FE-AFE transitions developed in Ch.2 and coincide with experimental data for other systems of solid solutions for which the presence of FE-AFE transitions has been uniquely established. In has to be especially emphasized that the difference between FC and ZFC behavior in PMN is also a consequence of FE-AFE transitions in external electric field.

Now let us formulate other statements concerning the behavior of PMN based on the generalized model of FE-AFE PT considered in detail in [20] (Ch.1 and 2), and then verify their validity in practice. Such statements are as follows.

Firstly, the AFE state which is realized in PMN at cooling differs from the one observed e.g. in PZT. This distinction consists in the fact that in PMN the mentioned state is not one-phase: it is characterized by the presence of FE phase domains. By analogy with PLZT or PLLZT, the size of the metastable phase domains must be of the order of 100 Å or smaller.

Secondly, since the equivalent crystal lattice sites (B-sites) are occupied by ions with different size, there should be observed the "stratification" of the substance along the interphase FE-AFE domain boundary and the formation of smaller composition domains rich either in niobium or in magnesium.

Thirdly, PMN will preserve its inhomogeneous structure at temperatures essentially higher than $T_m$ (but lower than 350°C). The said inhomogeneous structure has to lead to diffusive PE phase transition.

All these conclusions concerning the structural inhomogeneity of PMN which follow from the model of FE-AFE transformations [20] have been completely confirmed by experiments. But the described inhomogeneities, though experimentally revealed long ago, still have no clear interpretation by authors of the paper in publications.

Let us address studies of behavior of PMN in the presence of the hydrostatic pressure. It is known [6] that under the hydrostatic pressure the stability of AFE state relative FE one increases. In particular, in the presence of domains of these phases in the sample the increase of the pressure should lead to increase of the AFE phase share in the sample's volume and, finally, to single-phase AFE state. The data available in the literature completely confirm this conclusion.

Experiments on X-ray diffuse scattering in PMN under the high pressure were performed at room temperature in the 1 bar to 10 GPa range at the European Synchrotron Radiation Facility on the ID30 high-pressure beam line [105]. Bragg reflections, diffuse scattering, and weak superstructure reflections were observed at low pressure. The diffuse X-ray scattering (XDS) along with weak superstructure reflections corroborate the evidence for local deviations from average (cubic) structure in PMN. As rule, both structural deviations ($BO_6$ octahedra tilts or/and cation displacements) and a local substitution-type (chemical) disorder have to be considered as the origin of the XDS. As was shown in that paper, the pressure dependent changes of the XDS (Fig.8.5) cannot be explained by considering a chemical origin, which would be essentially pressure independent.

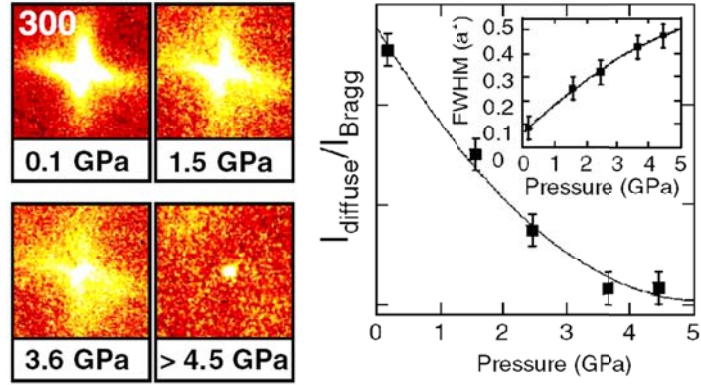

Fig.8.5. On the left - the illustration of the pressure dependence of the diffuse scattering for PMN around the 300 reflection. On the right – the evolution under pressure of the intensity and FWHM (inset) of the diffuse scattering around the 300 reflection [105].

A transition to a long-range phase occurs at the high pressure according to results of results of a high-pressure investigation by Raman spectroscopy [106]] that reveal a non cubic long-range structure taking place with increase of the pressure. The low pressure XDS the authors connect with two-phase structure of a substance. The pressure dependent evolution of $(135)_{1/2}$ superstructure and (400) and (220) Bragg reflections are illustrated in Fig.8.6 (taken from [105]).

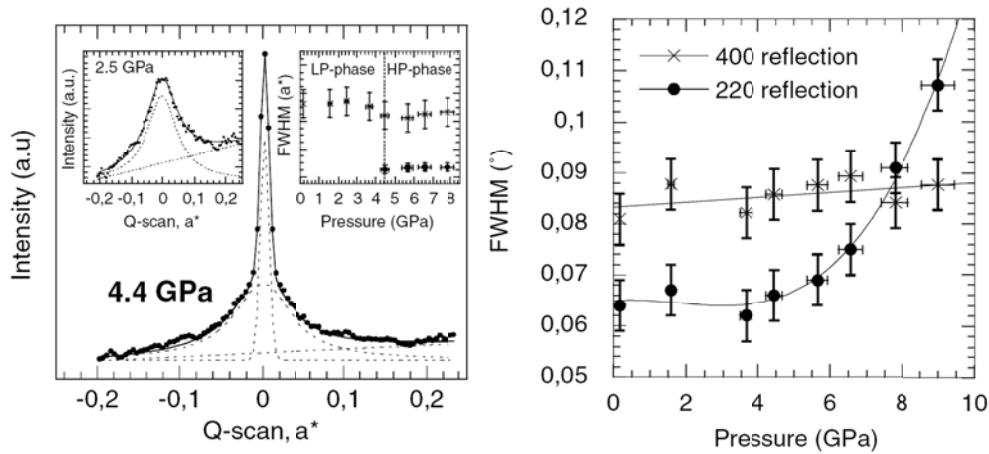

Fig.8.6. On the left: The pressure dependence of 1/2(135) superstructure reflection. The left inset illustrates the diffuse scattering for $p$ <4.5 GPa. The right inset presents the evolution of the FWHM of the diffuse and sharp features as a function of pressure. On the right – Pressure-dependent evolution of the FWHM of the (400) and (220) Bragg reflections [105].

The superstructure reflections appear to be composed of two components for pressure more than 4.0 GPa: one diffuse reflection (already present at ambient conditions) superimposed with a new sharp Bragg reflection (it is only one component of the said diffusive line for $P < 4.0$ GPa). As shown in the inset in Fig. 8.6 (on the left), the FWHM of the diffuse reflection is rather pressure independent, which adds further support to its chemical origin. On the other hand, the pressure-induced sharp reflection is related to a structural cell doubling, which might originate in antiphase tilts of octahedra and/or antiphase cation displacements. Such a superimposition of two components related to a unit cell doubling is reminiscent of the temperature-dependent investigation of 1/2($hkl$) reflections by Gosula [107], where a new sharp component has been observed at low temperature. However, Raman spectra at high pressure [106] and low temperature one are different, suggesting that the two phases are not identical. Details of this new structure were also observed by Raman spectroscopy [106] that showed that the high-pressure phase of PMN is characterized by nonpolar $B$ sites, i.e., the Nb and Mg cations are not displaced. Combining the complementary information, authors of [105] propose that the high-pressure phase of PMN has antiparallel displacements of the Pb cations.

The PMN and PMN:La(5 %) samples were studied by selected-area electron diffraction (SAED) [108], high-resolution electron (HREM) microscopy, and computer simulation. Exceptional super lattice diffraction spots (these is the name given by the authors of [108] at the positions $k/2\langle 110 \rangle$ and $h/3\langle 210 \rangle + m\langle 001 \rangle$ (where $k$, $h$ and $m$ are integer numbers) with diffusive scattering were observed at room temperature. SAED patterns of three different zone axis are shown in Fig.(8.7) ( taken from [108]). The diffraction sports from disordered PMN matrix are indexed in this figure with simple cubic indexes. The letter "O" denotes the $1/2\langle 110 \rangle$ superlattice diffraction spots from the 1:1 chemical ordering, which were well studied by many researches. The exceptional superlattice spots are indicated by letter "N".

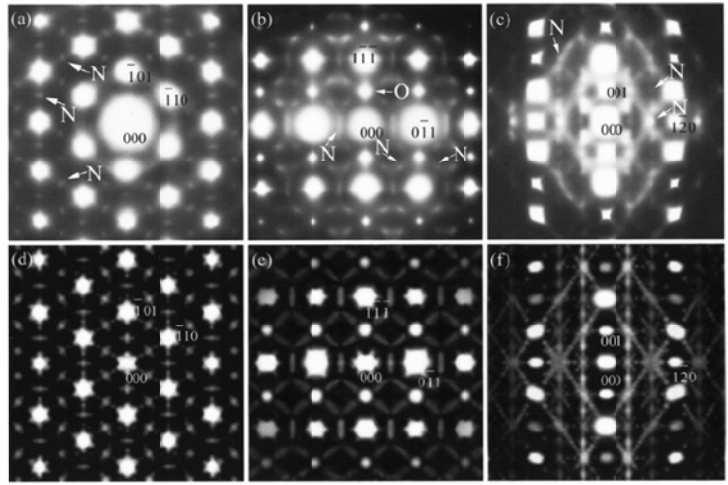

Fig.8.7. Experimental and simulated SAED patterns, indexed with simple cubic indices. (a) experimental [111] zone axis; (b) experimental [211] zone axis; (c) experiment [210] zone axis; (d) simulated [111] zone axis; (e) simulated [211] zone axis; (f) simulated [210] zone axis. [108.]

According to [108] the main features of these patterns can be summarized as follows:

-new spots at positions $k/2\langle 110\rangle$,

- new spots at positions $h/3\langle 210\rangle + m\langle 001\rangle$ with diffuse elongation along $\langle 001\rangle$,

- intensities fluctuation on "N" spots along $\langle 210\rangle$ and $\langle 001\rangle$, respectively,

- diffuse fringes along $\langle 213\rangle$ in $\langle 210\rangle$ zone-axis pattern.

The intensity of the "N" spots are very weak, therefore long exposure time and special contrast techniques were used, i.e. developing the strong and weak patters with different time for registration of strong "1:1" and weak "N" spots. Authors of the said paper shown that the "N" spots positions cannot be caused by the multiple scattering from strong diffracted beams (the latter can lead to extra diffraction spots in SAED patterns). Analyzing different structural factors that can lead to effects observed in this study, the authors came up with the possible pattern of structural deformations (Fig. 8.8) that can lead to appearance of local AFE ordering in PMN.

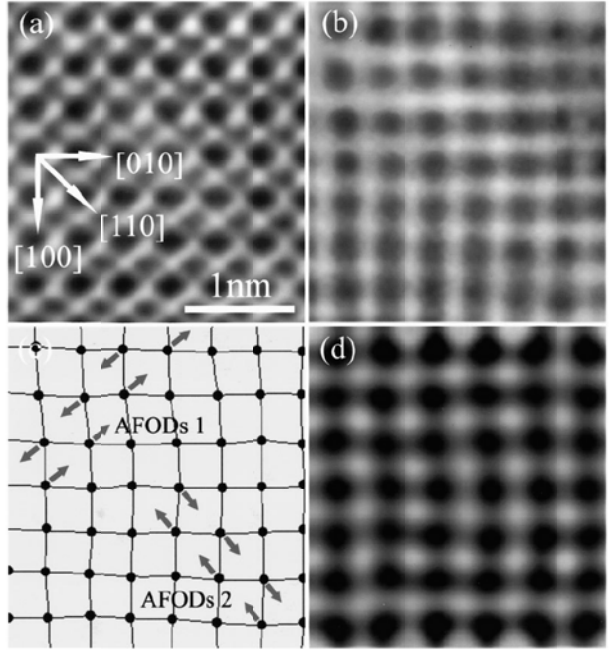

Fig.8.8. (a) HREM of PMN along [001] direction, Scherzer defocus 561 nm; (b) the fast Fourier transform (FFT) image of (a), the $(100)$, $(010)$, $(\bar{1}00)$, $(0\bar{1}0)$ and transmitted reflections are used; (c) a schematic diagram of the $Pb^{2+}$ displacements in (b); (d) the computer simulated image of (b). [108] AFOD – antiferroelectric ordered dipoles.

In our opinion the above-given experimental data uniquely indicate the composite structure that is present at the temperatures $T < T_{m(c)}$. This composite structure constitutes AFE matrix with inclusions that are domains of the FE phase. Peculiarities of behavior of this structure under external influences are determined by the structure of these domains of coexisting phases and their interactions. Effects caused by such internal structure of PMN samples will be considered below.

Similarly to PLLZT and PLZT solid solutions, which relaxor properties (and diffusive phase transitions) are caused by weak energy difference of FE and AFE states and the coexisting of the said states domains in the volume of samples (see Ch.6 and Ch.7 of the present paper and [20, 47, 109]), the relaxor properties of PMN and related materials are defined by the same mechanism, too. The corresponding data can be found in the literature.

The $(1 - y)PbTiO_3 - yPb(Ni_{1/3}Nb_{2/3})O_3$ system of solid solutions is identical to the $(1 - y)PbTiO_3 - yPb(Mg_{1/3}Nb_{2/3})O_3$ system. The phase Y-$T$ diagram of the former system is also identical to the one for the latter system (Fig.8.2). In both cases the solid solution with $y = 0.4$ is in FE state at temperatures $T < T_c$. The paraelectric PT for both solid solutions is sharp [110, 111]. According to the concept developed in the present paper, the increase of stability of the AFE state should lead to diffuse PT and to the manifestation of relaxor properties. In [111] such increase of stability of the AFE state was achieved by substitution of Zr ions for Ti ones. This predicted change in the behavior of $Pb[Ti_{0.4}(Ni_{1/3}Nb_{2/3})_{0.6}]O_3$ solid solution becomes apparent in experiment (see Fig.8.9). The discussed phenomenon corresponds completely to the model diagram in Fig.2.3. In this case the ion substitution takes place in B-sites of perovskite crystal lattice.

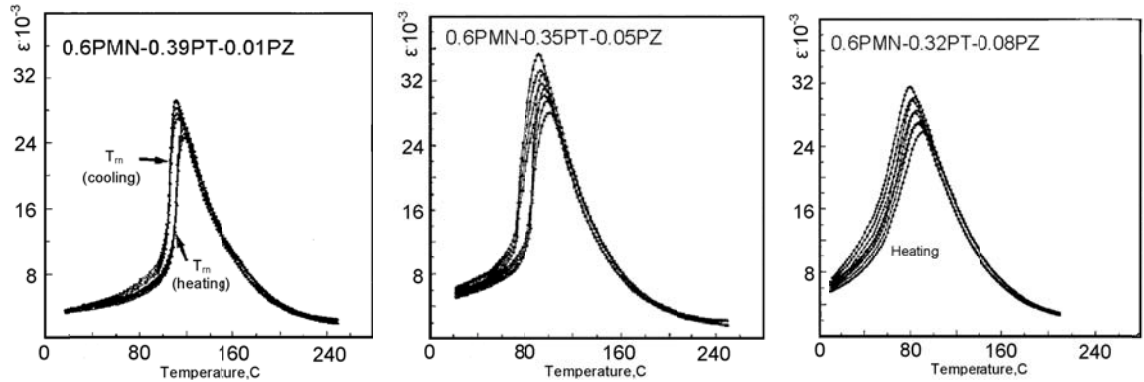

Fig.8.9. Temperature dependences of relative dielectric permittivity for 0.6PNN-(0.4-x)PT-xPZ solid solutions [111]. The measurement frequencies from the top curve to the bottom one are 0.1 kHz, 1 kHz, 10 kHz, and 100 kHz.

The inhomogeneous structure in domains of coexisting phases (that have close values of their thermodynamic potentials) in PMN has to be preserved up to the temperature of approximately 350 °C. The diffuse PT from PE phase is the consequence of the presence to this structure. Any influences leading to the change of the relative stability of the FE and AFE phases have to lead to the change of the shares of these phases in the sample's volume and as a consequence it will lead to the noticeable change in the diffuseness of the PT. The pressure is the main among these factors.

However, we must emphasize that there is a lack of the experimental results obtained using hydrostatic pressure for PMN and the system PMN – PT to compare the results obtained in the present paper with the experiment. On the other hand, the $Pb(Zn_{1/3}Nb_{2/3})O_3$ – $PbTiO_3$ (PZN – PT) system is the physical "twin" for the system PMN-PT. Both systems are characterized by identical phase diagrams "composition - temperature" $x – T$ diagrams. Results of studies of thermal phase transformations in the $Pb(Zn_{1/3}Nb_{2/3})O_3$ – $PbTiO_3$ (PZN – PT) system of solid solutions under the action of hydrostatic pressure and electric field were published in [112]. These results are in perfect agreement with the ideas developed in the present paper. In particular, the solid solution 0.905PZN – 0.095PT undergoes FE-PE phase transition. This transition is weakly diffused and so-called "relaxor properties" are practically absent (Fig.8.10).

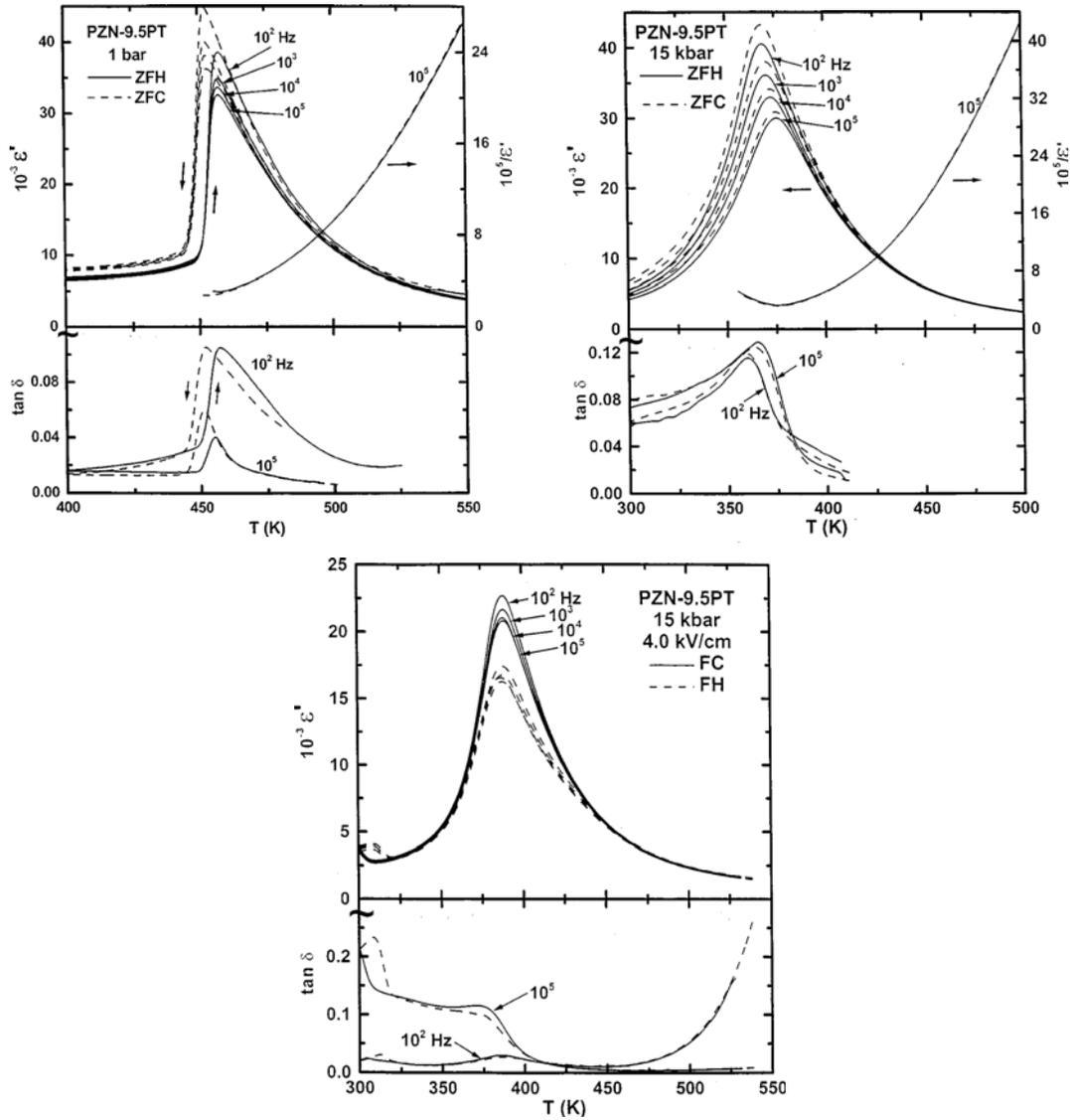

Fig. 8.10. Temperature dependences of the dielectric constant $\varepsilon'$ and dielectric loss $\tan\delta$ for PZN-9.5 PT [112]. The experimental conditions (DC bias and pressure) are explained in the upper corner of each picture.

Hydrostatic pressure leads to the rise of the AFE phase stability and displacement of the position of the above-said solid solution in the $x – T$ diagram to the region of AFE states. This means that the increase of the pressure results in the increase of AFE phase stability and coexistence of FE and AFE phase domains in the bulk for temperatures lower and higher Curie point. Because of this reason the paraelectric phase transition becomes diffused and the relaxor properties are found (Fig.8.10).

Then electric field is applied to the sample under the pressure. This action causes the return of the above-said solid solution into FE region of the x –T diagram. On this reason the stability of FE phase rises and the AFE phase stability decreases. So, the sample returns in that state which existed before hydrostatic pressure application (the initial uniform state). After the said procedure the phase transition into PE state becomes undiffused and relaxor properties disappears (Fig.8.10).

Now we would like to present and discuss the results of investigations of diffuse PT in the system PMN-PbZrO$_3$ [113]. In the PMN structure Zr$^{4+}$ ions substitutions for complex (Mg$_{1/3}$Nb$_{2/3}$)$^{4+}$ lead to increase of stability of the FE state. This means that the introduction of Zr ions into the crystal lattice has to lead to "transition" from the state of coexisting FE and AFE phases to the single-phase state. This transition has result in reduction of the degree of diffuseness of the PT. Temperature dependencies of dielectric permittivity for three solid solutions from the (1-x)PMN-xPbZrO$_3$ system with $x = 0$ (which is just PMN), c $x = 0.1$, and $x = 0.5$ are presented in Fig.8.11.

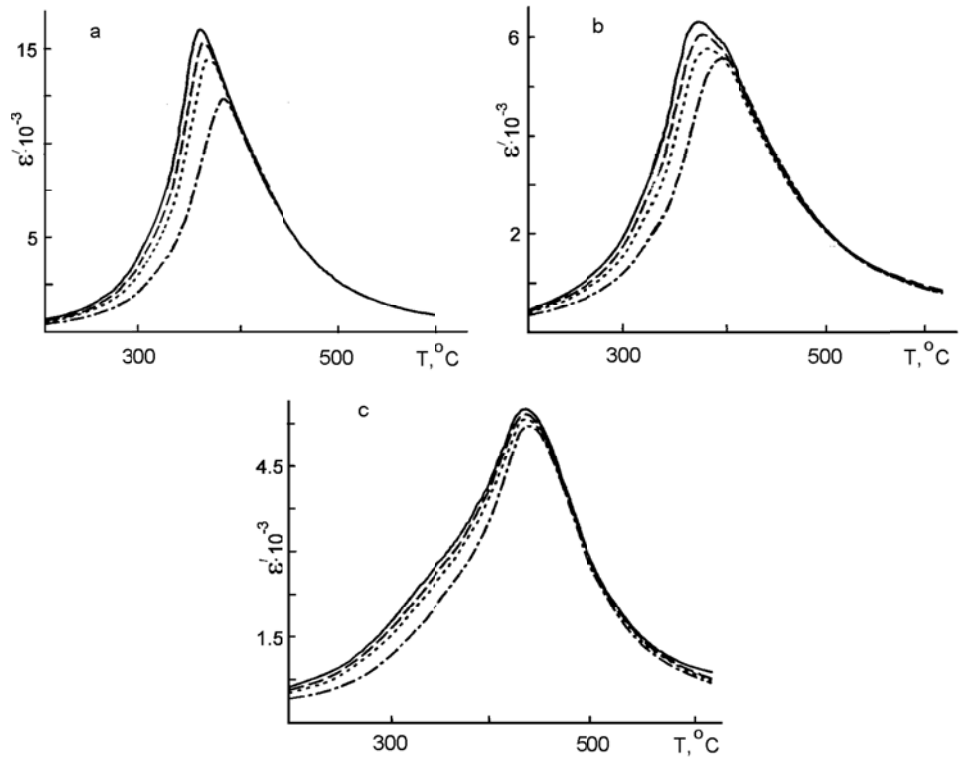

Fig.8.11. Temperature dependencies of $\varepsilon'$ for (1-x)PMN-xPbZrO$_3$ solid solutions [113]. $x$: a - 0, b – 0.1, c – 0.5.

Modification of these dependencies with increase of lead zirconate content, which can be clearly seen in this figure, completely confirms the conclusions of the model of coexisting FE and AFE phases in application to lead manganese niobate. Let us not that these conclusions were discussed in details in Chapter 6 of present paper.

In addition to above-given results let us not the well know result of experimental investigations that substitution of $Ti^{4+}$ ions for $(Mg_{1/3}Nb_{2/3})^{4+}$ complexes leads to increase of stability of the FE phase in lead manganese niobate (see diagram in Fig. 8.2) and transition of the substance toward the region of the single-phase FE state in the diagram of phase states (Fig. 8.3). As this takes place the degree of diffuseness of the PT decreases and relaxation properties of the (1-x)PMN-xPbTiO$_3$ solid solutions get suppressed.

All experimental data on studies of diffuse PE PT show that high-temperature state in the temperature interval $T_m < T < T_d \approx (300\text{-}350)$ $^oC$ are not homogeneous. However, the main result is that the degree of PT diffuseness and along with it the degree on inhomogeneity depend on the location of the composition of the concrete substance in the diagram of states. Everything is identical to the situation that was observed in the PLLZT solid solutions and was analyzed in Ch.6. That is why one can make a conclusion that the composite FE+AFE domains exist in the PE matrix of the PMN at the temperatures higher than $T_m$ (this excess can be as much as several hundred degrees).

Let us now consider the literature data that confirm existence of such two-phase FE+AFE domain in the PE matrix of the substance at the temperatures above $T_{m(c)}$. The results of investigations of PMN and PMN-based solid solutions up to the temperatures 600-700 $^oC$ are available in the literature.

There exist a large number of publications that contain substantial amount of data of investigations of diffuse X-ray and neutron scattering as well as on investigations of spectra of Raman and infrared light scattering along with data of acoustic studies by means of Brillouin scattering. Without going into details of all these studies we want to emphasize that all the authors pointed out the appearance (at the decrease of temperature) of peculiarities in behavior of PMN at the temperatures below $T_d \approx 350$ $^oC$ (Fig. 8.12 (a-c)).

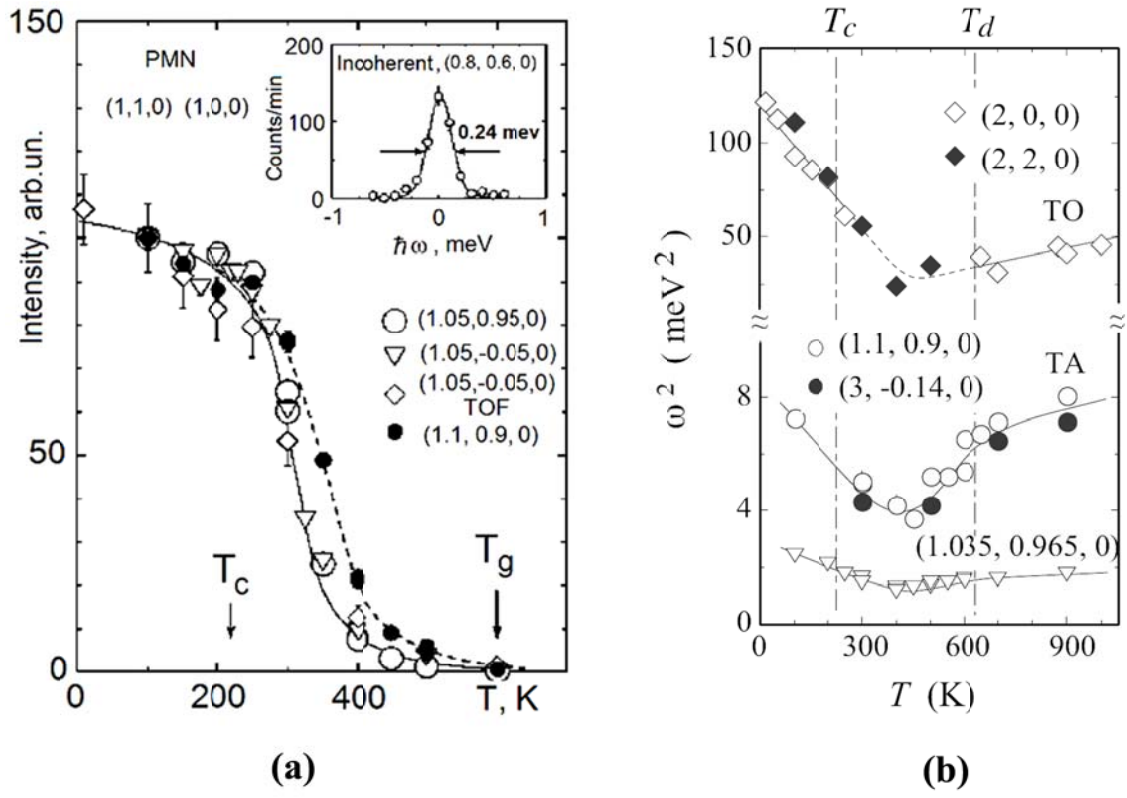

Fig. 8.12. (a) Temperature dependence of the elastic diffuse scattering intensity measured at $\zeta = 0.05$ and 0.1. Data are normalized at 100 K [114]. Open diamonds show the results from [115]. (b) The square of the phonon energy is plotted versus temperature for the zone-center TO modes [116] measured at (200) (open Diamonds) and (220) (solid diamonds) taken from [117] and [118], as well as for three TA modes measured at (1.035, 0.965, 0) (open triangles), (1.1, 0.9, 0) (open circles), and (3, −0.14, 0) (solid circles). Linear behavior is consistent with that expected for a conventional ferroelectric soft mode.

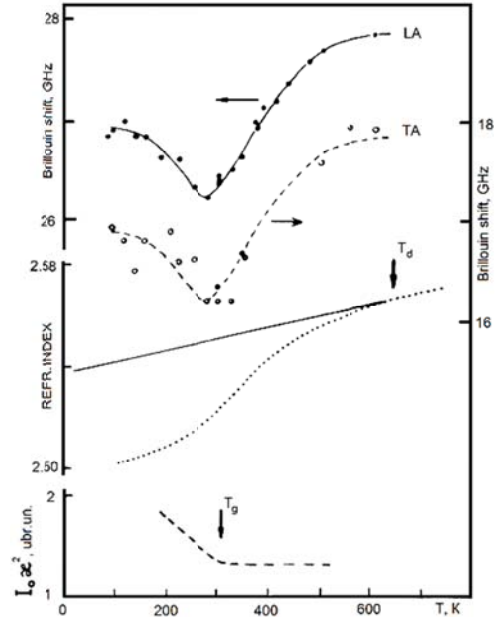

Fig. 8.12c. The temperature dependence of Brillouin shift for LA and TA acoustic phonons (upper part) [119], the refraction index [120-122] (in the middle) and characteristic value $J\chi^2$ from the neutron scattering [123] in PMN crystal.

In spite of a large number of experimental studies the nature of the phase state below this temperature is absolutely unclear.

Among all observed peculiarities of different physical characteristics we will dwell on peculiarities of the temperature dependencies of the frequencies of optic and acoustic oscillations. Both optic and acoustic modes undergo softening below the temperature $T_d$ and have pronounced minima (Fig. 8.12b and Fig.8.12c). Practically all the researchers are agreed now (it started since articles [120-122] were published) that the temperature $T_d$ determines the beginning of the process of formation on nuclei of new phase (with the nanoscale sizes) at cooling. In order to explain the difference between $T_d$ and the temperature that corresponds to minima of the frequencies of oscillation modes the opinion is expressed that these nuclei have a dynamic nature at the beginning and then become thermodynamically stable. The softening of the optic TO mode is attributed with the polar character of these nuclei. This explanation is clear and straightforward. However, in regard to the softening of the acoustic modes majority of researchers admit almost complete absence of understanding of physical mechanism of this phenomenon (we are not going to discuss different attempts to give the explanation because they are far from consistent explanation of this phenomenon).

As we have noted above the available data point to the composite structure of nuclei of the new phase in the PE matrix of the substance. These nuclei are two-phase FE+AFE domains. In Ch.6 experimental and theoretical justification of appearance of these domains in the PE matrix of the substances possessing the triple FE-AFE-PE point at the temperatures significantly higher than $T_c$ was discussed in details. The literature data given above in this chapter confirm the same character of high-temperature nuclei in PMN also. Based on such approach it is easy to explain the softening of acoustic modes. This softening is connected with the AFE component of two-phase high-temperature ($T_c < T < T_d$) domains. As we said above the relation between the shares of the FE and AFE phases inside the two-phase nucleus changes under external influences including heating and cooling. Such behavior of the shares of phases has to lead to changes in intensities of lines corresponding to optic and acoustic modes. The interaction of phases that is interaction of the order parameters inside domains described by the term $P^2\eta^2$ and interdomain interaction described by expressions (1.2) and (1.9) (see Ch. 1) has to lead to the changes in linewidth (damping of oscillations).

This approach is not something new and it was successfully applied for explanation of peculiarities of behavior of substances which the difference in energies of FE and AFE states is fairly small. This subject was discussed in detail for different situations in Ref. 6 where the literature references can be found. Here we want to point out the following. Studies of Mössbauer spectra in lead zirconate (PbZrO$_3$) [124] revealed at the temperatures above $T_c$ the peculiarities in behavior of PbZrO$_3$ are similar to the ones of PMN at the temperatures above $T_c$. Phase transition takes place at the temperature $T_c = 236$ °C, however, minimal value of intensity of Mössbauer line (decrease of the probability of Mössbauer effect) is noted at the temperature significantly higher than $T_c$ unlike the classic FE of the BaTiO$_3$-type. The decrease in the intensity of the line reaches 40% whereas in the lead titanate at the PE PT this decrease reaches only 10%. Naturally, that the authors relate the dielectric anomaly in lead zirconate with the displacement of transverse optic TO mode with $\vec{k}=0$. The longitudinal acoustic LA mode is responsible for the AFE transition. Lead ions in the lead zirconate are displaced in the direction [110] which is the result of the lattice instability with respect to the LA mode at the boundary of the Brillouin zone corresponding to [110] direction.

Possibility of existence and interaction of the FE and AFE oscillation modes in lead magnesium niobate should not cause any astonishment. The free energy of the FE and AFE states has small difference in complex oxides with perovskite structure. Typical result confirming this statement is the conclusions derived from calculations of stability of different dipole-ordered states in classic ferroelectric $BaTiO_3$ [125]. It was shown that small variation of parameters leads is enough for both FE and AFE ordering to be realized in barium titanate. On the other hand it was shown [126] that two types of ordering can be present in lead zirconate the AFE ordering (as it is in fact realized) and the FE ordering.

It has to be noted in connection with above-said that in the course of discussions of experimental results researches were reasonably close to accept the existence of the two-phase FE+AFE domains at the high temperatures and as a consequence the interaction between FE-active and AFE-active oscillation modes. For example, studies of the local lattice dynamics using the dynamic pair-density function determined by pulsed neutron inelastic scattering in PMN [127] demonstrated that the dynamic local polarization sets in around the so-called Burns temperature $T_d$ through the interaction of off-centered Pb ions with soft phonons. , and the slowing down of local polarization with decreasing temperature produces the polar nanoregions and the relaxor behavior below room temperature.

The major increase of the intensity of diffuse scattering that takes place more than 100 ºC below $T_d$ (see Fig. 8.12a) during the cooling of samples can be easily explained by the presence of the two-phase nuclei in the PE matrix of the PMN at the temperatures $T > T_c$. First two-phase FE+AFE nuclei appearing at the temperature $T_d$ have their own refractive index different from the of the PE matrix. The deviation of the linear temperature dependence of the refractive index in the PMN crystals is manifested exactly at the same temperature. The difference in the crystal lattice parameters for FE and AFE phases constituting the two-phase nucleus is not very pronounced at this stage. However, this difference increases at the decrease of the temperature. That is why the local decomposition (the separation of ions $Mg^{2+}$ and $Nb^{5+}$ having different ionic sizes) does not happen at this initial stage (see Ch. 6 for details). The difference in crystal lattice parameters of the FE and AFE phases constituting a nucleus increases at the further decrease of the temperature and at the same time the elastic energy in the vicinity of the boundary between these phases increases. This increase of elastic energy becomes sufficient to start the process of local decomposition in the vicinity of the boundary between FE and AFE phases. This process takes place at the temperatures considerably lower than the $T_d$ temperature. It is precisely the beginning of the local decomposition is accompanied by the increase of the diffuse X-ray or neutron scattering.

The detailed analysis of all peculiarities of behavior of the lead magnesium niobate and similar substances is not in the frames of the main goal of this article. That is why we will not dwell in detail on such phenomenon as the "waterfall" phenomenon in relaxors. The review of the last results of recent studies of this phenomenon one can find, for example in Ref. 128. In our opinion taking into account all above-said this phenomenon is connected with the build-up of the process of local decomposition in the temperature interval between $T_d$ and $T_c$. The "softening" of the crystal lattice and the damping of the oscillation modes take place along with the intense decomposition. The disruption of the translational symmetry occurs simultaneously not only in the vicinity of the FE-AFE interphase boundaries but in a whole volume of the crystal. The process of the ion redistribution among the lattice cites involves the whole crystal volume. It was discussed in Ch. 5 of this article (see Fig. 5.3 and Fig. 5.7). The intensity of the mentioned process is maximal at the maximum intensity of the local decomposition process. Outside the temperature interval within which the intensive decomposition takes place the "waterfall" phenomenon is absent.

There exist a lot of literature data on the shape of the lines of diffuse scattering and their differences during scanning in reciprocal space in the vicinity of different Bragg peaks. The relationship between the shares of the FE and AFE phases inside the two-phase nucleus and the orientations of the polar axis (for the FE part of the two-phase domain) and the anti-polar axis (for the AFE part of the domain) are given by the condition that this nucleus lead to the lowest possible value of elastic energy increase. Such advancement of the physical process has to lead to a very interesting effect. Let us remind that the FE and the AFE components of the two-phase domain are characterized by different sign of the changes on configuration volume) in comparison with the one in the PE state). The diffuse scattering in the (1-x)PMN-xPbTiO$_3$ solid solutions at the temperatures above the curie point is observed both for small values of *x* and for the big ones. In the first cans the low-temperature FE phase of the solid solutions is characterized by the rhombohedral type of distortions of the crystals lattice and in the second case these distortions are tetragonal. The nucleation of the two-phase domains is possible in both cases (see Fig. 6.2 and explanations in the text of Ch.6). However, the morphology of these domains will be different because they differ in orientations of the polar axes of the FE component of the domain in the first and second cases. This has to lead to a different shape of the profiles of diffuse lines during the scanning in the vicinity of the same Bragg peaks. The difference in shape of the profiles obtained during scanning in the same substance but in the vicinity of different Bragg peaks can be also explained by the difference in morphology of the two-phase domains. These effects are well studied experimentally.

Several different explanations of unusual development of the ferroelectric-like properties of PMN have been suggested, namely, a model based on compositional fluctuations, model of superparaelectric behavior, and a model of glass-like behavior. As a rule these models do not explain a whole set of properties and one has to use modifications of these models. All models as well as their modifications are based on chemical inhomogeneity (compositional fluctuations) of PMN which is given by the way of the sample preparation. That is why a method of fabrication of single crystalline and polycrystalline samples may significantly influence the results of studies. We want to point some technological factors, which can influence the results of experimental investigations, and subsequently the interpretation of these results. Herewith the greatest influence on the experimental results one may expect during high-temperature studies.

Review of experimental results on inelastic neutron scattering and model calculations for PMN one can find in Ref. 128 where the results obtained by different authors are considered and the comparison of measurements on different samples is made. It is natural that due to various kinds of inherent disorder (chemical, strain, polarization) in the so-called relaxor materials, they are not in a uniquely defined equilibrium state and, thus, their properties could depend on the specific crystal growth method and the sample history in a rather complicated way. For example, recently made comparison of PMN samples grown by different methods [129] allowed to show that some of them do not undergo the expected phase transition under electric field.

The damping of modes of inelastic neutron scattering spectra in PMN could be influenced by the geometric aspects of homogeneity regions, namely, by the sample's state. Such influence has been examined in [128]. In this paper, similar spectra obtained on different PMN samples in different experiments that were performed in the same way, namely, for the same orientation $\vec{q}\,||\,[100]$ of the phonon wave vector for (even/even/even)-type Brillouin zone are collected and directly compared. The results of such comparison for transverse modes are presented in Fig. 8.13.

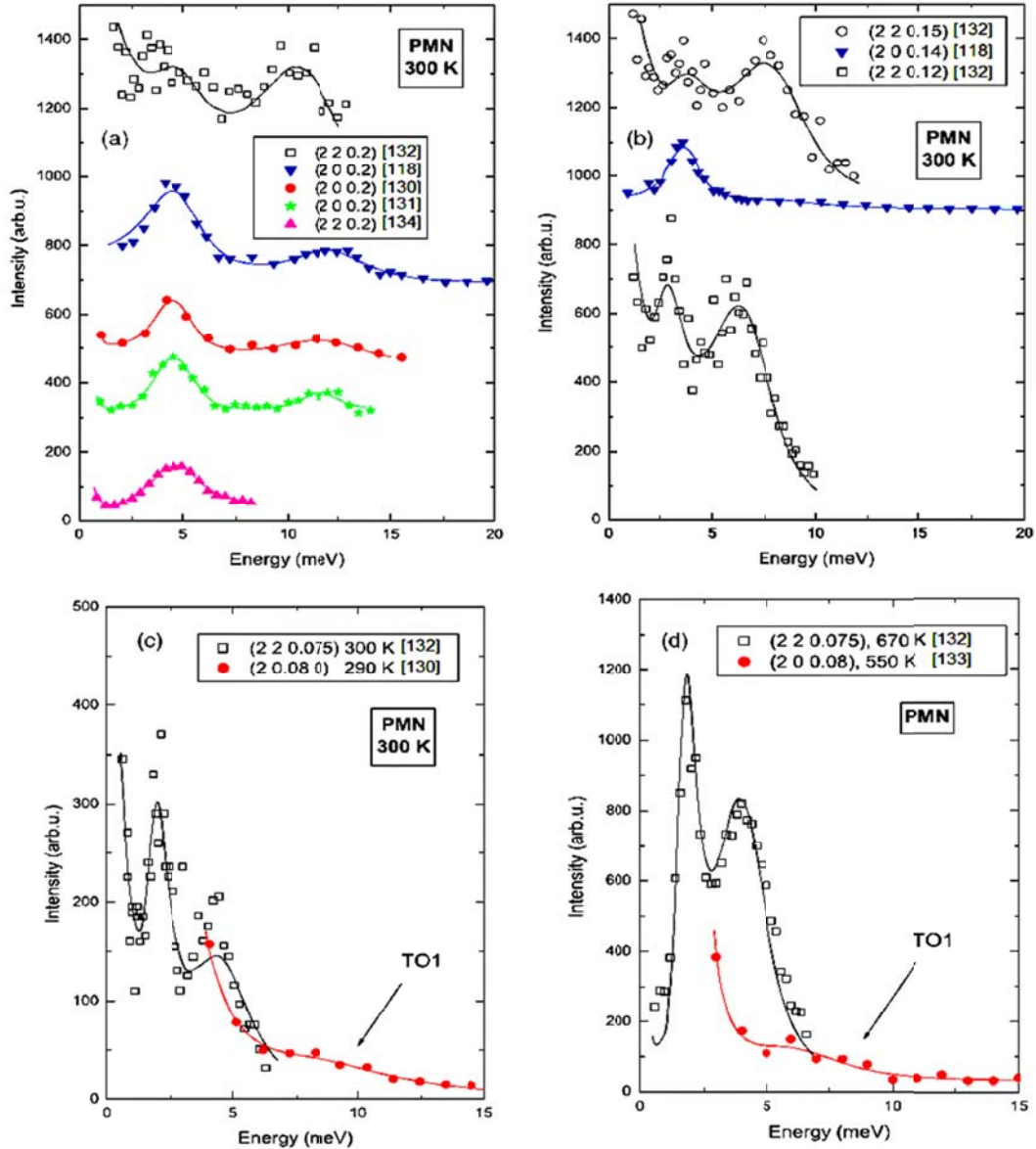

Fig.8.13 Spectra of inelastic neutron scattering with $\vec{q} \parallel [100]$ in (even/even/even)-type Brillouin zones in PMN obtained by different authors [128]. (a) spectra taken at $q = 0.2$ r.l.u.; (b) spectra taken at $q = 0.12$-$0.15$ r.l.u.; (c, d) spectra taken at $q = 0.075$-$0.08$ r.l.u. Spectra (a, b, c) are taken at room-temperature, (d) are taken at higher temperatures around 600 K. The intensity scale for the experimental data and original fitting curves in (a) were adapted so that the TA mode intensity is similar in all scans.

Clearly, all spectra show one mode below 5 MeV and another one above 10 MeV, in agreement with the expectations for rough positions of the TA and lowest frequency TO modes for this phonon wave vector. Samples used in these measurements had different origin, for example, the sample of Ref. [118] was grown by the modified Bridgman method, the sample of Ref. [130] by the top-seeded solution method from the PbO flux, and the sample of Ref. [131] by the Czochralski method. Despite the different origin of these samples, the differences in their spectra at this wave vector ($\vec{q} = 0.2\vec{c}^*$) are rather minor. At the same time, a closer look suggests that the upper peak in the spectra corresponding to the no-waterfall case [132] is indeed somewhat less damped, more intense and has a noticeably lower frequency.

This difference becomes more apparent at lower wave vectors. At wave vectors $\vec{q} = (0.12 - 0.15)\vec{c}^*$ the position of the second peak in the 'no-waterfall' experiment strongly shifts downwards in a similar way as the acoustic mode does, so that it keeps (due to the Bose-Einstein thermal factor) practically the same intensity as the acoustic one, while in the other experiments the upper mode does not show such strong dispersion (Fig.8.13d). The spectra taken at even smaller wave vectors ($\vec{q} = (0.075 - 0.08)\vec{c}^*$, Fig.8.13c) clearly show that the position of the upper peak in the two experiments differs by more than 3 MeV. Similar discrepancy is also observed at higher temperatures (Fig.8.13d), because the TO mode frequency in the experiment of Ref. [132] is almost temperature independent. Such a huge variation in the TO mode frequency could be comparable to changes associated with some structural phase transition or anomalous crystal structure in the specimens

Many researchers refer the presence of structural inhomogeneities in PMN to technological factors. Hence there may be drawn the following conclusion: the inhomogeneities formed in the process of crystal growth (or the process of manufacturing of ceramic samples), i.e. at temperatures of 1000° C and higher, must be preserved at all low temperatures. However, the model of FE-AFE states for PMN yields quite different statement: since the structural and chemical inhomogeneities are caused by the coexistence of FE and AFE phase domains, they will be preserved till such coexistence takes place and will disappear when the coexistence of phases is absent. For PMN the highest temperature at which such inhomogeneities are observed should correspond to 300-350°C (and this is fully analogous to the situation observed in PLLZT solid solutions by X-ray method [13] and discussed in Ch. 6 of the present paper). In the course of thermocycling carried out at the temperatures close to the above-mentioned one such inhomogeneities must be formed at cooling and be completely broken down at heating (thus making the structure homogeneous). Naturally, this process may be somewhat protracted, since at the said temperatures diffusion of elements is rather slow. So, we have formulated the first basic test for the applicability of the model of the FE-AFE transformation to PMN. Contemporary methods of transmission electron microscopy allow direct verification of presence or absence of domains of the coexisting FE and AFE phases in the high-temperature region as well as verification of the presence of the local segregations of ions of different sizes in the vicinity of boundaries between domains of above-mentioned phases.

Now let us point to the second experiment for checking the validity of the developed ideas for the explanation of PMN properties. As mentioned above, PMN must be characterized by *P-T* diagrams similar to the ones presented in Fig.2.2c and in its turn it unambiguously calls for the *E-T* diagrams and their evolution given in Fig.2.2i. This experiment may be easily realized.

Here the discussion of the problem of phase transformations in PMN may be finished. Based on available experimental data we claim that all the peculiarities in the behavior of PMN are caused by FE-AFE-PE transformations taking place under the action of external factors. Other states do not participate in phase transitions since they are absent in this substance. However, each of the mentioned states has peculiarities bound up with the interaction of the coexisting FE and AFE phase domains.

## 8.2. FE-AFE nature of phase transitions and physical behavior in $Pb(In_{1/2}Nb_{1/2})O_3$ and other $Pb(B'_{1/2}B''_{1/2})O_3$ compounds

On the one hand, consideration of PT peculiarities in lead indium niobate is impeded, since up to now very few investigations of this substance have been performed using pressure and high-intensity electric field. Moreover, the number of different types of phase diagrams to be examined while discussing the investigation results is also very limited.

On the other hand, such a consideration is somewhat promoted by the fact that, as established reliably and unambiguously, low-temperature state in lead indium-niobate (and in related substances such as $Pb(B'_{1/2}B''_{1/2})O_3$) is either ferroelectric or antiferroelectric. For lead indium niobate the type of dipole ordering in low-temperature phase depends on the presence or absence of spatial ordering in the arrangement of indium and niobium ions (for the said ions the existence of far composition ordering is defined by the conditions under which heat treatment of the substance has been realized).

Fig.8.14 presents the phase diagram of lead indium niobate taken from [135]. As is seen, the spatially-ordered (order parameter s ~ 1) and spatially-disordered arrangement (order parameter s ~ 0) of B-ions correspond to AFE and FE state, respectively. In the completely ordered crystals the transition into PE state is well-defined and not diffused.

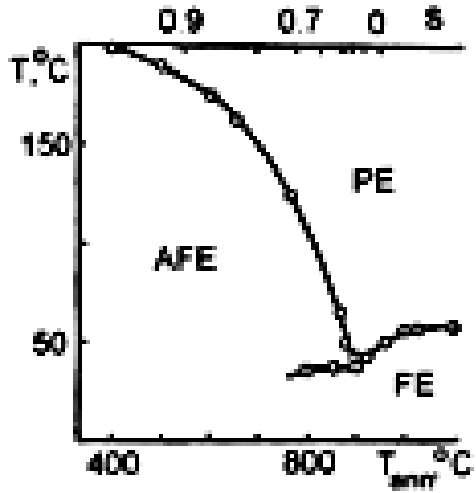

Fig.8.14. Phase diagram of lead indium-niobate [135].

There are practically no distinctions in the dependences ε(*T*) measured at different frequencies of AC electric field. In the crystals with the minimal degree of composition ordering paraelectric PT is diffused, the temperature dependences of permittivity manifests the relaxor character completely.

The above mentioned behavior of permittivity is usually interpreted on the base of different versions of the model of composition fluctuations (the said model is generally used while attempting to explain the nature of the relaxor ferroelectric state). However, if applied to the description of FE-PE or AFE-PE transformations in lead indium niobate, this approach has more drawbacks than advantages. Some experimental results of studies of PT may be explained in the frame of the theory of two-component ordered alloys of (B′B″) type (see [6], Sect. 6.5). However, in this case the displacements of ions provided with changes in the degree of compositional order must be realized at a distance not exceeding the lattice constant. Naturally, under such conditions the regions with predominating content of certain ions ($In^{3+}$ or $Nb^{5+}$) cannot be formed.

Another discrepancy of the model of composition fluctuations consists in the following. A strong diffuse character of the paraelectric PT and relaxor behavior are observed after high-temperature ($T > 1000°C$) annealing, whereas a well-defined sharp PT (without diffuse character) occurs after annealing at temperatures from 400 to 600°C. The annealing carried out at so low temperatures will hardly increase the degree of crystal homogeneity, in view of the fact that the said cannot be achieved in the process of high-temperature annealing.

In our opinion, a high degree of diffuseness of the PE PT in lead indium niobate is connected with a small difference in the free energies of the FE and AFE states which must inevitably lead to the coexistence of these phases even in the absence of composition fluctuations. Changes in the degree of composition ordering in the process of heat treatment result in the displacement of the energy equilibrium between FE and AFE states and the rise of the stability of one of the phases [6]. The latter factor leads to an essential change of the degree of PT diffuseness [20, 73]. The phase diagram of lead indium niobate (Fig.8.14) clearly shows that a high degree of diffuseness is peculiar only to the substances displaced by heat treatment towards the vicinity of the boundary which separates the regions of FE and AFE ordering.

Now let us experimentally verify our model of the diffuse PT for lead indium niobate and for the general case of $Pb(B'_{1/2}B''_{1/2})O_3$-type compounds. Unfortunately, up to now nobody has attributed the diffuseness of the transitions (and the nature of the so-called relaxor state) to the location of the substance in the diagram of phase states. Therefore, the required experimental data to be taken from the literature are very few and obtained for different substances. Though it is difficult to compare such results we shall try to perform this comparison using the scheme tested above on PZT- and PMN-based solid solutions.

For the $Pb(Lu_{1/2}Nb_{1/2})O_3 - PbTiO_3$ system of solid solutions (the phase diagram is presented in Fig.8.1) the first state of the substance located below the Curie point is an AFE with a high degree of composition ordering of Lu and Nb ions. When the content of lead titanate in the solid solution is increased the low-temperature AFE ordering is substituted by FE ordering. The boundary concentration region separating those of the FE and AFE states in the Y-$T$ phase diagram is located in the vicinity of 16-20 mol% of $PbTiO_3$. The results of X-ray and dielectric investigations of the samples of this solid solution system presented in [90] corroborate the model proposed in this section which relates the degree of diffuseness (and the so-called relaxor behavior) to the relative stability of the FE and AFE phases. In $Pb(Lu_{1/2}Nb_{1/2})O_3$ the PE PT is clearly non-defuse one. The increase of the lead titanate content leads to the diffuseness of the PT and relaxor behavior. In the solid solution with 10% $PbTiO_3$ such effects manifest themselves weakly, whereas at the 20% content they are maximally vivid. With further growth of lead titanate content PE the PT becomes non-diffuse again (Fig.6 and 8 in [90] ).According to the X-ray analysis data, the degree of composition ordering Lu and Nb ions constantly diminishes as the concentration of $PbTiO_3$ in the solid solution increases.

The results of studies of the $(1-x)Pb(In_{1/2}Nb_{1/2})O_3 - xPbTiO_3$ solid solutions are presented in [136]. Though the investigation of PT has not been a goal of this paper, its results show that the degree of diffuseness of the PE PT has its maximum alue for $x \approx 0.38$ (it is the boundary between the regions of the FE and AFE states in the phase diagram) and decreases with the increase of the content of lead titanate and when the solid solution passes to the region of the uniform FE state (far from $x \approx 0.38$) in the phase x-$T$ diagram (Fig.2 in [136]). For $x \rightarrow 0$ the degree of diffuseness of the PE phase transition is low also. Here we would like to emphasize that in the papers on relaxor ferroelectrics the introduction of an additional ion into the B-site of perovskite crystal lattice is considered to be a factor which promotes the increase of both the composition disorder and the degree of the relaxor behavior. However, this example shows that the so-called relaxor behavior is defined not by the mentioned factor, but by the location of the substance in the diagram of phase states and by the ratio of free energies of the FE and AFE states.

The boundary between the FE and AFE phases in x-T phase diagram of the $(1-x)Pb(Yb_{1/2}Nb_{1/2})O_3 - xPbTiO_3$ system is placed near $x = 0.2$ [88] (see Fig.8.1). The solid solution with $x = 0.2$ manifests relaxor properties (Fig.4 in [88]). Solid solutions which are located far from $x = 0.2$ in x-T diagram are ordinary ferroelectrics ($x \geq 0.5$) or antiferroelectrics ($x < 0.1$). Similar results were obtained in [137] (see Fig.6 in [137]) for the $Pb(Yb_{1/2}Nb_{1/2})O_3-PbZrO_3$ based solid solutions at the substitution of the complex $(Yb_{1/2}Nb_{1/2})$ by titanium ions, though the mentioned system was more complex: the solid solutions also contained 0.5 mol% of lead zirconate.

In the above example the introduction of titanium ions into $Pb(In_{1/2}Nb_{1/2})O_3$ shifts the solid solutions located in the x-T phase diagram towards the region of spontaneous FE phase (away from the FE-AFE phase boundary), thus resulting in suppression of the relaxor properties and promoting manifestation of the properties of ordinary FE. In [138] the properties of the $Pb(In_{1/2}Nb_{1/2})O_3 - Pb(Yb_{1/2}Nb_{1/2})O_3$ system of solid solutions are considered. In this case the increase of the content of the second component leads to the displacement of the solid solutions from the FE-AFE phase boundary towards the region of spontaneous AFE (note that Pb(Yb$_{1/2}$Nb$_{1/2}$)O$_3$ is in the AFE state at temperatures lower than ~ 302 °C). As seen from Fig.2, 3 and 4 in [138], the so-called relaxor properties are being suppressed in the process of moving away from the FE-AFE phase boundary in the diagram of phase states.

The behaviour typical for the so-called relaxors and the diffuse phase transitions near the FE-AFE phase boundary in the T-x diagram had been reported in [139] for the Pb(Fe$_{1/2}$Nb$_{1/2}$)O$_3$ - Pb(Yb$_{1/2}$Nb$_{1/2}$)O$_3$ system of solid solutions. The relaxor properties of this system of solid solutions are suppressed when position of the concrete solid solution in the phase diagram goes away from the said boundary.

It should be emphasized that in all considered examples the relaxor properties are suppressed despite the presence of an additional ion (Ti or Yb) in the B-sites of the crystal structure. In the approaches based on the model of relaxor ferroelectrics [86] it is assumed that the introduction of an additional ion into a crystal lattice site increases the composition disorder and strengthens the relaxor properties. As one can see, such interpretation contradicts the presented experimental results.

Thus, the considered data show that the so-called relaxor character of the behavior of Pb(B′$_{1/2}$ B″$_{1/2}$)O$_3$-type compounds is defined by the location of the given solid solutions in the diagram of phase states. The degree of manifestation of the said behavior depends on the position of the given compound with respect to the boundary which separates the regions of the FE and AFE states in the phase diagram. Therefore, the relative stability of the FE and AFE states of the substance in question may be considered a decisive factor for its relaxor behavior. If the stability of these states is the same (this signifies the coexistence of the FE and AFE domains in the sample's volume), the relaxor properties manifest themselves completely. Further in the present paper we will confirm this statement by experiments performed using hydrostatic pressure which effectively changes the relative stability of the FE and AFE states (in detail the change of the relative stability of FE and AFE states under external actions is discussed in [6]).

Now let us consider the few results obtained in experiments using hydrostatic pressure. In [138] the properties of the (1-x)Pb(In$_{1/2}$Nb$_{1/2}$)O$_3$ − xPb(Yb$_{1/2}$Nb$_{1/2}$)O$_3$ solid solutions are investigated. It is shown that at x ≈ 0.20-0.21 the FE relaxor state of lead indium niobate is superseded by the AFE state of lead ytterbium niobate. The dependences ε′(T) obtained at hydrostatic pressures up to 0.6 GPa and measuring field frequencies varying from 1 to 100 kHz for the solid solution of the said series with x = 0.16 are investigated in [140]. The solid solution with x = 0.16 is located in the FE part of the phase diagram (in the vicinity of the FE-AFE boundary). The dependences ε′(T) show pronounced "relaxor" character at the atmospheric pressure: $T'_m$ rises with the increase of the frequency and $\varepsilon'_m$ decreases (Fig.4 in [140]). At a pressure of 0.6 GPa both $T'_m$ and $\varepsilon'_m$ are independent of the field frequency. The so-called relaxor properties practically are not observed. In the frame of the developed model of the FE-AFE transformations applied to lead indium and lead ytterbium niobates this fact means that the pressure has increased the stability of the AFE state and "shifted" the initial position of the solid solution in the x-T diagram to the region of uniform AFE states. The share of unstable FE state domains has sharply decreased, and the oscillations of interphase domain boundaries no more contribute to the value of permittivity. The relaxation properties are lost.

Authors of [141, 142] reported that in disordered (s ~ 0.4) lead indium niobate (phase diagram of $Pb(In_{1/2}Nb_{1/2})O_3$ is in Fig.8.14) the pressure-induced structural phase transition from the FE relaxor state to the AFE one takes place at the pressure $P = 0.4$ Gpa (see. Fig. 8.15).

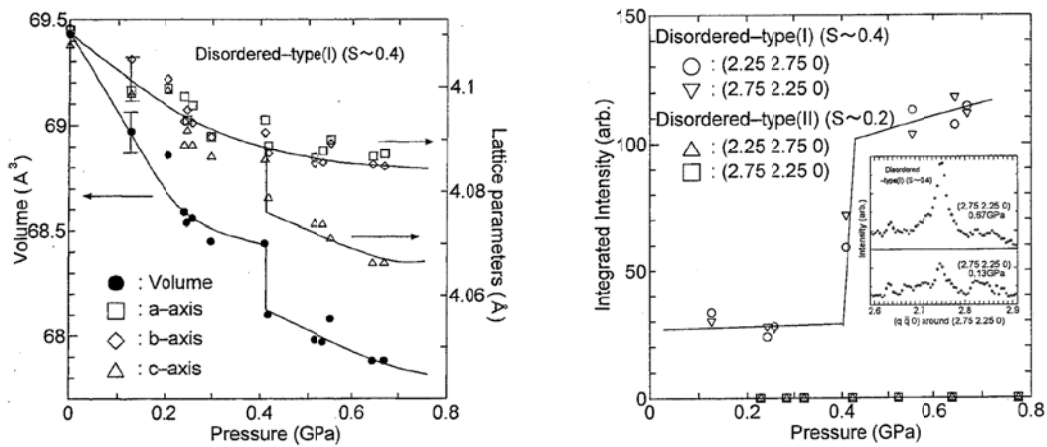

Fig.8.15. The pressure dependencies of the crystal sell parameters for $Pb(In_{1/2}Nb_{1/2})O_3$ disorder-type crystals at room temperature (on the left), and of the integrated intensities of (h/4 k/4 0) reflections, associated with the shift of lead cations in antiparallel (on the right) (Figures 3 and 5 of [141]).

Pressure induced changes of dielectrics properties for this crystal one can find in [142]. As one can see from Fig.8.16, relaxor properties are preserved up to pressures of 0.4 GPa until the crystal's position in the phase diagram is located in the vicinity of the energy equilibrium for the FE and AFE phases.

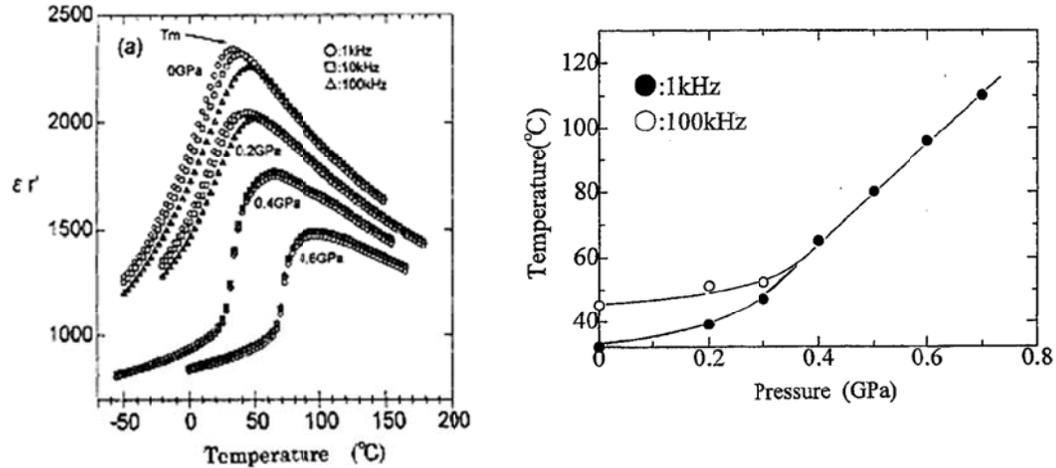

Fig.8.16. The temperature dependencies of dielectric constant at different frequencies and pressure for disordered crystal $Pb(In_{1/2}Nb_{1/2})O_3$ [142] (on the left). The pressure dependence of $T_m$ indicating $\varepsilon'_m$ for 1 and 100 kHz (on the right).

Concluding the present section of the paper let us consider the $(1-x)Pb(Fe_{1/2}Nb_{1/2})O_3 - xPb(Mg_{1/2}W_{1/2})O_3$ system of solid solutions. The solid solutions of this system that are rich in $Pb(Fe_{1/2}Nb_{1/2})O_3$ are ferroelectrics at low temperatures, on the other hand, they are antiferroelectric, when they are rich in $Pb(Mg_{1/2}W_{1/2})O_3$ [143, 144]. The boundary between the regions of the FE and AFE states on the x-T phase diagram is located near a concentration of 55% of $Pb(Mg_{1/2}W_{1/2})O_3$ (see Fig. 9.7). The maximum diffuseness of the paraelectric PT corresponds to this region of concentrations. In solid solutions removed from the said borderland in x-T diagram the PE transition is not diffuse.

To complete the discussion of the problem in question let us dwell on some properties of $Pb(Co_{1/2}W_{1/2})O_3$. At lowering the temperature, the AFE state is realized at first (at temperatures close to 298° C) and then the FE state is observed (at ~ 235°) in this substance (the AFE state is somewhat complicated by incommensurate modulation of the structure, however, this is insignificant for our analysis). The behavior of this compound is investigated under the hydrostatic pressure in [145-146]. The *P-T* phase diagram of $Pb(Co_{1/2}W_{1/2})O_3$ obtained in these papers is given in Fig.8.17. It is similar to that of 6/100-Y/Y PLZT series (Ch.3 and [3, 6]) (naturally, one must take into account the shift of the diagram along the axis of pressures when Y is changed in PLZT).

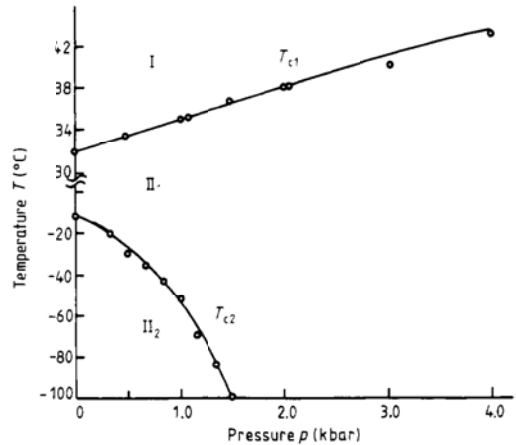

Fig.8.17. Pressure-temperature phase diagram of ceramic Pb(Co$_{1/2}$W$_{1/2}$)O$_3$ [145].

At the pressures of the order of 1.25 – 1.6 kbar the sequence of transitions and their dependence of the sample's history in Pb(Co$_{1/2}$W$_{1/2}$)O$_3$ are analogous to those in the PLZT solid solutions which are called relaxors - (6-9)/65/35. Fortunately, the *P-T* phase diagram for lead cobalt tungstate is available, and on its base the phase states can be identified at any variation of external parameters. Otherwise, this substance should have been given a strange unscientific name.

Considerable attention is paid to the spontaneous transition from the so-called relaxor state to the FE state during discussions of properties of the Pb(B′$_{1/2}$B″$_{1/2}$)O$_3$ type compounds in the literature. Such transition is considered in detail for the PLZT system of solid solutions [6, 20, 42]. In the compounds discussed in the present paper its nature is the same: a weak difference between the thermodynamic minima of the FE and AFE states.

All presented results corroborate the proposed model based on coexistence of domains of the FE and AFE phases. The authors have not found other systematic data concerning the behavior of the discussed substances.

# 9. COEXISTENCE OF FERROELECTRIC AND ANTIFERROELECTRIC STATES AND "DIPOLE-GLASS" BEHAVIOR

## 9.1. Model concepts

The concept of "dipole glasses" was introduced by analogy with the concept of "spin glasses" in physics of magnetic phenomena in the middle of seventies [147-149]. It started to be used in discussions of experimental results of investigations of the "relaxor ferroelectrics".

The substances which are usually attributed to "dipole glasses" possess the following main properties:

1) An essential dispersion of dielectric permittivity in the region of diffuse maximum of the dependences $\varepsilon'(T)$ and $\varepsilon''(T)$ and fulfilment of the Fogel-Fulcher law: $\omega = \left(\dfrac{1}{\tau_0}\right)\exp\left[-\dfrac{\Delta}{k(T_m - T_f)}\right]$, where $T_m$ is the temperature of $\varepsilon'(T)$ and $\varepsilon''(T)$ maximum ($T'_m$ or $T''_m$), $T_f$ is an effective temperature different for ($T'_m$ and $T''_m$);

2) An increase of the frequency of the AC measuring field causes a decrease of the maximum value of the real component of permittivity, whereas for the maximum value of the imaginary component increases, the temperature of $\varepsilon'(T)$ maximum being always higher than that of $\varepsilon''(T)$ maximum;

3) An increase of the measuring field amplitude $E_0$ leads to linear decrease of $T'_m$ and to increase of $\varepsilon'(T'_m)$; the tangent of the $T'_m(E_0)$ dependence decreases with increasing AC field frequency;

4) Hysteresis loops of these substances have a specific form: narrow dielectric hysteresis loops with a small residual polarization and narrow quadratic loops of electrooptic hysteresis;

5) The presence of effects which point to the existence of polar phase micro/nano domains at temperatures essentially higher than $T'_m$ and $T''_m$;

6) The dependence of properties of these substances on the sample's history;

7) The long-duration relaxation;

8) The high degree of diffuseness of the paraelectric phase transition.

The above-listed properties may be somewhat varied, or may manifest themselves not in the complete set. As a rule, the fulfilment of the first, sixth and seventh condition is dominating.

As was shown in the Ch.1-8 of the present paper (see papers [3, 42, 55, 58, 73 85] also), the above set of properties is typical for the substances in which FE and/or AFE ordering may be realised, and the energy difference for the said states is small (i.e. under the action of external factors such as temperature, field or pressure, FE-AFE phase transition may take place).

In this section our main attention will be focused on the process of the long-time relaxation of properties and physical characteristics of these substances after their state of thermodynamic equilibrium was disturbed by external influences. This long-time relaxation along with the pronounced frequency dependence of parameters (for example, dependence of dielectric or magnetic characteristics on the frequency of measuring field) is considered as a feature of the spin or dipole glass. The substances manifesting such properties are considered as "glasses" practically immediately

We will demonstrate that the above noted peculiarities of physical behavior are manifested in the substances in which FE and AFE phases are realized and the difference between their free energies is small. After that we will discuss the possible phase diagrams of substances that can be misguidedly attributed as "glasses". By no means have we wanted to cast doubt on the existence of "dipole glasses" in the nature. Our main purpose is only to call attention to the fact that one has to be really careful during the interpretation of experimental results in substances in which the inhomogeneous states with negative energies of interphase boundaries can take place. The peculiarities of behavior of such system can be really misleading during interpretation. More over the existence of such states (stable in the absence of external influences) was considered impossible in a wide class of substances until recently.

In Ch.5, 6 we have shown that the formation of the heterophase structure of coexisting FE and AFE domains is accompanied by the emergence of the chemical inhomogeneity of the substance. Here we are to strictly emphasize that it is just phase inhomogeneity (coexisting of FE and AFE phases) that leads to chemical segregations, but not on the contrary.

The process of chemical segregation in the vicinity of the FE-AFE phase boundary can take place only at those temperatures when the dipole ordered FE and AFE states exist and coexist. At high temperatures (when the uniform PE state is present in the substance) driving forces of chemical segregation are absent, and the process of the substance annealing leads to the chemical homogeneity. Thus, the considered process of chemical segregation in the vicinity of the FE-AFE phase boundary is reversible in the course of temperature cycling during cooling and heating.

If the IDW is displaced under the action of electric field (or appears after cooling from PE state), then the process of chemical segregation will occur in the vicinity the position of this new IDW. "Old" chemical segregations will be cleaned out. However, it is not only IDW itself that controls the process of chemical segregation in the vicinity of its location. There exists an influence opposite to the chemical segregation on the IDW during its motion (see Eq.7.1). The said segregation becomes a mobile defect of the crystal lattice. The interaction between the segregation and the IDW should be the most pronounced when the formation of such segregation is accompanied with a violation of local electroneutrality.

The mentioned interrelation between the IDW and the formation of the mobile defects of the crystal lattice lead to peculiarities of the IDW dynamics. At the temperatures, at which FE and AFE states are realized, the rate of ionic diffusion is low. In Ch.7 we have discussed the relaxation dynamic of IDW and obtained the Eq.7.8 describing the motion of IDW. This equation contains the time-function $G(t) = 1 - \exp(-t/\tau)$. In Ch.7 we adopted condition $G = 1$ (for $t \to \infty$). In view of the fact that the above mentioned process of chemical segregation is long-duration one, the value of $G = 1$ in (7.8) is to be substituted by the time function $G(t)$, since the condition $t \to \infty$ is not fulfilled in real experiment. In this case more complicated expressions for $\varepsilon'$ and $\varepsilon''$ are obtained instead (7.12) [3]:

$$\varepsilon'(\omega, T) = \varepsilon_0 \frac{1 + \omega^2 \tau^2 [1 + \eta G(t)]}{1 + \omega^2 \tau^2 [1 + \eta G(t)]^2}, \qquad (9.1a)$$

$$\varepsilon''(\omega, T) = \varepsilon_0 \frac{\omega \tau \eta G(t)}{1 + \omega^2 \tau^2 [1 + \eta G(t)]^2}. \qquad (9.1b)$$

Under the conditions of the experiment (when the period of oscillations of the measuring field is much shorter than characteristic time of diffusion processes) $\omega \tau \to \infty$ we have:

$$\frac{\varepsilon'}{\varepsilon_0} = \frac{1}{1 + \eta G(t)}. \qquad (9.2)$$

As one can see from this expression, the dielectric constant reaches its equilibrium value during a long period of time after the action of any factor leading to the shift the IDWs (or after their appearance during cooling from high temperatures) because for the temperatures $T < T_c$ coefficients of diffusion are small. For a particular case of $\eta = 1$ Fig.9.1 shows the time dependence of $(\varepsilon' / \varepsilon_0)$ for some values of $\tau$.

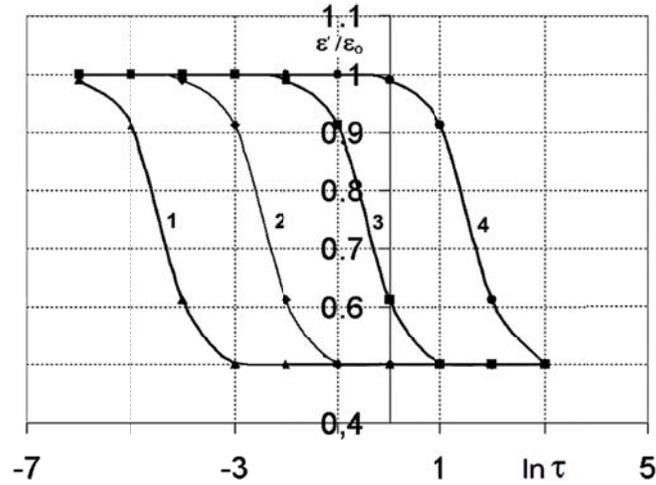

Fig.9.1. Time dependence of the function $\varepsilon' / \varepsilon_0 = \left[1 + (1 - \exp(-t/\tau))\right]^{-1}$, for different values of $\tau$: 1 – 0.0001; 2 – 0.01; 3 – 1.0; 4 – 100.0.

Let us remind (see eq. 7.13) that the relaxation time $\tau$ is determined by the activation energy $\Delta$ for the processes under our consideration. The extended in time process of change of the chemical composition near the "bare" IDW takes place during the formation of segregates. It is naturally that the change of the activation energy ($\Delta = \Delta(t)$) takes place along with the process of segregates formation. Therefore, the characteristic relaxation time also depends on time ($\tau = \tau(t)$). A small number of experiments on the establishment of dependency $\tau(t)$ has been done up to now. We will use the results of investigation of long-time relaxation given in the Ch. 6 (see Fig. 5.4 and 5.5). The process of formation of segregates along the IDWs takes place during more than 30 hours.

Time dependencies of the intensities of diffuse X-ray lines for the 15/77/23 PLLZT and the 6/73/27 PLZT solid solutions for this time interval are presented in Fig. 9.2 in logarithmic scale along both axes.

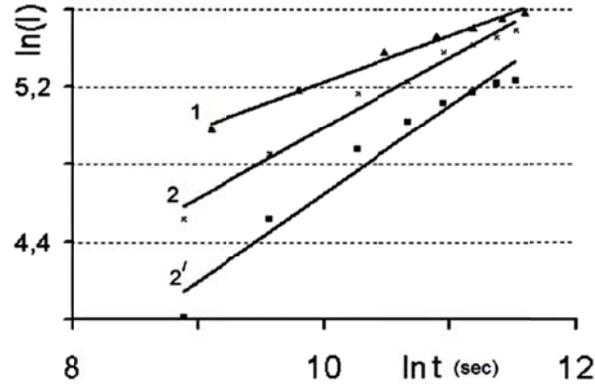

Fig.9.2. Time dependences of integral intensities of diffusive X-ray lines for 15/77/23 PLLZT (1) and 6/73/27 PLZT (2 and 2′) solid solutions.

As one can see in this figure the linear dependence is observed with the high level of accuracy. It clearly demonstrates that then time dependence of the relaxation time has a power character $\tau(t) = At^n$ with $n < 1$. Hence the time dependence of the dielectric permittivity in the process of aging (in the course of formation of segregates in the process of their growth) is given by the formula

$$\frac{\varepsilon'}{\varepsilon_0} = \frac{1}{1+\eta\left[1-\exp(-t^{1-n}/A)\right]} \quad (n<1). \tag{9.3.}$$

The dependencies described by the formula (9.3) are presented in Fig. 9.3.

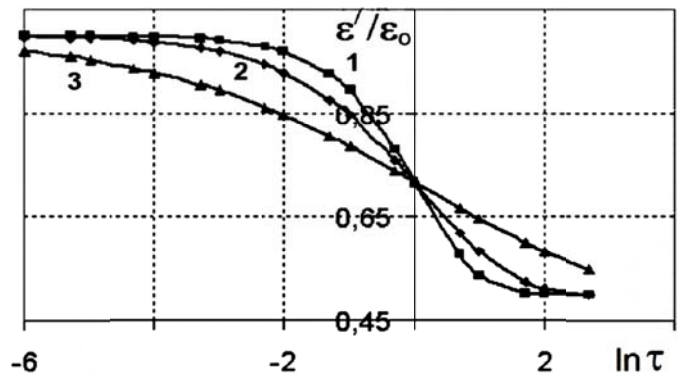

Fig.9.3. Time dependence of function $\varepsilon'/\varepsilon_0 = \left[1+\left(1-\exp(-t^{(1-n)}/\tau)\right)\right]^{-1}$ for different $n$: 1 - 0.4, 2 – 0.6, 3 – 0.8.

Relaxation curves typical for spin glass systems are presented in Fig. 9.4a and 9.4b. These curves were obtained in the aging experiments measured after the switching of the external field and after heating up (the data is taken from the review [150]).

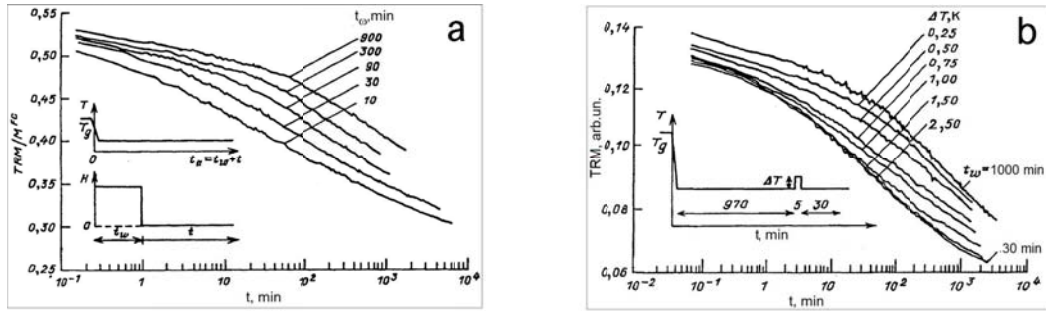

Fig.9.4. (a) – Magnetic relaxation after the magnetic field was switched off [150];
(b) – Magnetic relaxation in aging experiments with heating cycle [150].

As one can see the dependence (9.3) completely describes the time behavior of the systems which are referred to as glasses in the process of the aging. It has to be stressed that the local decomposition of the solid solution caused by local mechanical stresses in the vicinity of IDWs was the physical factor that determined the shape of the dependency (9.3).

## 9.2. Experimental results.

All results of experimental studies presented in this section were obtained for substances phase diagram of which contain regions of coexisting domains of FE and AFE phases. These phase diagrams are given in previous sections of this paper and will be used without references in what follows unless otherwise specified.

The aging of PLZT ceramic samples with composition 9.5/65/35 after quenching from high temperatures was investigated in [151]. We selected this study as an example due to the following reason. It was unambiguously shown that in the samples analogous in composition and location in the phase diagram (see Ch. 5 of this paper) the local decomposition of the solid solution and the formation of segregates in the vicinity of the IDWs separating domains of the FE and AFE phases takes place after quenching from high temperatures. As one can see in Fig. 9.5 the real aging process may well be described by means of the formula (9.3). More over we have to ascertain that the expression (9.3) can describe even more protracted relaxation processes.

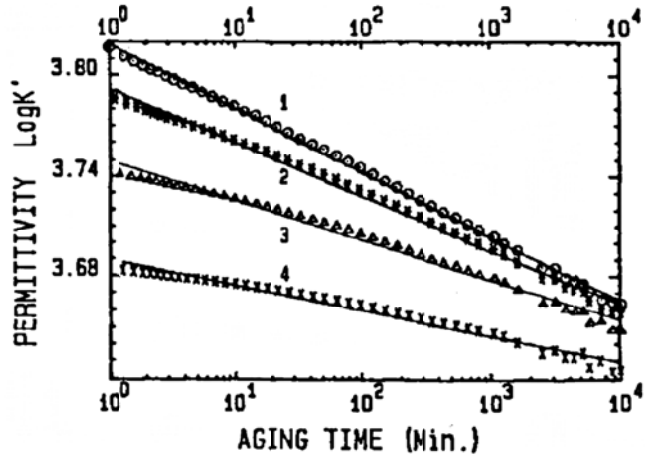

Fig.9.5. The permittivity as function of aging time for quenched (400°C) 9.5/65/35 PLZT samples [151].

Phenomena of the so-called thermal dielectric, field, or mechanical memory that were experimentally observed in "relaxor ferroelectrics" of the PLZT or PMN type are even more interesting from the physical point of view. All above mentioned effects have common physical basis, namely, the local decomposition in the vicinity of IDWs. Therefore, we will discuss in detail only the phenomenon of thermal dielectric memory. This effect implies that when the samples are cooled to some particular temperature $T = T_{age}$ after annealing at high temperatures and are subjected to aging (during the time interval of 20 hours and longer) at this temperature a specific characteristic feature is observed on the ε(T) dependencies at following cycling. This feature consist in a "drop-shaped" behavior of the dependence ε(T) in the vicinity of the aging temperature $T_{age}$. Typical results of experiments are presented in Fig. 9.6 [151, 152].

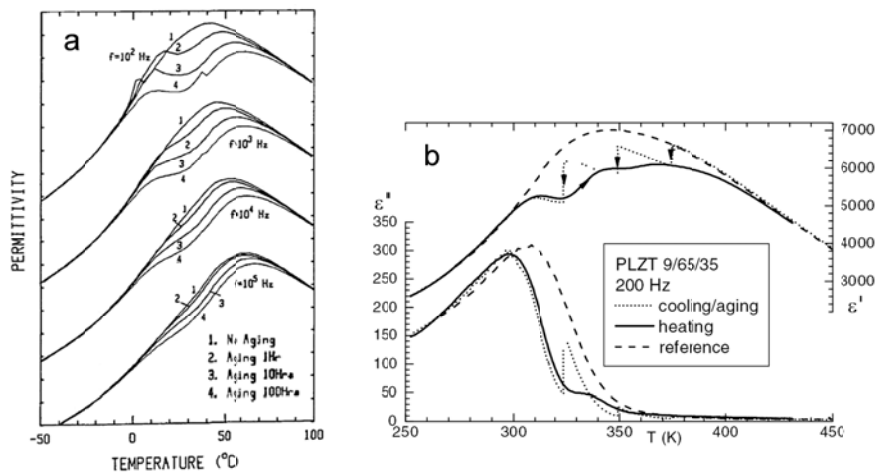

Fig.9.6. (a) – Temperature dependences of dielectric constant for different aging time at ~ 23 °C for 9.5/65/35 PLZT samples annealed at 400 °C [151]; (b) – Dielectric permittivity of 9/65/35 PLZT samples at 200 Hz during multiple aging stages (24 h each) and subsequent heating curves with memory, compared with the reference curve measured on continuous cooling [152]

All presented results can have simple unambiguous explanation if one takes into account the inhomogeneous structure of coexisting domains of the FE and AFE phases present in the sample. Domain structure is formed in the bulk of the sample after the high-temperature annealing. The shares of each phase (and consequently the sizes of domains) are determined by the relation between free energies of these phases. Diffusive local decomposition of the solid solution constantly takes place in the vicinity of the interdomain boundaries. However, this is the long time process and it is practically not manifested in material parameters. This influence is extremely small due to the fact that the characteristic time of measurement (during the temperautre measurements), which is of the order of several degrees Celcius per minute, is much shorter than the characteristic relaxation times of diffusion processes (which are of the order of tens of hours). The same is true for the external field or mechanical stress measurements. During aging at the temperature $T_{age}$, which lasts for long time (tens of hours), the local decomposition of the solid solution takes place and the structure of the formed segregates repeats the structure of the interphase boundaries. The longer is the the aging time the more pronounced is the spatial structure (the distinctive network) of segregates and the larger vloume of the sample is occupied by segregates. The dielectric permittivity is reduced as a result (Fig.9.6). The formed spatial structure of segregates is conserved during the temperature cycling (with the several degrees per minute rate of the temperature change) following the aging.

The equilibrium relation between the shares of each phase and the sizes of domains of these phases are constantly changing in the process of temperature cycling. The sizes and the structure of domains coincide with the preserved spatial structure of segregates when the temperature $T_{age}$ is achieved. At this moment the effective pining of interphase boundaries takes place and the contribution of the oscillations of interdomain boundaries which determine the main contribution to the dielectric permittivity decreases (Ch. 7). This pining of the interdomain boundaries leads to the appearance of the "drop-shaped" feature on the dependence ε(*T*) (see Fig 9.6).

The nature of the effect of dielectric memory at the temperatures both above $T_m$ and below $T_m$ is the same as it clearly seen from the results of [152] (see also [156]). However, the authors of numerous papers who were trying to connect this phenomenon with relaxor ferroelectrics were forced to come up with different mechanisms for these two cases. One does not need to do that using the model suggested here. As shown in Ch. 5, 6 the structure of coexisting domains of the FE and AFE phases exists both above $T_m$ and below $T_m$ (let us remind that the X-ray studies carried out above $T_m$ demonstrated the coexisting domains of the FE and AFE phases constitute the two-phase FE+AFE domain. Thus, the IDW structure that determines the effect of the dielectric memory exists both below $T_m$ and above $T_m$.

The shares of the AFE and FE phases and, thus, the spatial structure of IDWs can be also changed by means of other thermodynamic parameters such as external field and mechanical stress. The long time aging of the samples at some nonzero values of the above-mentioned parameters will lead to formation of the new spatial structure of segregates and the effects of the filed or mechanical dielectric memory will be manifested during the cyclic changes of these parameters after the aging [153-155]. These effects are presented in Fig.9.7 and Fig. 9.8.

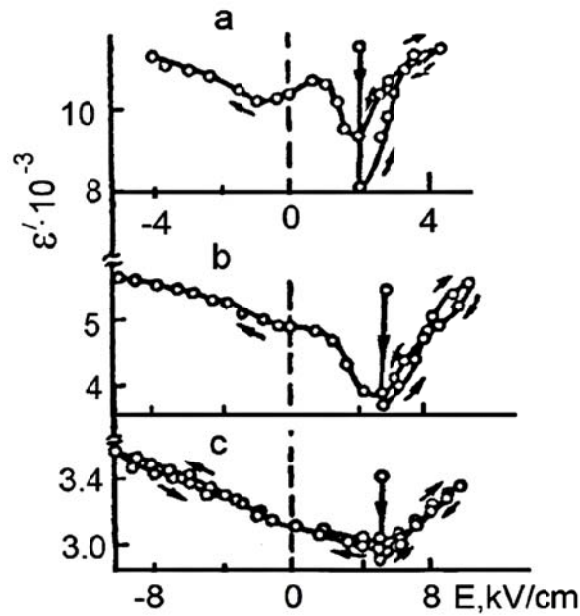

Fig.9.7. Dependences of dielectric constant on DC electric field in the PLZT solid solutions after aging during 20 h [153]. The solid solutions compositions: a – 8/65/35, $T_{age}$ = 57 °C; b – 11/65/35, $T_{age}$ = 22 °C; c – 13.5/65/35, $T_{age}$ = 22 °C.

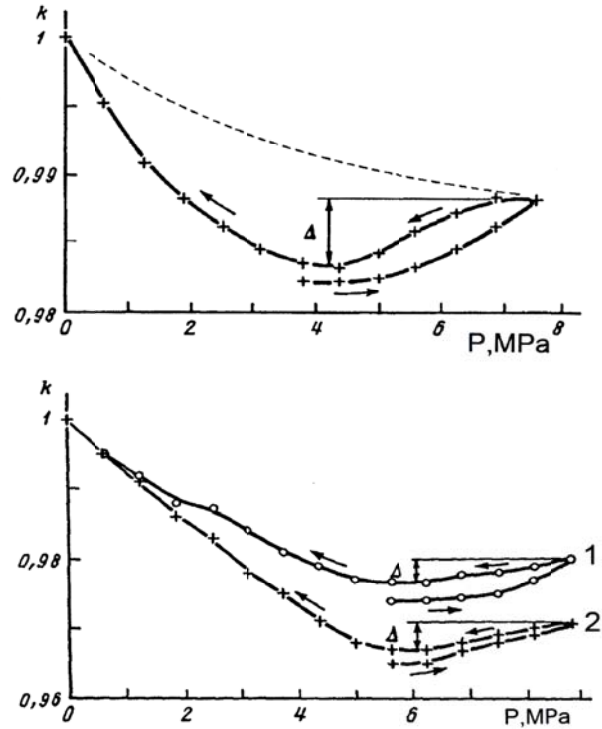

Fig.9.8. Dependencies of the normalized dielectric constant on uniaxial stress after aging during 20 h for 8/65/35 PLZT [154] (on the top). Dashed line shows the dependence, obtained without aging (added by authors of the present paper). Dependencies of the normalized dielectric constant on uniaxial stress after aging during 48 h (1) and 120 h (2) in DC electric field $E = 1850$ $V/cm$, ($T_{age} = 27\ ^{\circ}C$, $P_{age} = 3.8$ MPa) for 8/65/35 PLZT (on the bottom).

The effects discussed above are as a rule attributed to the so-called dipole-glass-like behavior in relaxor systems. However, based on this dipole-glass approach it is impossible to describe and explain all manifestation of other phenomena (see previous chapters). We demonstrated that rather simple explanation can be given if one takes into account coexistence of the AFE and FE phases.

We have considered above the effect of dielectric memory using the PLZT family of solid solutions as example. Similar effect is also characteristic of PMN based solid solutions. The result obtained for single crystals with composition $(1-x)Pb[(Mg_{1/3}Nb_{2/3})O_3 - xPbTiO_3$ with $x = 0.10$ and $0.12$ one can find in [156] (phase diagram in presented in Fig. 8.3. They are analogous to the result for PLZT given above. Some examples for different aging temperatures (both above and below $T_m$) are shown in Fig.9.9. The physical mechanism of the effect is the same as in PLZT.

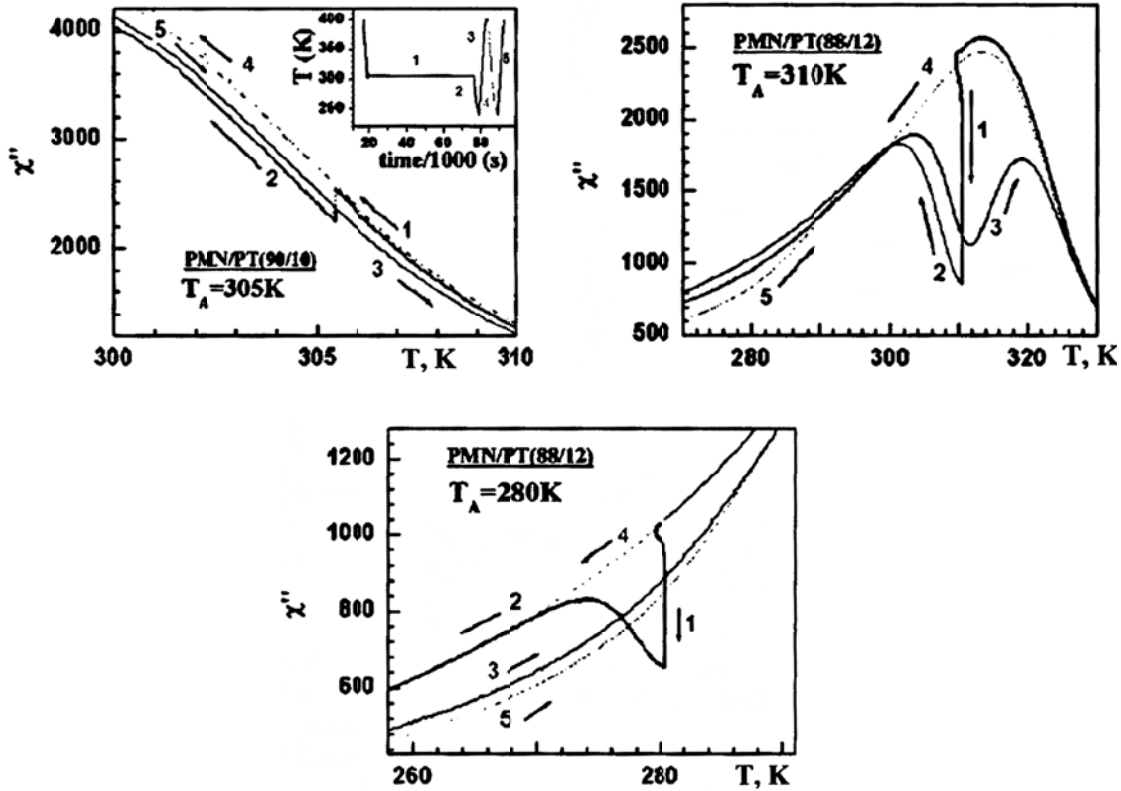

Fig.9.9. Dielectric memory effect in $(1-x)Pb[(Mg_{1/3}Nb_{2/3})O_3 - xPbTiO_3$ solid solutions [156]. In a typical aging memory experiment (see inset for temperature history profile) the sample is cooled to $T_A$ (curve 1). After aging at $T_A$, the sample is cooled to a lower excursion temperature $T_{EX}$ (curve 2) and then immediately reheated past $T_A$ (curve 3). For comparison, reference starting curves (4 and 5) are also taken at the same sweep rates without stopping at $T_A$.

## 9.3. Peculiarities of x-T diagrams and order-disorder systems

We have demonstrated that taking into account the coexistence of the FE and AFE phases one can give consistent explanation of all effects which were attempted to attribute to the dipole-glass state in the so-called relaxor ferroelectrics. In what follows we will try to show how this approach can be slightly extended.

Let us consider experimental "composition – temperature" phase diagrams of the solid solutions in which one of the components is a ferroelectric and the other is an antiferroelectric. The phase diagrams of the solid solutions for which FE and AFE nature of the low-temperature state is unambiguously identified are shown in Fig. 9.10 [88, 90, 143, 144].

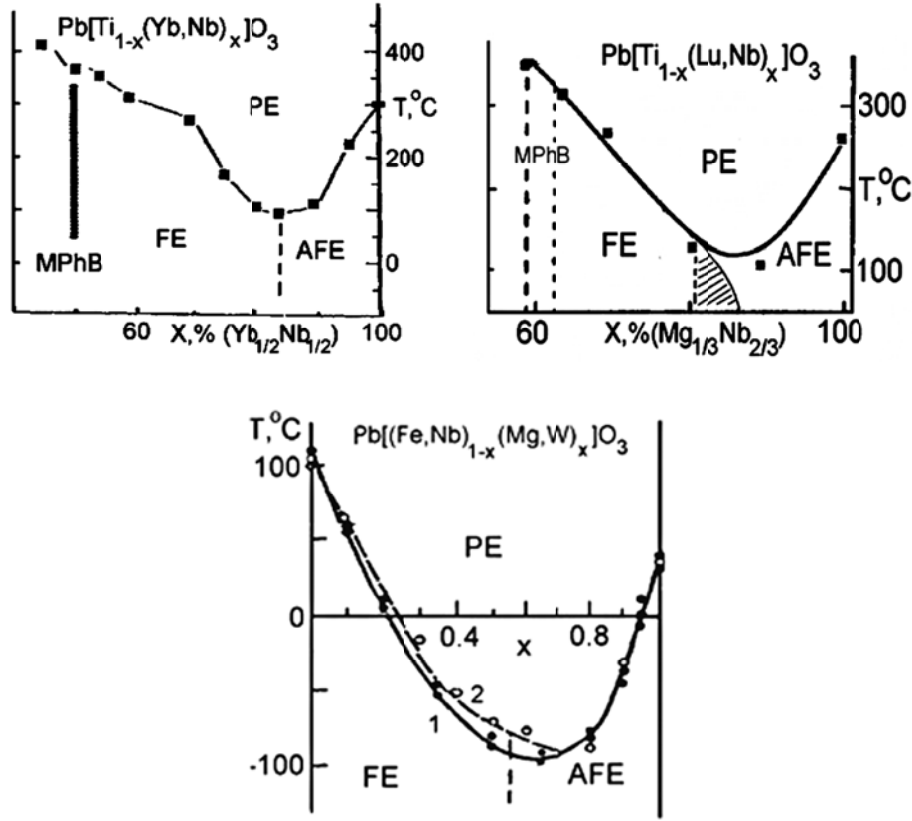

Fig.9.10. Phase diagrams of solid solutions with perovskite structure obtained by the substitution of titanium by the complex: $(Yb_{1/2}Nb_{1/2})$ [88], $(Lu_{1/2}Nb_{1/2})$ [90], $Pb(Mg1/2W1/2)O3$ [143, 144].

The presence of a sharp drop on the dependences $T_c(x)$ observed for the solid solutions located near the boundary which separates the regions with FE and AFE ordering is a common feature of all the presented phase diagrams. Such drop in the value of $T_c$ may reach 200°C and more. It should be also noted that for all the solid solutions which phase diagrams are shown in Fig.9.10 the diffuseness of PE PT noticeably increases as the solid solution composition approaches the FE-AFE phase boundary in the diagram. As shown in Ch.6 [3, 13, 73], such behavior is caused by the existence of two-phase (FE+AFE) domains in the PE matrix of the substance at $T > T_c$ and the presence of ion segregations in the vicinity of the interdomain boundaries.

In all above-considered cases of oxides with perovskite structure (Fig.9.10) the solid solution components had rather high proper values of the Curie point. Due to this the minima on the dependence $T_c(x)$ do not reach the zero of the Kelvin temperature scale. The phase diagrams of the solid solution with components having low values of the Curie temperatures are schematically shown in Fig.9.11 (at the top). The dependence of the segregation temperature on composition $T_{seg}(x)$ is denoted by dotted line. $T_{seg}$ is the temperature at which for appearance of two-phase FE+AFE domains in PE matrix of the solid solution start to appear. At the same time at this temperature the segregates start to emerge when temperature of the sample is decreasing from high temperatures. The bell-like dependence of the diffuseness parameter $\delta(x)$ (see Ch.6) is typical for all discussed solid solutions and promotes such a behavior of $T_{seg}(x)$ dependence.

To conclude the consideration of the diagram shown at the top of Fig.9.11, we would like to discuss possible nature of the line $T_{quant}(x)$ in the phase diagram. In our opinion, this feature is clear. To begin with, consider the solid solutions of (1-x)BaTiO$_3$ – xSrTiO$_3$ system (a lot of attention is has been paid lately to SrTiO$_3$-based solid solutions as an example of quantum ferroelectrics). The Curie point of these solutions lowers with the increase of content of the strontium titanate.

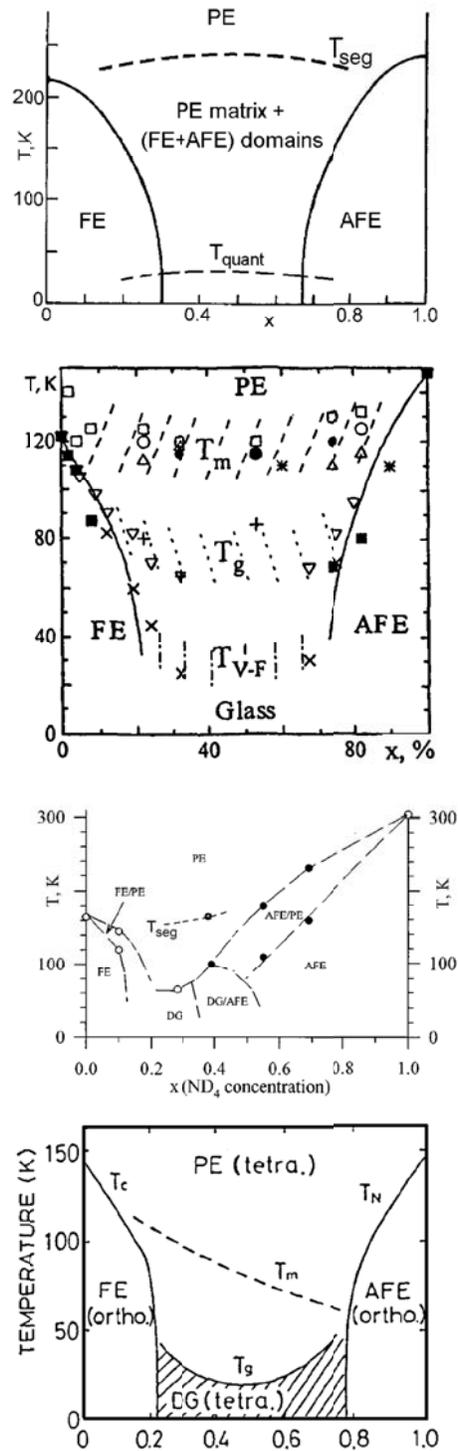

Fig.9.11. Phase diagrams of solid solutions with order-disorder phase transitions. From top to bottom: model diagram; $K_{1-x}(NH_4)_xH_2PO_4$ [157]; $Rb_{1-x}(NH_4)_xD_2AsO_4$ [160] (the line $T_{seg}$ is added by authors of the present paper based on [158]); $Rb_{1-x}(NH_4)_xH_2PO_4$ [162].

However, in these solid solutions the PE phase transitions typical for ordinary ferroelectrics do not take place at changing the temperature. When the dependence $T_c(x)$ reaches the region of about (20-30) *K* further decrease of the Curie point is not observed, and the character of the dependence $\varepsilon(E)$ changes. Such behavior is explained by the quantum effects contribution which determines the nature of the line $T_{quant}(x)$ in the phase diagram presented at the top of the Fig.9.11.

The majority of studies devoted to investigations of the dipole-glass state are performed on series of solid solutions $K_{1-x}(NH_4)_xH_2PO_4$, $Rb_{1-x}(NH_4)_xH_2PO_4$, $Rb_{1-x}(ND_4)_xD_2AsO_4$ that fall into the class order-disorder ferroelectrics. Let us remind that the first component of these solid solutions is a ferroelectric and the second is an antiferroelectric. Experimental phase diagrams of $K_{1-x}(NH_4)_xH_2PO_4$, $Rb_{1-x}(NH_4)_xH_2PO_4$ and $Rb_{1-x}(ND_4)_xD_2AsO_4$ solid solutions are presented in Fig.9.11. These diagrams were built on the base of the results of measurements carried out by different methods and presented in [157, 160, 162]. One can clearly see that these experimental diagrams coincide practically completely with the model one in the top of the Fig.9.11. However, the latter is typical for those substances in which the phases with FE and AFE ordering are realized and domains of these phases coexist, while the experimental diagrams belong to the substances classified nowadays as "dipole glasses".

For this reason, it seems expedient to analyze some experimental results obtained during investigations of phase transitions in the $K_{1-x}(NH_4)_xH_2PO_4$ and $Rb_{1-x}(NH_4)_xH_2PO_4$ systems of solid solution (and other related substances). We are going to consider the results that could not find their explanation in the scope of traditional approach but can be explained from the viewpoint of the effects discussed in the present research (and earlier in the papers [3, 6, 20, 42, 73]). We would like to emphasize that in most cases such experimental results are consistent with the recently developed ideas about FE-AFE phase transformations with taking into account the interactions between coexisting phases (interaction between the domains of these phases). Moreover, there exist some results which interpretation from another viewpoint seems to be artificial.

In this respect, we would like to note the experimental results of studies of the diffuseness of the PE transition in $K_{1-x}(NH_4)_xH_2PO_4$ ($0 \leq x \leq 0.24$) solid solutions as an example [163]. These results are presented in Fig.9.12.

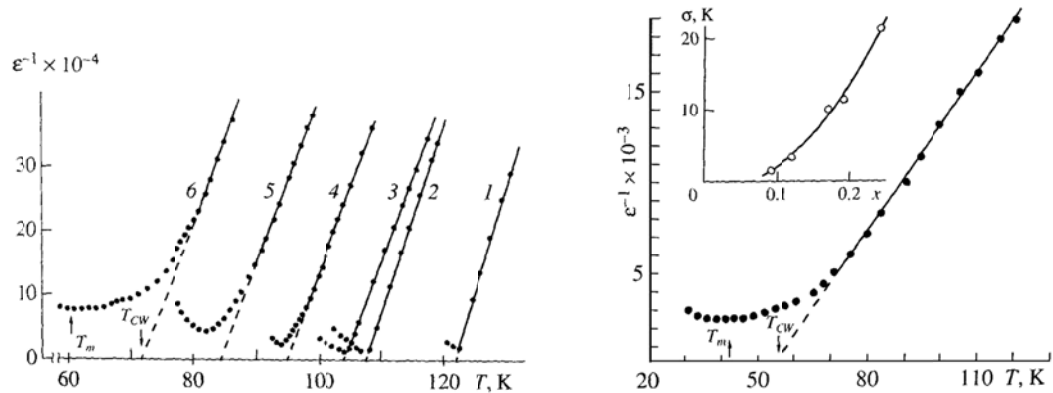

Fig.9.12. On the left: Dependencies $\varepsilon^{-1}(T)$ for the crystals of the $K_{1-x}(NH_4)_xH_2PO_4$ series. x, 1 – 0; 2 – 0.04; 3 – 0.05; 4 – 0.09; 5 – 0.12; 6 – 0.19 [163]. On the right: Dependencies $\varepsilon^{-1}(T)$ for the $K_{0.76}(NH_4)_{0.24}H_2PO_4$ crystal. Dependence of the diffuseness parameter on the composition is given in the insert [163].

The dependencies $\varepsilon^{-1}(T)$ for solid solutions with $x$ in the interval from 0 to 0.24 are presented here (let us remind that the phase boundary between the "dipole-glass" state and AFE state (Fig.9.11) is at x = 0.20). As one can see the diffuseness parameter σ($x$) increases with increase of $x$ and does not have any peculiarities at x = 0.20. It is an interesting detail that the temperatures at which the deviation from the Curie-Weiss law starts ($T_{seg}(x)$ in our notations) are very close to the line $T_m(x)$ in the diagram for $K_{1-x}(NH_4)_xH_2PO_4$ in Fig. 9.11. Both these experimantal results are in complete agreement with the concept about coexisting domains of the AFE and FE phses that were presented above and discussed in the course of this article. We would like to call ones attention to the fact that the character of dependencies ε($T$) does not change in any way when the composition of the solid solution crosses the phase boundary x = 0.20 in the phase diagram of $K_{1-x}(NH_4)_xH_2PO_4$ solid solutions. It is apparent that the mechanism of diffuseness of the phase transition for the solid solutions with x < 0.20 and x > 0.20 is the same.

The results of studies of diffuseness of the phase transition in the presence of DC electric field applied to samples are more interesting. The degree of the diffuseness parameter reduces as the field intensity increases (Fig.9.13) for $K_{0.76}(NH_4)_{0.24}PO_4$ solid solutions [163].

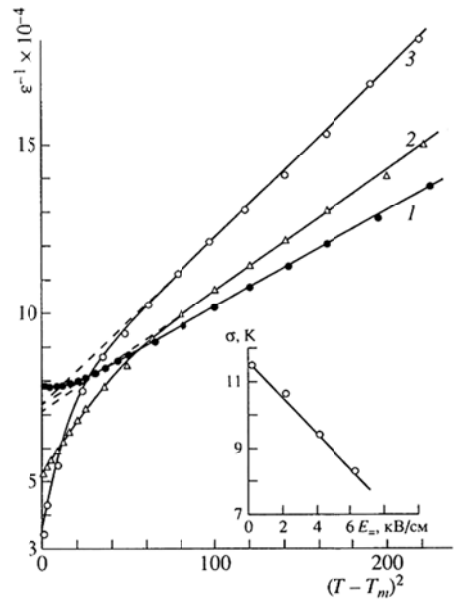

Fig.9.13. Temperature dependencies of $\varepsilon^{-1}$ for the solid solution $K_{1-x}(NH_4)_xH_2PO_4$ with x = 0.19 in external electric field *E*. *E* (*kV/mm*): 1 – 0; 2 – 4; 3 – 6. Dependence of the diffuseness parameter on the electric field *E* is given in the insert [163].

An electric field leads to increase of the degree of diffuseness of the ordinary FE-PE phase transformations. However, in those substances for which FE and AFE orderings have a small energy difference and the domains of the FE and AFE phases coexist in the bulk of the samples, the described behavior is normal. The stability of the FE state increases as against that of the AFE state when the field intensity increases. This is equivalent to shift of the position of the state of the sample out of the FE-AFE-PE triple point in the diagram of phase states (from the point 2 to the point 1 in Fig.6.2). As seen from the upper part of this figure, the parameter of PT diffuseness decreases. This fact was revealed in experiments on $K_{0.76}(NH_4)_{0.24}PO_4$.

The described above experiments can be unambiguously explained if one takes into account the coexistence of the domains of the AFE and FE phases in those solid solutions in which one of the components is a ferroelectric and the other is an antiferroelectrics. Both the increase of the electric field intensity and decrease of *x* lead to the rise of stability of the FE phase relatively to the AFE phase, which lead to the reduction of the share of the AFE phase in the sample's volume and as a consequence to the reduction of the diffuseness of the phase transition (see Ch. 6 and [6, 20, 73] for details).

Presented results point to the essential influence of the coexistence of phases on the kinetics of the phase transition in order-disorder type substances in which one component is a ferroelectric and the other one is an antiferroelectric. The most complete presentation of results of investigation of phase coexistence in the systems under discussion is presented in [158-161, 164] with the $Rb_{1-x}(ND_4)_xD_2AsO_4$ solid solutions as an example (phase diagram is presented in Fig.9.11). Before further discussions of this topic we believe that it necessary to point out the resemblance of the latter phase diagram and diagrams depicted in the Fig. 9.10. The authors of above mentioned publications have clearly demonstrated the existence of broad regions in the vicinity of the FE-PE and the AFE-PE phase transition lines. These regions are characterized by the coexistence of domains of ordered phases in the paraelectric matrix of the samples. Although, in our opinion this is not the main circumstance

The temperature dependencies of the LA[100] Brillouin backscattering phonon spectra [158] and the Raman vibration modes [159] have been studied in the mixed FE-AFE system of the $Rb_{1-x}(ND_4)_xD_2AsO_4$ solid solutions with compositions corresponding to the FE side of the phase diagram. The same investigations for this system of solid solutions with compositions corresponding to the AFE side of diagram have been carried out in [160]. Authors attributed the anomalies of physical properties (frequency shift of the line of Brillouin phonon spectra and Half-width at half maximum) of the solid solutions with x = 0.10 near the temperature 160 K to the onset of short-range AFE order caused by the freezing-in of $(ND)_4$ reorientations and imply a growth of local structure competition (between FE and AFE ordering). With respect to the opinion of the authors of [158, 159], one can expect that such FE-AFE ordering competition, which can suppress a long-range-order FE transition, is responsible for both phase coexistence and presence of the broad damping peak in the Brillouin backscattering spectrum centered at $T \sim$ 146 K (Fig.9.14 in the left). The rapid growth of FE ordering near 130 K is responsible for the Landau-Khalatnikov-like maximum.

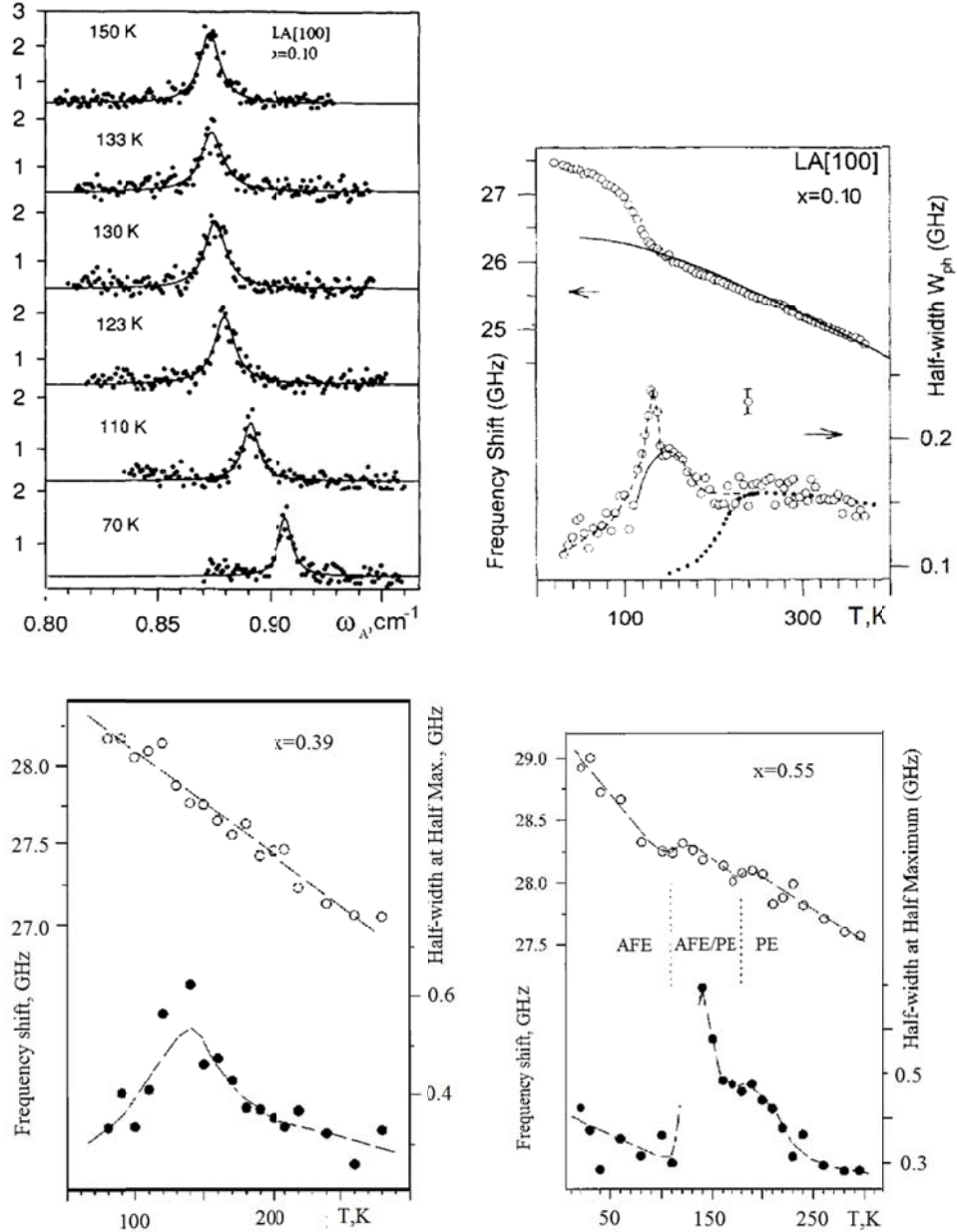

Fig.9.14. Anti-Stockes component of LA [100] Brillouin frequency shift for temperatures around the maximum value of half-width (for x = 0.10 as example) in $Rb_{1-x}(ND_4)_xD_2AsO_4$ mixed crystals [158], and Brillouin shift (open circle) and half width (solid circle) vs. temperature of the LA[100] phonons for (a) x = 0.10 [158], (b) $x$ = 0.55, and (c) $x$ = 0.39 [160].

Analogous results were also obtained for compositions from the AFE region of the phase diagram for the $Rb_{1-x}(ND_4)_xD_2AsO_4$ solid solutions. Temperature dependencies of the same parameters for the solid solution with x = 0.55 are given in the Fig.9.14 on the right for illustration purpose [160]. As one can see that in both cases it may be safely suggested that the inhomogeneous states of the complex FE+AFE domains exist in the paraelectric matrix of the $Rb_{1-x}(ND_4)_xD_2AsO_4$ system of solid solutions with compositions both from FE part and AFE part of the phase diagram. These inhomogeneous states exist in wide intervals of thermodynamic variables (such as temperature and composition). It looks like nobody took into account the possibility of these states until the present time. The behavior of the diffuseness parameters of the phase transitions discussed above for $K_{1-x}(NH_4)_xH_2PO_4$ solid solutions has the same nature.

Let us now discuss the situation that takes place in the $Rb_{1-x}(ND_4)_xD_2AsO_4$ series of solid solutions with compositions that are characterized by approximately equal stability of the FE and AFE states. Temperature dependencies of parameters of the LA[100] Brillouin backscattering for composition with x = 0.39 are presented in the lower left part of Fig.9.14. The temperature dependence of the half width of the anti-Stokes line of the Brillouin phonon spectra manifests a wide and intensive maximum in the vicinity of $T = 140\ K$. As one can see from the phase diagram of $Rb_{1-x}(ND_4)_xD_2AsO_4$ (in Fig.9.11) the line on which the point with this temperature has to be located is absent. It seems likely that the nature of this anomaly is not clear to the authors of mentioned studies. The satisfactory explanation for it can be found if to use the concept of coexisting domains of the FE and AFE phases both above and below $T_C$.

In the substances, with the small difference in energies of the FE and AFE phases the domains of the FE and AFE phases coexist in the sample. The dimensions of these domains are determined by a number of factors. The most decisive factor, in our opinion, is the difference of the interplane distances of the crystal lattice in these phases. As noted above, the IDW which separates the FE and AFE phase domains is coherent. The conjugation of the crystallographic axes along these boundaries is accompanied with elastic stresses which increase the total energy of the system. Therefore, the smaller is the difference in the crystal lattice parameters for the coexisting FE and AFE phases, the larger IDW area is allowed and the smaller dimensions of FE and AFE phase domain may be.

Hence it is extremely difficult to identify these states by means of X-ray method or different spectroscopy methods. We as well as other authors encounter this difficulty during investigations of the FE-AFE solid solutions with perovskite structure. Form the physics standpoint such situation may also take place in the system of solid solutions under discussion. In this case the anomaly at the temperature 140 K in $Rb_{0.61}(ND_4)_{0.39}D_2AsO_4$ (see Fig.9.14) may be related to the formation of the complex FE+AFE domains in the paraelectric matrix of the substance (see Ch.6). The dependence of the Brillouin shift on the temperature in this case has to be similar to the one found in [160] and shown in Fig. 9.14.

**CONCLUSION**

In our opinion, all known approaches to investigation of phase states in solids can be subdivided into two extensive groups. To understand an essential difference between them one should refer to Fig.C.1.

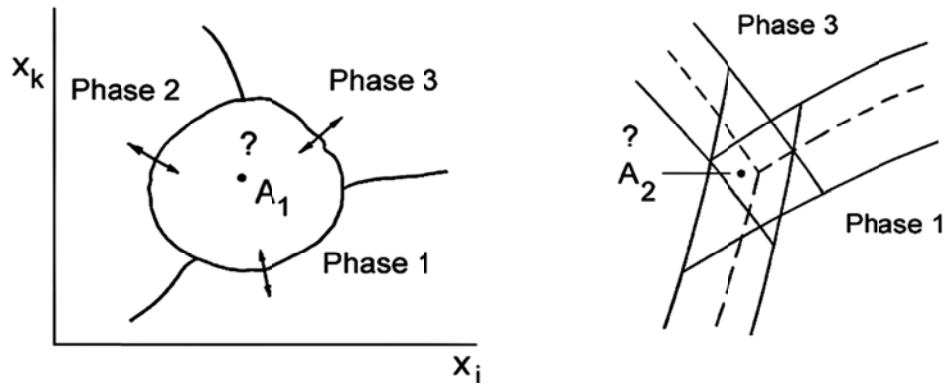

Fig. C.1. Schematic diagram of phase states for many-phases system.

The set of possible states in the ($x_i$, $x_k$) space of thermodynamic variables (that is in the diagram of phase states of the substance) is schematically shown at the left. In particular, such a set of thermodynamic variables comprises the set of the following parameters temperature, hydrostatic pressure or unilaxial mechanical stresses, field, and composition. The situation may exist when a state to be identified borders those already known and the nature of which does not give rise to doubts.

One of these groups of investigations requires experiments to be performed by means of numerous sets of up-to-date equipment at a single point or in a limited space of thermodynamic variables $x_i$. In the simplest way such investigations can be realized under normal conditions (room temperature, atmospheric pressure, and in the absence of external fields).

The investigations for which the nature of an unknown state can be established by studying phase transformations between already identified states and the given unknown one belong to the other group. Certainly, in most cases both types of investigation methods are combined. However, practically always preference is given to the first-group investigations, the second-group ones being auxiliary. Many items of advanced measuring equipment are available nowadays, and researchers seem to believe in their potentials only.

The second-group investigations are more laborious, and not always the samples under study withstand all the experimental conditions: they may be either damaged by high-intensity fields, or be broken while using uniaxial mechanical stresses or when subjected to heating-cooling in the vicinity of phase transformation. However, such studies are most informative and yield more food for the mind.

Let us now assume that the experimental conditions correspond to the point $A_1$ in the space of $x_i$ variables. If such a point lies inside a region of state which that needs to be identified (far from phase boundaries), the first-group investigations and methods are surely efficient and, as a rule, allow to solve the problem stated. But the situation is quite different if under varying external conditions the state of a substance in the phase diagram is near a phase boundary.

The case (most difficult for investigation) when the state of a substance corresponds to the region near the triple point where the lines of phase transitions of the first order converge is presented in the right part of Fig.C.1. Under such conditions, any of the three states may be realized either at the point $A_2$ or in its vicinity due to hysteresis phenomena accompanying this type of phase transitions. Moreover, as a rule, the difference between the energies of the phases is not large within the limits of the hysteresis region, and all the three phases may coexist. In this case the coexisting phases influence each other (the simplest and most investigated case is the appearance of mechanical stresses at different configuration volumes of the phases). Such interactions eliminate clear distinctions between individual phases. What is more, in the vicinity of interphase boundaries the local violations from the stoichiometric composition may arise. The latter fact makes the observed situation entirely tangled. Therefore, while using the first-group investigation methods such composite state of substance cannot be correctly identified, so new terms are devised, in particular, penferroelectricity, quasi-ferroelectricity, relaxor ferroelectrics, etc. However, while studying phase transformations between each two phases one should consider regions located far from the triple point approaching the latter step by step. This will allow to determine a state (or states) which is realized at the point $A_2$ and in its vicinity. Only then model ideas can be formulated.

Results presented in this review demonstrate that variety of physical effects, which accompany the FE-AFE phase transition, are much wider than the class of effects that described in the frames of approach developed in [48-51]. As well as a class of substances in which the transitions between the above-mentioned phases take place is also wider than considered in [48-51].

The essential dispersion of dielectric permittivity in the vicinity of the PE phase transition (and the fulfillment of the Vogel-Fulcher law as a consequence) has to be specially highlighted among other physical phenomena that are discussed in the literature but are not associated with the stability of the FE and AFE phases and phase transitions between them.

As it is demonstrated in this article (see Ch.7 for details) the above mentioned effect is a consequence of specific dynamic behavior of IDWs in the substances that are characterized by a small difference in energies of the FE and AFE phases. The account of "lagging effects" for the IDW boundary dynamics in real substances proves to be sufficient for obtaining a wide set of physical properties for which special terms that we have mentioned before have been devised in the literature. However, introduction of these new terms does not allow to solve any scientific problems. On the contrary, in some cases excessive number of new terms leads to a deadlock in the research itself. As an example and review of similar situations, we would like to mention a remarkable book [165] where such situations typical of physics as a whole, are discussed.

Nevertheless, the relaxation dynamics of IDW is not the most significant consequence of the new approach to studies of the relative stability of FE and AFE states. We have begun the present discussion with this phenomenon only in view of the fact that it has been paid much attention in papers on ferroelectricity published during last decade. In our opinion, the possibility of the presence of the inhomogeneous state of coexisting domains of the FE and AFE phases is the most notable factor which defines all other phenomena. Such inhomogeneous state turns out to be more advantageous from the viewpoint of energy then any homogeneous state. Paradoxical as it sounds, the energy of the interphase boundary separating the domains with the FE and AFE ordering may be negative. Since our university days we have got accustomed to the fact that "this cannot take place at all", and therefore it is very difficult to interpret such experimental results obtained for the first time while investigating one or another substance.

The coexistence of domains of the FE and AFE phases defines the ion segregation along IDWs, specific loops of the dielectric and electrooptical hysteresis. Moreover, the phase coexistence and the interphase interaction define the two-phase nucleation at the temperatures essentially higher than $T_c$ and, as a consequence, the diffuseness of the paraelectric PT.

To establish boundaries, within which the above-mentioned phenomena may manifest themselves, let us again refer to the generalized phase diagram presented in Fig.2.3. In accordance with our results, two substances located in different parts of the same region of the phase diagram, e.g. at points 3 and 4 of the FE region, will have quite different properties. This also applies to the substances that belong to the AFE region of the phase diagram and are located near the points 1 and 2. A small energy difference between the FE and AFE states and the coexistence of domains of these phases in the volume of the sample defines a "special set of properties" for the substances located near the points 2 and 3 of the phase diagram. Strictly speaking the FE-AFE phase transformation itself may not be even revealed in experiments.

In order to build the complete state diagram it is necessary to have the complete set of experimental phase diagrams such as *P-T*, *E-T*, *P-E*, etc. Starting from these phase diagrams and taking into account their evolution caused by changes of the other thermodynamic variables one can determine the order parameters, which describe the corresponding types of ordering, and can build the free energy for each substance. In Ch.7 the evolution of different types of phase diagrams is investigated, and the physical factor which gives rise to the appearance of phase diagrams with a characteristic form is found.

In our opinion, during the past 20-22 years the first-group investigations dominated in FE physics at studying the so-called relaxor ferroelectrics (Fig.D.1 and comments to it). On the base of the results of few measurements attempts were made to formulate model ideas (often contradicting each other). In our opinion, such a situation developed after the appearance of the paper [86] in 1987. As mentioned in [6], PLZT solid solutions had been investigated since 1969. Particular attention was paid to the processes taking place at the PT. However, this period was also characterized by sufficient quantity of the first-group investigations. In the late eighties the problem of PT in PLZT seemed to be close to successful solution, though many inconsistent statements had been declared. In particular, all the ideas discussed in [20, 47, 59, 85] and in the present paper had already been considered in the monograph [6], though the arguments advanced in the said work were not substantiated enough. But the paper [86] completely reversed the picture.

As it is shown both in [20, 47, 59, 85] and the present paper, absolutely all peculiarities in the behavior of substances which fall into the class of relaxor ferroelectrics can be easily explained and interpreted from the viewpoint of physics of the FE-AFE phase transitions. The approach proposed for this purpose is based on the fact that all the mentioned substances are oxygen-containing octahedron compounds. This thesis is complemented by the following one: if one of the possible types of dipole ordering is realized in such substances at lowering the temperature the FE and AFE states have close values of the free energy. Here we would like to emphasize once more that the latter fact is proved in a number of theoretical and experimental studies.

For many years researches considered phase transformations between the FE and AFE phases (or between the said states and the PE one) to proceed according to the model developed in [48-51]. Now it is clear that these papers describe only the simplest particular case. In the general case the FE-AFE transformations have turned out to proceed in another way. Up to now many researches have not yet accustomed to this fact.

In our opinion, traditions existing in each branch of physics influence apprehension of experimental results. Such a situation took place in physics of magnetic phenomena in 1955-1975 while studying phase transitions in MnBi and in alloys based on it. It was shown later for MnBi (in which the ferromagnetic and the antiferromagnetic orderings are possible) that phase *P-T* diagrams are analogous to those for ferroelectric and antiferroelectric substances, which manifest relaxor properties (see Fig.6 in [20], and also Fig.2.2 in Ch.2 of the present paper). Different types of phase diagrams have been constructed and thoroughly investigated during the above-mentioned time period. There have been many difficulties in the interpretation of results that, nonetheless, have been overcome without introduction of new states. In the long run, a genuine scientific result was obtained. It has to be noted that these investigations continue and are developing at present using our ideas about the behavior of systems with a small energy difference between the phases that participate in phase transition. In particular, the processes caused by local violation of stoichiometry in the vicinity of ferromagnetic-antiferromagnetic interphase boundaries have been studied.

In this paper we have repeatedly emphasized that it is necessary to have a wide set of phase diagrams for the substances classified as relaxor ferroelectrics in accordance with [86]. However, such diagrams are absent in the literature devoted to the topic under consideration, though the number of the corresponding papers exceeds that of works on any other problem of ferroelectric physics.

Why are we so concerned about the complete set of the phase diagrams? To answer this question let us dwell on the PMN compound that is most often discussed in the literature. According to a large number of papers, the high-temperature phase of this substance is the PE state. The presence of the FE phase is unambiguously established among its low-temperature states (in particular, it can be induced by electric field). It is clear that for consistent description of the phase transformations in PMN one must take into account at least two order parameters in the expression for the free energy. In the frame of the modern form of the Landau's theory of PTs, the coefficients in the free energy expansion in terms of order parameters are treated as the functions of pressure and temperature. This necessitates availability of "pressure-temperature" experimental phase diagrams. Comparing the experimental diagrams and the calculated model ones it is easy to identify phase states in PMN (or in any other substance) completely and self-consistently. Moreover, it is not difficult to identify phase states of the substance subjected to any sequence of actions of external factors, i.e. to determine peculiarities of behavior of PMN that depend on its prehistory. However, these studies have not been reflected in the literature so far. The examples considered both in [6, 20, 47, 59, 85] and in the present paper show the results which follow from such an approach.

There is one more circumstance that justifies the necessity of obtaining the complete set of experimental phase diagrams for the substances under. At present the only criterion for referring substances to the relaxor ferroelectrics is the existence of the dispersion of permittivity in the region of maximum of the $\varepsilon(T)$-dependence and the presence of the dependence of $T_m$ on the frequency of a measuring field. Thus, it is evident that using such approach, the substances in which the mentioned facts have different nature can be also included to this group. We have considered the substances that manifest the so-called relaxor behavior which is caused by the presence of domains of coexisting FE and AFE phases due to a small energy difference between the FE and AFE states in oxygen-containing octahedral crystalline structures. However, in a number of cases the same behavior, namely, relaxational dynamics, occurs because of the presence of domains of other coexisting phases in the volume of the samples. One of such possible examples is discussed in remarks of [85]. Under these circumstances the nature of the observed phenomena also cannot be identified without the analysis of complete set of phase diagrams.

The problem under discussion has another aspect to be dwelt on. In the Pb(In$_{1/2}$Nb$_{1/2}$)O$_3$-type substances the stability of the FE and AFE phases is connected with the absence or presence of the composition ordering for B-ions. The energy advantage of the domains of coexisting FE and AFE phases in some substances must give rise to the appearance of domains with/without composition ordering. The structure of such domains is stable at the temperatures below $T_c$ and this is caused by the decrease of the free energy of the whole system due to the interaction between domains with the FE and AFE ordering (see eq.(1.13) and (1.16) in Ch.1).

After heating the samples at the temperatures ~ $T_C$ + (100-200)°C the "motive (driving)" force of such a process reduces to zero, and the whole of the substance must become homogeneous with respect to the degree of composition ordering. Under subsequent cooling down to $T < T_c$ the domains with different degree of composition ordering will be formed again.

We would like to emphasize that from the viewpoint of the present-day ideas, the existence of domains which have different degree of the composition ordering is predetermined in the process of annealing and is not connected with the stability of the FE and AFE states.

The authors have been concerned with pure academic topics of studies of the substances called relaxor ferroelectrics. These compounds become increasingly more and more important for practical applications. We are convinced that our approach to understanding the physical nature of phenomena observed in these substances will essentially promote development of new materials (belonging to the said class) with predetermined properties.

The whole set of physical phenomena which manifest themselves in the relaxor ferroelectrics is interpreted on the base of data about FE-AFE transformations both in [6, 20, 47, 59, 85] and the present paper. As mentioned earlier, the authors are not aware of experimental results tan basically contradict this approach. Imaginary differences arising in some cases have been discussed and explained in the frames of these ideas. The authors are eager to collaborate with anybody who can point to discrepancy between experimental results and the statements of developed approach.